\documentclass[lengthcheck,superscriptaddress,showpacs,letterpaper,nofootinbib]{revtex4}
\pdfoutput=1
\usepackage[utf8]{inputenc}
\usepackage{ifpdf}
\usepackage{color}
\usepackage{graphicx}  
\usepackage{float}
\usepackage{amsmath}
\usepackage{acronym}
\usepackage{multirow}
\usepackage{xspace}
\usepackage{units}
\usepackage{aas_macros}
\usepackage{hyperref}
\usepackage{amsfonts}
\usepackage{bm}
\usepackage{braket}
\usepackage{cleveref}

\def\thercsid{\relax}
\def\rcsid#1{\def\next##1#1{\def\thercsid{##1}}\next}
\rcsid$Id: s6pe.tex,v 1.216 2013/03/18 21:12:42 vivien Exp $
\renewcommand{\today}{\number\day\space\ifcase\month\or
  January\or February\or March\or April\or May\or June\or
  July\or August\or September\or October\or November\or December\fi
  \space\number\year}

\newcommand{\li}{\texttt{LALInference}}
\newcommand{\LAL}{\texttt{LAL}}
\newcommand{\Msun}{\ensuremath{\mathrm{M}_\odot}}
\newcommand{\Mc}{\ensuremath{\mathcal{M}}}
\newcommand{\data}{\ensuremath{\bm{d}}}
\newcommand{\h}{\ensuremath{\bm{h}}}
\newcommand{\n}{\ensuremath{\bm{n}}}
\renewcommand{\vec}[1]{\ensuremath{\bm{#1}}}
\newcommand{\pvec}{\ensuremath{\bm{\theta}}}
\newcommand{\pnophi}{\ensuremath{\bm{\Omega}}}
\newcommand{\nvec}{\ensuremath{\bm{\eta}}}
\newcommand{\dL}{\ensuremath{d_L}}
\newcommand{\D}{\ensuremath{\mathrm{d}}} 

\newacro{BNS}{binary neutron star}
\newacro{NSBH}{neutron star -- black hole binary}
\newacro{BBH}{binary black hole}
\newacro{SNR}{signal-to-noise ratio}
\newacro{PDF}{probability density function}
\newacro{PSD}{power spectral density}
\newacro{GW}{gravitational wave}
\newacro{CBC}{compact binary coalescence}

\begin{document}
\title{Parameter estimation for compact binaries with ground-based gravitational-wave
observations using \textsc{LALInference}}

\author{J.~Veitch}\affiliation{School of Physics and Astronomy, University of Birmingham, Birmingham, B15 2TT, UK}\affiliation{Nikhef, Science Park 105, Amsterdam 1098XG, The Netherlands}

\author{V.~Raymond}\affiliation{LIGO, California Institute of Technology, Pasadena, CA 91125, USA}

\author{B.~Farr}\affiliation{Enrico Fermi Institute, University of Chicago, Chicago, IL 60637, USA}\affiliation{Center for Interdisciplinary Exploration and Research in Astrophysics (CIERA) \& Dept. of Physics and Astronomy, 2145 Sheridan Rd, Evanston, IL 60208, USA}

\author{W.~Farr}\affiliation{School of Physics and Astronomy, University of Birmingham, Birmingham, B15 2TT, UK}

\author{P.~Graff}\affiliation{NASA Goddard Space Flight Center, 8800 Greenbelt Rd, Greenbelt, MD 20771, USA}

\author{S.~Vitale}\affiliation{Massachusetts Institute of Technology, 185 Albany St, Cambridge, Massachusetts 02138 USA}

\author{B.~Aylott}\affiliation{School of Physics and Astronomy, University of Birmingham, Birmingham, B15 2TT, UK}

\author{K.~Blackburn}\affiliation{LIGO, California Institute of Technology, Pasadena, CA 91125, USA}

\author{N.~Christensen}\affiliation{Physics and Astronomy, Carleton College, Northfield, MN 55057 USA}

\author{M.~Coughlin}\affiliation{Department of Physics, Harvard University, Cambridge, MA 02138, USA}

\author{W.~Del~Pozzo}\affiliation{School of Physics and Astronomy, University of Birmingham, Birmingham, B15 2TT, UK}

\author{F.~Feroz}\affiliation{Astrophysics Group, Cavendish Laboratory, J.J. Thomson Avenue, Cambridge CB3 0HE, UK}

\author{J.~Gair}\affiliation{Institute of Astronomy, University of Cambridge, Madingley Road, Cambridge, CB3 0HA, UK}

\author{C.-J.~Haster}\affiliation{School of Physics and Astronomy, University of Birmingham, Birmingham, B15 2TT, UK}

\author{V.~Kalogera}\affiliation{Center for Interdisciplinary Exploration and Research in Astrophysics (CIERA) \& Dept. of Physics and Astronomy, 2145 Sheridan Rd, Evanston, IL 60208, USA}

\author{T.~Littenberg}\affiliation{Center for Interdisciplinary Exploration and Research in Astrophysics (CIERA) \& Dept. of Physics and Astronomy, 2145 Sheridan Rd, Evanston, IL 60208, USA}

\author{I.~Mandel}\affiliation{School of Physics and Astronomy, University of Birmingham, Birmingham, B15 2TT, UK}

\author{R.~O'Shaughnessy}\affiliation{University of Wisconsin-Milwaukee, Milwaukee, WI 53201, USA}\affiliation{Rochester Institute of Technology, Rochester, NY 14623, USA}

\author{M.~Pitkin}\affiliation{SUPA, School of Physics and Astronomy, University of Glasgow, University Avenue, Glasgow, G12 8QQ, UK}

\author{C.~Rodriguez}\affiliation{Center for Interdisciplinary Exploration and Research in Astrophysics (CIERA) \& Dept. of Physics and Astronomy, 2145 Sheridan Rd, Evanston, IL 60208, USA}

\author{C.~R\"{o}ver}\affiliation{Max-Planck-Institut f\"{u}r Gravitationsphysik
(Albert-Einstein-Institut), Callinstra{\ss}e~38, 30167~Hannover, Germany}
\affiliation{Department of Medical Statistics, University Medical Center
G\"{o}ttingen, Humboldtallee~32, 37073~G\"{o}ttingen, Germany}

\author{T.~Sidery}\affiliation{School of Physics and Astronomy, University of Birmingham, Birmingham, B15 2TT, UK}

\author{R.~Smith}\affiliation{LIGO, California Institute of Technology, Pasadena, CA 91125, USA}
\author{M.~Van~Der~Sluys}\affiliation{Department of Astrophysics/IMAPP, Radboud University Nijmegen, P.O. Box 9010, 6500 GL Nijmegen, The Netherlands}

\author{A.~Vecchio}\affiliation{School of Physics and Astronomy, University of Birmingham, Birmingham, B15 2TT, UK}

\author{W.~Vousden}\affiliation{School of Physics and Astronomy, University of Birmingham, Birmingham, B15 2TT, UK}

\author{L.~Wade}\affiliation{University of Wisconsin-Milwaukee, Milwaukee, WI 53201, USA}

\date{\today}

\begin{abstract}

The Advanced LIGO and Advanced Virgo \ac{GW} detectors will begin
operation in the coming years, with compact binary coalescence events a likely
source for the first detections. The gravitational waveforms emitted directly
encode information about the sources, including the masses and spins of the
compact objects. Recovering the physical parameters of the sources from the
\ac{GW} observations is a key analysis task.
This work describes the \li\ software library for Bayesian parameter estimation
of compact binary signals, which builds on several previous methods to
provide a well-tested toolkit which has already been used for several studies.

{We show that our implementation is able to correctly recover the parameters of compact binary signals
from simulated data from the advanced \ac{GW} detectors.}
We demonstrate this with a detailed comparison on three compact binary systems:
a \ac{BNS}, a \ac{NSBH} and a \ac{BBH}, where we show a cross-comparison of
results obtained {using three independent sampling algorithms}.
These systems were analysed with non-spinning, aligned spin and generic
spin configurations respectively, showing that consistent results can be obtained
even with the full 15-dimensional parameter space of the generic spin configurations.

We also demonstrate statistically that the Bayesian credible intervals
we recover correspond to frequentist confidence intervals under
correct prior assumptions by analysing a set of 100 signals drawn from the prior.

We discuss the computational cost of these algorithms, and describe the
general and problem-specific sampling techniques we have used to
improve the efficiency of sampling the \ac{CBC} parameter space.

\end{abstract}


\maketitle

\section{Introduction} 

The direct observation of \acp{GW} and the study of relativistic
sources in this new observational window is the focus of a growing effort with
broad impact on astronomy and fundamental physics. The network of \ac{GW} laser interferometers -- LIGO~\cite{Abbott:2007kv},
Virgo~\cite{Acernese:2008} and GEO\ 600~\cite{Grote:2010zz} -- completed science
observations in initial configuration in 2010, setting new upper-limits on a
broad spectrum of \ac{GW} sources. At present, LIGO and Virgo are being upgraded to
their advanced configurations~\cite{AdvLIGO, AdvVirgo}, a new Japanese interferometer,
KAGRA (formerly known as the Large-Scale Gravitational-wave Telescope,
LCGT)~\cite{Kuroda:2011zz} is being built, and plans are underway to relocate
one of the LIGO instruments upgraded to Advanced LIGO sensitivity to a site in
India (LIGO-India)~\cite{2013IJMPD..2241010U}. Advanced LIGO is currently on track to resume science
observations in 2015 with Advanced Virgo following soon
after~\cite{gw_commission_observe}; 
around the turn of the decade 
LIGO-India and KAGRA should also join the network of
ground-based instruments.

Along with other possible sources, advanced ground-based interferometers are expected to detect \acp{GW} generated during the last seconds to minutes of life of extra-galactic
compact binary systems, with neutron star and/or black hole component masses in
the range $\sim 1\,\Msun - 100\,\Msun$. The current uncertainties on some of the
key physical processes that affect binary formation and evolution are reflected
in the expected detection rate, which spans three orders of magnitude.
However, by the time interferometers operate at design sensitivity, between one observation per few years and hundreds of observations per year are anticipated~\cite{gw_commission_observe, ratesdoc}, opening new avenues for studies of compact objects in highly relativistic conditions.

During the approximately ten years of operation of the ground-based \ac{GW}
interferometer network, analysis development efforts for binary coalescences 
have been focused on the
detection problem, and rightly so: how to unambiguously identify a binary
coalescence in the otherwise overwhelming instrumental noise. The most sensitive compact binary searches are based on matched-filtering techniques, and are designed to keep up
with the data rate and promptly identify detection candidates~\cite{ihope,
Cannon2011Early}. A confirmation of the performance of detection pipelines 
has been provided by the ``blind
injection challenge" in which a synthetic compact binary coalescence signal  was
added (unknown to the analysis teams) to the data stream and successfully
 detected~\cite{Colaboration:2011np}. 
 
Once a detection candidate
has been isolated, the next step of the analysis sequence is to extract full
information regarding the source parameters and the underlying physics. With the
expected detection of 
\acp{GW} in the coming years, this part of the analysis has become the focus of a
growing number of studies.

The conceptual approach to inference on the \ac{GW} signal is deeply rooted in the
Bayesian framework.  This framework makes it possible to evaluate the marginalized posterior
probability density functions (PDFs) of the unknown parameters that describe a given model
of the data and to compute the so-called evidence of the model itself. It is well known
that Bayesian inference is computationally costly, making the efficiency of the PDF and evidence calculations an important issue.   For the case of coalescing binary
systems the challenge comes from many fronts: the large number of unknown
parameters that describe a model (15 parameters to describe a gravitational
waveform emitted by a binary consisting of two point masses in a circular orbit assuming that general relativity is accurate, plus other model
parameters to account for matter effects in the case of neutron stars, the
noise, instrument calibration, etc.), complex multi-modal likelihood functions,
and the computationally intensive process of generating waveforms. 

Well known stochastic sampling
techniques -- Markov chain Monte Carlo~\cite{Christensen:2001cr,MCMC:2004,2006CQGra..23.4895R,
2007PhRvD..75f2004R, 2008ApJ...688L..61V, 2008CQGra..25r4011V,
2009CQGra..26k4007R, 
2009CQGra..26t4010V, 2010CQGra..27k4009R}, Nested Sampling~\cite{Skilling:2006,
Veitch:2010} and {\sc
MultiNest}/BAMBI~\cite{Feroz:2008,Feroz:2009,Feroz:2013, Graff:2012}  -- have
been used in recent years to develop algorithms for Bayesian inference on GW
data aimed at studies of coalescing binaries. An underlying theme of this work
has been the comparison of these sampling techniques and the cross-validation of results with independent algorithms.  In parallel, the inference effort has benefited from a number of
advances in other areas that are essential to maximise the science exploitation
of detected \ac{GW} signals, such as waveform generation and standardised
algorithms and libraries for the access and manipulation of \ac{GW} data. The initially
independent developments have therefore progressively converged towards
dedicated algorithms and a common infrastructure for Bayesian inference applied
to \ac{GW} observations, specifically for coalescing 
binaries but applicable to other sources.  These algorithms and infrastructure are now contained in a dedicated software package: \texttt{LALInference}.

The goal of this paper is to describe \texttt{LALInference}.
We will cover the details of our implementation, designed to
overcome the problems faced in performing Bayesian inference for \ac{GW}
observations of \ac{CBC} signals.
This includes three independent sampling techniques which were cross-compared
to provide confidence in the results that we obtain for \ac{CBC} signals,
and compared with known analytical probability distributions.
We describe the post-processing steps involved in converting the output of
these algorithms to meaningful physical statements about the source parameters
in terms of credible intervals. We demonstrate that these intervals are well-calibrated
measures of probability through a Monte Carlo simulation, wherein we confirm
the quoted probability corresponds to frequency under correct prior assumptions.
We compare the computational efficiency of the different methods and mention
further enhancements that will be required to take full advantage of the advanced
\ac{GW} detectors.

The \texttt{LALInference} software consists of a C library and several
post-processing tools written in python. It leverages the existing
LSC Algorithm Library (LAL) to provide
\begin{itemize}
\item Standard methods of accessing \ac{GW} detector data, using LAL methods
        for estimating the \ac{PSD}, and able to simulate stationary Gaussian noise with a
        given noise curve.
\item the ability to use all the waveform approximants included in
LAL that describe the evolution of point-mass binary systems, and waveforms
under development to account for matter effects in the evolution of binary
neutron stars and generalisations of waveforms beyond general relativity;
\item Likelihood functions for the data observed by a network of ground-based laser
interferometers given a waveform model and a set of model parameters;
\item Three independent stochastic sampling techniques of the parameter space to
compute PDFs and evidence;
\item Dedicated ``jump proposals'' to efficiently select samples in parameter space that take
into account the specific structure of the likelihood function;
\item Standard post-processing tools to generate probability credible regions for any set
of parameters.
\end{itemize}

During the several years of development, initial implementations of these
Bayesian inference algorithms and \texttt{LALInference} have been successfully
applied to a variety of problems, such as the impact of different network
configurations on parameter estimation~\cite{2012PhRvD..85j4045V}, the ability
to measure masses and spins of 
compact objects~\cite{RodriguezEtAl:2013,2008ApJ...688L..61V,PhysRevLett.112.251101}, to
reconstruct the sky location of a detected \ac{GW} binary~\cite{2009CQGra..26k4007R,
Blackburn:2013ina, Grover:2013sha} and the equation of state of neutron
stars~\cite{DelPozzo:2013ala}, the effects of calibration errors on information
extraction~\cite{2012PhRvD..85f4034V} and tests of general
relativity~\cite{2011PhRvD..83h2002D, 2012PhRvD..85h2003L, Agathos:2013upa}.
Most notably \texttt{LALInference} has been at the heart of the study of
detection candidates, including the blind injection, during the last LIGO/Virgo
science run~\cite{2013PhRvD..88f2001A}, and has been used for the Numerical
INJection Analysis project NINJA2~\cite{Aasi:2014tra}.
It has been designed to be flexible in the choice of signal model, allowing it
to be adapted for analysis of signals other than compact binaries, including
searches for continuous waves~\cite{Pitkin:2012yg}, and comparison of
core-collapse supernova models based on~\cite{Logue:2012zw}.

The paper is organised as follows: Section~\ref{sec:bayesian_analysis} provides
a summary of the key concepts of Bayesian inference, and specific discussion
about the many waveform models that can be used in the analysis and the relevant
prior assumptions. In Section~\ref{sec:algorithms} we describe the conceptual
elements concerning the general features of the sampling techniques that are
part of \texttt{LALInference}: Markov chain Monte Carlo, Nested Sampling and
{\sc MultiNest}/BAMBI. Section~\ref{sec:post_processing} deals with the problem of 
providing marginalized probability functions and (minimum) credible
regions at a given confidence level from a finite number of samples, as is
the case of the outputs of these algorithms. In Section~\ref{sec:validation} we
summarise the results from extensive tests and validations that we have carried
out by presenting representative results on a set of injections in typical
regions of the parameter space, as well as results obtained by running the 
algorithms on 
known distributions. This section is complemented by
Section~\ref{sec:performance} in which we consider efficiency issues, and we
report the run-time necessary for the analysis of coalescing binaries in
different cases; this provides a direct measure of the latency timescale over
which fully coherent Bayesian inference results for all the
source parameters will be available after a detection candidate is identified. Section~\ref{sec:conclusions} contains our
conclusions and pointers to future work.

\section{Bayesian Analysis} %
\label{sec:bayesian_analysis}

We can divide the task of performing inference about the \ac{GW} source into two
problems: using the observed data to constrain or estimate the unknown
parameters
of the source~\footnote{The whole set of unknown parameters of the model can
also contain parameters not 
related to the source, such as noise and calibration parameters~\cite{RoeverMeyerChristensen2011,Littenberg:2013,Cornish:2014kda,Littenberg:2014oda}. } under a fixed model of the waveform (parameter estimation), and deciding which
of several models is {more probable} in light of the observed data, and by how
much (model selection).
We tackle both these problems within the formalism of Bayesian inference, which
describes the state of
knowledge about an uncertain hypothesis $H$ as a probability, denoted
$P(H)\in[0,1]$, or about an unknown
parameter as a probability density, denoted $p(\theta|H)$, where $\int
p(\theta|H)\D\theta=1$.
Parameter estimation can then be performed using Bayes' theorem, where a prior
probability distribution $p(\theta|H)$
is updated upon receiving the new data $d$ from the experiment to give a
posterior distribution $p(\theta|d,H)$,
\begin{align}
p(\theta|d,H)=\frac{p(\theta|H)p(d|\theta,H)}{p(d|H)}.\label{eq:Bayes}
\end{align}
Models typically have many parameters, which we collectively
indicate with $\pvec=\{\theta_1,\theta_2,\ldots,\theta_N\}$.
The joint probability distribution on the multi-dimensional space
$p(\pvec|d,H)$ describes the collective knowledge about all parameters as well as
their relationships.
Results for a specific parameter are found by marginalising over the unwanted
parameters,
\begin{align}
p(\theta_1|d,H)=\int \D\theta_2\ldots \D\theta_N p(\pvec|d,H)\,.
\end{align}
The probability distribution can be used to find the expectation of various
functions
given the distribution, e.g. the mean
\begin{align}
\left<\theta_i\right>=\int \theta_i
p(\theta_i|d,H)\D\theta_i\,,\label{eq:expectations}
\end{align}
and credible regions, an interval in parameter space that containing a given
probability (see Section ~\ref{sec:post_processing}).

Model selection is performed by comparing the fully marginalized likelihood, or
`evidence',
for different models. The evidence, usually denoted $Z$, is simply the integral
of the likelihood, $L(d|\pvec)=p(d|\pvec,H)$, multiplied by the prior over all
parameters
of the model $H$,
\begin{align}
\label{eq:evidence}
Z=p(d|H)=\int  \D\theta_1\ldots \D\theta_N~p(d|\pvec,H)p(\pvec|H).
\end{align}
This is the normalisation constant that appears in the denominator of
Eq.~\eqref{eq:Bayes} for a particular model.
Because we cannot exhaustively enumerate the set of exclusive models describing
the data, we typically
compare two competing models.
To do this, one computes the ratio of posterior probabilities
\begin{align}
        O_{ij}=\frac{P(H_i|d)}{P(H_j|d)}=\frac{P(H_i)}{P(H_j)}\times \frac{Z_i}{Z_j}
\end{align}
where $B_{ij}=Z_i / Z_j$ is the `Bayes Factor' between the two competing models
$i$ and $j$, which shows
how much more likely the observed data $d$ is under model $i$ rather than model
$j$.

While the Bayesian methods described above are conceptually simple, the
practical details of performing an analysis depend greatly
on the complexity and dimensionality of the model, and the amount of data that
is analysed. The size of the parameter space and the amount of data to be
considered mean that the resulting probability distribution cannot tractably be
analysed through a fixed sampling of the parameter space. Instead, we have
developed methods for stochastically sampling the parameter space to solve
the problems of parameter estimation and model selection, based on the Markov
chain Monte Carlo (MCMC) and Nested Sampling techniques, the details of which are
described in section \ref{sec:algorithms}.
Next we will describe the models used for the noise and the signal.

\subsection{Data model}\label{sec:data}
The data obtained from the detector is modelled as the sum of the compact
binary coalescence signal $\h$ (described in section \ref{sec:waveforms}) and
a noise component $\n$, 
\begin{align}
\data = \h + \n.
\label{eq:data}
\end{align}
Data from multiple detectors in the network are analysed coherently, by calculating the strain
that would be observed in each detector:
\begin{align}
\h = F_+(\alpha, \delta, \psi) \h_+ + F_\times(\alpha, \delta, \psi) \h_\times
\label{eq:strain}
\end{align}
where $\h_{+,\times}$ are the two independent \ac{GW} polarisation amplitudes and
$F_{+,\times}(\alpha, \delta, \psi)$ are the antenna response functions (\cite[e.g.][]{Anderson:2000yy}) that depend
on the source location and the polarisation of the waves.
Presently we ignore the time dependence of the antenna response
function due to the rotation of the Earth, instead assuming that it is
constant throughout the observation period. This is justifiable for the short
signals considered here. Work is ongoing to include this time
dependence when analysing very long signals with a low frequency cutoff below $40$\,Hz,
to fully exploit the advanced detector design sensitivity curves.
The waveforms $\h_{+,\times}$ are described in Section~\ref{sec:waveforms}. 

As well as the signal model, which is discussed in the next section, we must include a
description of the observed data, including the noise, which is used to create
the likelihood function. This is where knowledge of the detectors' sensitivity
and the data processing procedures are folded into the analysis.

We perform all of our analyses using the calibrated strain output of the \ac{GW}
detectors, or a simulation thereof.
This is a set of time-domain samples $d_i$ sampled uniformly at times $t_i$, which we
sometimes write as a vector $\data$ for convenience below.
To reduce the volume of data, we down-sample the data from its original sampling
frequency ($16384$\,Hz) to a lower rate $f_s \ge 2 f_{\rm max}$, which is high
enough to contain the maximum frequency $f_{\rm max}$ of the lowest mass signal
allowed by the prior, typically $f_s=4096$\,Hz when analysing the inspiral part
of a \ac{BNS} signal. To prevent aliasing the data is first low-pass filtered with a 20th
order Butterworth filter with attenuation of 0.1 at the new Nyquist frequency,
using the implementation in \LAL~\cite{LAL}, which preserves the phase of the
input.
We wish to create a model of the data that can be used to perform the analysis.
In the absence of a signal, the simplest model which we consider is that of
Gaussian, stationary noise with a certain power spectral density $S_n(f)$ and
zero mean. 
$S_n(f)$ can be estimated using the data adjacent to the segment of interest,
which is normally selected based on the time of coalescence $t_c$ of a candidate
signal identified by a search pipeline.
The analysis segment $\data$ spans the period $[t_c-T+2,t_c+2]$, i.e. a time $T$
which ends two seconds after the trigger
(the 2\,s safety margin after $t_c$ allows for inaccuracies in the trigger time
reported by the search, and should encompass
any merger and ringdown component of the signal).
To obtain this estimate, by default we select a period of time (1024\,s
normally, but shorter if less science data is available) from before the time of
the trigger to be analysed, but ending not later than $t_c-T$, so it should not
contain the signal of interest.
This period is divided into non-overlapping segments of the same duration $T$ as
the analysis segment, which are then used to estimate the PSD.
Each segment is windowed using a Tukey window with a 0.4\,s roll-off, and its
one-sided noise power spectrum is computed. For each frequency bin the median
power over all segments is used as an estimate of the PSD in that bin. We follow
the technique of \cite{Allen:2005fk} by using the median instead of the mean to
provide some level of robustness against large outliers occurring during the
estimation time.

The same procedure for the PSD estimation segments is applied to the analysed
data segment before it is used for inference, to ensure consistency.

For each detector we assume the noise is stationary, and characterised only by
having zero mean and a known variance (estimated from the power spectrum).  Then 
the likelihood function for the noise
model is simply the product of Gaussian distributions in each frequency bin
\begin{align}
        p(\data|H_N,S_n(f)) = \exp \sum_i \left[
-\frac{2|\tilde{d}_i|^2}{TS_n(f_i)} - \frac{1}{2}\log(\pi T S_n(f_i) /2) \right],
\end{align}
where $\tilde{\data}$ is the discrete Fourier transform of $\data$
\begin{align}
        \tilde{d}_j=\frac{T}{N}\sum_k d_k \exp (-2\pi ijk/N).\label{eq:DFT}
\end{align}
The presence of an additive signal $\h$ in the data simply adjusts the mean
value of the distribution, so that the likelihood including the signal is
\begin{align}
        p(\data|H_S,S_n(f),\pvec) &= \exp\sum_i \left[
-\frac{2|\tilde{h}_i(\pvec)-\tilde{d}_i|^2}{TS_n(f_i)}\right.\nonumber\\
 &- \left.\frac{1}{2}\log(\pi T S_n(f_i) /2) \right].\label{eq:L}
\end{align}
To analyse a network of detectors coherently, we make the further assumption
that the noise is uncorrelated in each. This allows us to write the coherent
network likelihood for data obtained from each detector as the product of the
likelihoods in each detector~\cite{Finn:1996}.
\begin{align}\label{eq:LikelihoodAllDetectors}
        p(\data_{\{H,L,V\}}|H_S,{S_n}_{\{H,L,V\}}(f))=\prod_{i\in\{H,L,V\}}p(\data_i|H_S,{S_n}_i(f))
\end{align}
This gives us the default likelihood function which is used for our analyses,
and has been used extensively in previous work.

\subsubsection{Marginalising over uncertainty in the PSD estimation}
Using a fixed estimate of the PSD, taken from times outside the segment being
analysed, cannot account for slow variations in the shape of the spectrum over
timescales of minutes. We can model our uncertainty in the PSD estimate by
introducing extra parameters into the noise model which can be estimated
along with the signal parameters; we follow the procedure described in \cite{Littenberg:2013}.
We divide the Fourier domain data into
$\sim 8$ logarithmically spaced segments, and in each segment $j$, spanning $N_j$ frequency bins,
introduce a
scale parameter $\eta_j(f_i)$ which modifies the PSD such that $S_n(f_i)\rightarrow
S_n(f_i)\eta_j $ for $i_j < i \leq i_{j+1}$, where the scale parameter is constant within a frequency segment.
With these additional degrees of freedom included in our model, the likelihood becomes
\begin{align}
        p(\data|H_S,S_n(f),\pvec,\nvec) &= \exp\sum_i\left[
        -\frac{2|\tilde{h}_i(\pvec)-\tilde{d}_i|^2}{T\eta(f_i)S_n(f_i)}\right.\nonumber\\
& \left.- \frac{1}{2}\log(\pi \eta_i T S_n(f_i) /2)\right].
\end{align}
The prior on $\eta_j$ is a normal distribution with mean 1 and
variance $1/N_j$.  In the limit $N_j\rightarrow 1$
(i.e., there is one scale parameter for each Fourier bin), replacing the
Gaussian prior with an inverse chi-squared distribution and integrating
$p(d|H_S,S_n(f),\pvec,\nvec)\times p(\pvec,\nvec|H_S,S_n(f))$ over $\nvec$, we
would recover the Student's t-distribution likelihood considered for \ac{GW} data
analysis in \cite{RoeverMeyerChristensen2011,Roever:2011}.  For a thorough discussion of the relative merits
of Student's t-distribution likelihood and the approach used here, as
well as examples which show how including $\nvec$ in the model improves the
robustness of parameter estimation and model selection results,
see~\cite{Littenberg:2013}.  In summary, the likelihood adopted here offers more
flexibility given how much the noise can drift between the data used for
estimating the PSD and the data being analysed. Further improvements
on this scheme using more sophisticated noise models are under active development.

\subsection{Waveform models}\label{sec:waveforms} 

There are a number of different models for the \ac{GW} signal that is expected to be
emitted during a compact-binary merger.  These models, known as waveform
families, differ in their computational complexity, the physics they simulate,
and their regime of applicability.  \li~  has been designed to easily interface
with arbitrary waveform families.  

Each waveform family can be thought of as a function that takes as input a
parameter vector $\pvec$ and produces as output $\h_{+,\times}(\pvec)$, either a
time domain $h(\pvec;t)$ or frequency-domain $h(\pvec;f)$ signal.  The parameter
vector $\pvec$ generally includes at least nine parameters:
\begin{itemize}
\item Component masses $m_1$ and $m_2$.  We use a reparametrisation of
        the mass plane into the chirp mass,
\begin{align}
\Mc = (m_1 m_2)^{3/5} (m_1+m_2)^{-1/5}
\label{e:mchirp}
\end{align}
and the asymmetric mass ratio
\begin{align}
q= m_2/m_1,
\label{e:q}
\end{align}
as these variables tend to be less correlated and easier to sample.
We use the convention $m_1 \geq m_2$ when labelling the components.
The prior is transformed accordingly (see figure \ref{fig:m1m2Prior}).
Another possible parametrisation is the symmetric mass ratio 
\begin{align}
\eta= \frac{(m_1 m_2)}{(m_1+m_2)^2}
\label{e:eta}
\end{align}
although we do not use this when sampling the distribution since the Jacobian of the transformation
to $m_1,m_2$ coordinates becomes singular at $m_1=m_2$.
\item The luminosity distance to the source $d_L$;
\item The right ascension $\alpha$ and declination  $\delta$ of the source;   
\item The inclination angle $\iota$, between the system's orbital angular
        momentum and the line of sight. For aligned- and non-spinning systems
        this coincides with the angle $\theta_{JN}$ between the total angular
        momentum and the line of sight (see below). We will use the more general
        $\theta_{JN}$ throughout the text.
\item The polarisation angle $\psi$ which describes the
        orientation of the projection of the binary's orbital momentum vector onto the plane on the sky, as defined in \cite{Anderson:2000yy};
\item An arbitrary reference time $t_c$, e.g. the time of coalescence
of the binary;
\item The orbital phase $\phi_c$ of the binary at the reference time $t_c$.
\end{itemize}
Nine parameters are necessary to describe a circular binary consisting of
point-mass objects with no spins. If spins of the binary's components are
included in the model, they are described by six additional parameters, for a
total of 15: 
\begin{itemize}
\item dimensionless spin magnitudes $a_i$, defined as 
$a_i \equiv {|\vec{s_i}|}/{m_i^2}$
and in the range $[0,1]$, where $\vec{s_i}$ is the spin vector of the object
$i$, and 
\item two angles for each $\vec{s_i}$ specifying its orientation with respect to
the plane defined by the line of sight and the initial orbital angular momentum.
\end{itemize}
In the special case when spin vectors are assumed to be aligned or anti-aligned
with the orbital angular momentum, the four spin-orientation angles are fixed,
and the spin magnitudes alone are used, with positive (negative) signs
corresponding to aligned (anti-aligned) configurations, for a total of 11
parameters.  
In the case of precessing waveforms, the \textit{system-frame} parametrisation
has been found to be more efficient than the radiation frame typically employed
for parameter estimation of precessing binaries.  The orientation of the system
and its spinning components are parameterised in a more physically intuitive way
that concisely describes the relevant physics, and defines evolving quantities
at a reference frequency of $100$ Hz, near the peak sensitivity of the
detectors~\cite{Farr:2014qka}:
\begin{itemize}
\item $\theta_{JN}$: The inclination of the system's total angular momentum with
respect to the line of sight;
\item $t_1,t_2$: Tilt angles between the compact objects' spins
and the orbital angular momentum;
\item $\phi_{12}$: The complimentary azimuthal angle separating the spin
vectors;
\item $\phi_{\rm JL}$: The azimuthal position of the orbital angular momentum on
its cone of precession about the total angular momentum.
\end{itemize}
Additional parameters are necessary to fully describe matter effects in
systems involving a neutron star, namely the equation of state~\cite{2013PhRvL.111g1101D},
or to model deviations from the post-Newtonian expansion of the waveforms~\cite[e.g.][]{YunesPretorius:2009,2012PhRvD..85h2003L},
but we do not consider these here.
Finally, additional parameters could be used to describe waveforms from
eccentric binaries~\cite{Konigsdorffer:2006zt} but these have not yet
been included in our models.

GWs emitted over the whole coalescence of two compact objects produce a
characteristic ``chirp" of increasing amplitude and frequency during the
adiabatic inspiral phase, followed by a broad-band merger phase and then damped
quasi-sinusoidal signals during the ringdown phase. 
The characteristic time and frequency scales of the whole
inspiral-merger-ringdown are important in choosing the appropriate length of the
data segment to analyse and the bandwidth necessary to capture the whole
radiation.  At the leading Newtonian quadrupole order, the time to coalescence
of a binary emitting GWs at frequency $f$ is~\cite{Allen:2005fk}:
\begin{align}
\tau = 93.9
\left(\frac{f}{30\,\mathrm{Hz}}\right)^{-8/3}\,\left(\frac{\Mc}{0.87\,M_\odot}
\right)^{-5/3}\,\mathrm{sec}\,.
\end{align}
Here we have normalised the quantities to an $m_1 = m_2 = 1\,M_\odot$ equal mass
binary. 
The frequency of dominant mode gravitational wave emission at the innermost stable circular orbit
for a binary with non-spinning components is~\cite{Allen:2005fk}:
\begin{align}
f_\mathrm{isco} & = \frac{1}{6^{3/2} \pi (m_1 + m_2)} = 4.4
\left(\frac{M_\odot}{m_1 + m_2}\right) \,\mathrm{kHz}\,,
\end{align}

The low-frequency cut-off of the instrument, which sets the duration of
the signal, was 40 Hz for LIGO in initial/enhanced configuration and 30 Hz for
Virgo. When the instruments operate in advanced configuration, new
suspension systems are expected to provide increased low-frequency sensitivity
and the low-frequency bound will progressively move towards $\approx 20$ Hz. The
quantities above define therefore the longest signals that one needs to consider
and the highest frequency cut-off.  The data analysed (the `analysed
segment') must include the entire length of the waveform from the desired
starting frequency.

Although any waveform model that is included in the LAL libraries can be readily
used in \li, the most common waveform models used in our previous studies
\cite[e.g.,][]{S6pepaper} are:
\begin{itemize}
\item Frequency-domain stationary phase inspiral-only
post-Newtonian waveforms for binaries with non-spinning components, particularly
the TaylorF2 approximant \cite{Buonanno:2009zt};
\item Time-domain inspiral-only post-Newtonian waveforms that allow for
components with arbitrary, precessing spins, particularly the SpinTaylorT4
approximant \cite{BuonannoChenVallisneri:2003b};
\item Frequency-domain inspiral-merger-ringdown phenomenological waveform model
calibrated to numerical relativity, IMRPhenomB, which describes systems with
(anti)aligned spins \cite{Ajith:2009bn};
\item Time-domain inspiral-merger-ringdown effective-one-body model calibrated
to numerical relativity, EOBNRv2 \cite{Pan:2011gk}.
\end{itemize}
Many of these waveform models have additional options, such as varying the
post-Newtonian order of amplitude or phase terms.  Furthermore, when exploring
the parameter space with waveforms that allow for spins, we sometimes find it
useful to set one or both component spins to zero, or limit the degrees of
freedom by only considering spins aligned with the orbital angular momentum.

We generally carry out likelihood computations in the frequency domain, so
time-domain waveforms must be converted into the frequency domain by the
discrete Fourier transform defined as in eq.~\eqref{eq:DFT}.
To avoid edge effects and ensure that the templates and data are treated
identically (see Section \ref{sec:data}), we align the end of the time-domain waveform to
the discrete time sample which is closest to $t_c$ and then taper it in the same way as the
data (if the waveform is non-zero in the first or last $0.4$s of the buffer),
before Fourier-transforming to the frequency domain and applying any finer
time-shifting in the frequency domain, as described below.

Some of the parameters, which we call intrinsic parameters (masses and spins),
influence the evolution of the binary.  Evaluating a waveform at new values of
these parameters generally requires recomputing the waveform, which, depending
on the model, may involve purely analytical calculations or a solution to a
system of differential equations.  On the other hand, extrinsic parameters (sky
location, distance, time and phase) leave the basic waveform unchanged, while
only changing the detector response functions $F_+$ and $F_\times$ and shifting
the relative phase of the signal as observed in the detectors.
This allows us to save computational costs
in a situation where we have already computed the waveform and are now
interested in its re-projection and/or phase or time shift; in particular, this
allows us to compute the waveform only once for an entire detector network, and
merely change the projection of the waveform onto detectors.  We typically do
this in the frequency domain.

The dependence of the waveform on distance (scaling as $1/d_L$), sky location
and polarisation (detector response described by antenna pattern functions
$F_{+,\times}(\alpha,\delta,\psi)$ for the $+$ and $\times$ polarisations, see eq.~\eqref{eq:strain}) and phase ($\tilde{h}(\phi_c)=\tilde{h}(\phi=0)
e^{i\phi_c}$) is straightforward.  A time shift by $\Delta t$ corresponds to a multiplication 
$\tilde{h}(\Delta t)=\tilde{h}(0) e^{2\pi i f \Delta t}$ in the frequency domain;
this time shift will be different for each detector, since the arrival time of a
GW at the detector depends on the location of the source on the sky and the
location of the detector on Earth.

The choice of parameterization greatly influences the efficiency of posterior
sampling.  The most efficient parameterizations minimize the correlations
between parameters and the number of isolated modes of the posterior.  For the
mass parameterization, the chirp mass $\mathcal{M}$ and asymmetric mass ratio
$q$ achieve this, while avoiding the divergence of the Jacobian of the symmetric
mass ratio $\eta$ at equal masses when using a prior flat in component masses. 
With generically oriented spins comes precession, and the evolution of angular
momentum orientations.  In this case the structure of the posterior is
simplified by specifying these parameters, chosen so that they evolve as little as possible, at a reference frequency of $100$ Hz near the peak sensitivity of the detector \cite{Farr:2014qka}.

\subsubsection{Analytic marginalisation over phase}
The overall phase $\phi_c$ of the \ac{GW} is typically of no astrophysical interest,
but is necessary to fully describe the signal. When the signal model includes
only the fundamental
mode ($l=m=2$) of the \ac{GW} it is possible to analytically marginalize over
$\phi_c$,
simplifying the task of the inference algorithms in two ways. Firstly, the
elimination of one
dimension makes the parameter space easier to explore; secondly the marginalized
likelihood function over the remaining parameters has a lower dynamic range than the original likelihood.
The desired likelihood function over the remaining parameters $\pnophi$ is
calculated by marginalising Eq.~\eqref{eq:L},
\begin{align}
        p(\data|H_S,S_n(f),\pnophi)=\int p(\phi_c|H_S) p(\data|\pvec,H_S,S_n(f))
\D\phi_c\label{eq:LmargOverPhi}
\end{align}
where $p(\phi_c|H_S)=1/2\pi$ is the uniform prior on phase.

Starting from Eq.~\ref{eq:LikelihoodAllDetectors}
we can write the likelihood for multiple detectors indexed $j$ as
\begin{align}
        p(\data_{j}|H_S,{S_n}_j(f),\pvec)&\propto\exp\left[-\frac{2}{T}\sum_{i,j}\frac{|\tilde{h_0}_{ij}|^2+|d_{ij}|^2}{{S_n}_j(f_i)}\right]\nonumber\\
                                         & \times \exp\left[\frac{4}{T}\Re\left(\sum_{i,j}\frac{\tilde{h_0}_{ij}e^{i\phi_c}d^*_{ij}}{{S_n}_{j}(f_i)}\right)\right]
\end{align}
where $\h_0$ is the signal defined at a reference phase of $0$.
Using this definition, the integral of Eq.~\eqref{eq:LmargOverPhi} can be cast
into a standard form to yield
\begin{align}
        p&(\data_j|H_S,{S_n}_j(f),\pnophi)=\nonumber\\
        \exp&\left[-\frac{2}{T}\sum_{i,j}\frac{|\tilde{h_0}_{ij}|^2+|d_{ij}|^2}{{S_n}_j(f_i)}\right]\mathrm{I}_0\left[\frac{4}{T}\left|\sum_{i,j}\frac{\tilde{h_0}_{ij}d^*_{ij}}{{S_n}_j(f_i)}\right|\right]
\label{eq:Lmargphi}
\end{align}
in terms of the modified Bessel function of the first kind $\mathrm{I}_0$.
Note that the marginalised likelihood is no longer expressible as the product of likelihoods in each
detector.
We found that using the marginalized phase likelihood could reduce the
computation time of a nested sampling analysis by a factor of up to $4$,
as the shape of the distribution was easier to sample, reducing the
autocorrelation time of the chains.

\subsection{Priors}\label{sec:prior} 

\begin{figure*}
\centering
\includegraphics[width=\columnwidth]{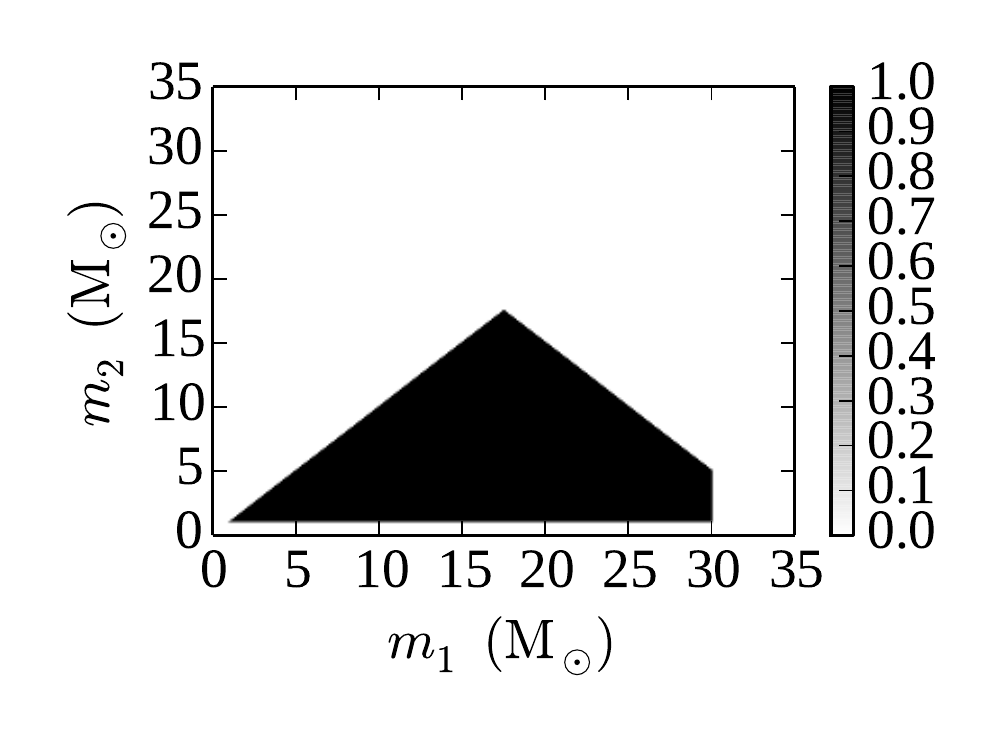}
\includegraphics[width=\columnwidth]{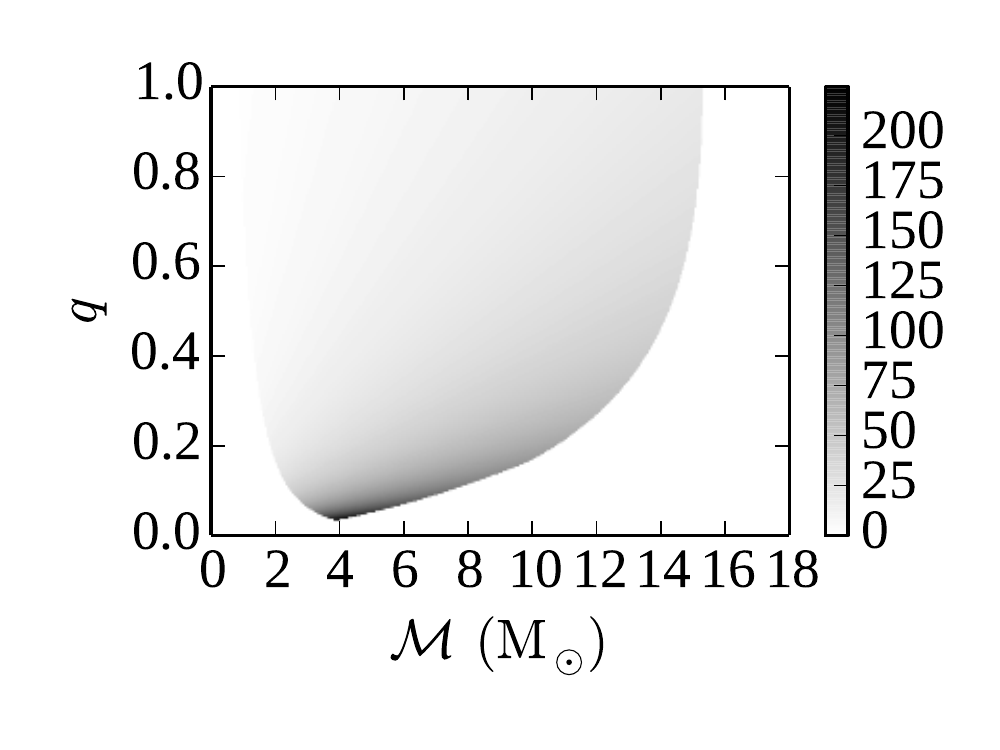}
\caption{\label{fig:m1m2Prior} Prior probability $p(m_1,m_2|H_S)$, uniform in
component masses within the bounds shown (left), and the same distribution
transformed into the $\Mc$,$q$ parametrization used for sampling.}
\end{figure*}

As shown in~Eq.~\eqref{eq:Bayes}, the posterior distribution of $\theta$ (or
$\pvec$) depends both on the likelihood and prior distributions of $\theta$. \li~allows for flexibility in the choice of priors.  For all analyses described here, we
used the same prior density functions (and range).  For component masses, we used uniform priors
in the
component masses with the range $1\,\Msun\le m_{1,2}\le 30\,\Msun$, and with the
total mass constrained by $m_1 + m_2 \le 35\,\Msun$, as shown in Fig.
\ref{fig:m1m2Prior}. This range encompasses the low-mass search range used in
\cite{Colaboration:2011np} and our previous parameter estimation report
\cite{S6pepaper}, where $1\,\Msun\le m_{1,2}\le 24\,\Msun$ and $m_1 + m_2 \le
25\,\Msun$. When expressed in the sampling variable $\Mc,q$ the prior
is determined by the Jacobian of the transformation,
\begin{align}
        p(\Mc,q|I)\propto \Mc m_1^{-2}
\end{align}
which is shown in the right panel of figure \ref{fig:m1m2Prior}.

The prior density function on the location of the source was taken to be
isotropically distributed on the sphere of the sky, with $p(d_L|H_S)\propto
{d_L}^2$, from $1$\,Mpc out to a maximum distance
chosen according to the detector configuration and the source type of interest.
We used an isotropic prior on the orientation of the binary
to give $p(\iota,\psi,\phi_c|H_S)\propto\sin \iota$.
For analyses using waveform models that account for possible spins, the prior on
the spin magnitudes, $a_1,a_2$, was taken to be uniform in the range $[0,1]$
(range $[-1,1]$ in the spin-aligned cases), and the spin
angular momentum vectors were taken to be isotropic.

The computational cost of the parameter estimation pipeline precludes us from
running it on all data; therefore, the parameter estimation analysis relies on
an estimate of the coalescence time as provided by the detection pipeline
\cite{Colaboration:2011np}.  In practice, a 200\,ms window centered on
the trigger time is sufficient to guard against the uncertainty and bias in the
coalescence time estimates from the detection pipeline, see for instance
\cite{Brown:2004vh,ihope}. 
For the signal-to-noise ratios (SNRs) used in this paper, our posteriors are much
narrower than our priors for most parameters.

\section{Algorithms}\label{sec:algorithms}

\subsection{MCMC} 
\label{sec:MCMC}

Markov chain Monte Carlo methods are designed to estimate a posterior by stochastically wandering through
the parameter space, distributing samples proportionally to the density of the
target posterior distribution. Our MCMC implementation uses the
Metropolis--Hastings algorithm~\citep{1953JChPh..21.1087M,HASTINGS01041970},
which requires a proposal density function $Q(\pvec'|\pvec)$ to generate a new
sample $\pvec'$, which can only depend on the current sample $\pvec$. Such a
proposal is accepted with a probability $r_s = \text{min}(1, \alpha)$, where
\begin{equation}
  \label{eqn:acceptance}
  \alpha =
\frac{Q(\pvec|\pvec')p(\pvec'|\data,H)}{Q(\pvec'|\pvec)p(\pvec|\data,H)} \;.
\end{equation}
If accepted, $\pvec'$ is added to the chain, otherwise $\pvec$ is repeated.

Chains are typically started at a random location in parameter space, requiring
some number of iterations before dependence on this location is lost. 
Samples from this \emph{burn-in} period are not guaranteed to be draws from the
posterior, and are discarded when estimating the posterior. 
Furthermore, adjacent samples in the chain are typically correlated, {which is undesirable
as we perform Kolmogorov-Smirnov tests of the sampled distributions, which requires independent samples}.
To remove this correlation we thin each chain by its integrated autocorrelation time (ACT) $\tau$, defined
defined as
\begin{align}\label{eq:ACL}
\tau &= 1+2\sum_t\hat{c}(t),
\end{align}
where $t$ labels iterations of the chain and $\hat{c}(t)$ is the Pearson correlation coefficient between the chain of samples and itself shifted by $t$ samples~\cite{Gelman:1996}
The chain is thinned by using only every $\tau$-th sample, and the samples remaining
after burn-in and ACT thinning are referred to as the \emph{effective samples}.
This is necessary for some post-processing checks which assume that the samples
are statistically independent.

The efficiency of the Metropolis--Hastings algorithm is largely dependent on the
choice of proposal density, since that is what governs the acceptance rates and
ACTs. 
The standard, generically applicable distribution is a Gaussian centered on
$\pvec$, the width of which will affect the acceptance rate of the proposal. 
Large widths relative to the scale of the target posterior distribution
will lead to low acceptance rates with many repeated samples, whereas
small widths will have high acceptance rates with highly correlated samples,
both resulting in large ACTs.
For a simplified setting of a unimodal Gaussian posterior, the optimal acceptance
rate can be shown to be ~$0.234$~\cite{CJS:CJS5}.
Though
our posterior can be more complicated, we find that targeting this acceptance
rate gives good performance and consistent ACTs for all posteriors that we have
considered.
Therefore, during the first 100,000 samples of a run, we adjust the 1D Gaussian
proposal widths to achieve this acceptance rate.  This period of adjustment is 
re-entered whenever the sampler finds a log likelihood ($\log L$) that is $N/2$ larger than 
has been seen before in a run, under the assumption that this increase in likelihood
may indicate that a new area of parameter space is being explored.

When the posterior deviates from a unimodal Gaussian-like distribution, using
only the local Gaussian proposal becomes very inefficient.  The posteriors
encountered
in \ac{GW} data analysis typically consists of multiple isolated modes, separated by
regions of lower probability.
To properly weigh these modes, a Markov chain must jump between them frequently,
which is a very unlikely process when using only a local Gaussian proposal.
In section \ref{sec:proposals} we describe the range of jump proposals more
adept at sampling the parameter space of a compact binary inspiral. We also
describe the technique
of parallel tempering, which we employ to ensure proper mixing of samples
between the modes.

\subsubsection{Parallel Tempering}

\emph{Tempering}~\cite{GilksRichardsonSpiegelhalter,earl05} introduces an inverse ``temperature''~$1/T$ to the standard
likelihood function, resulting in a modified posterior

\begin{equation}
  \label{eqn:PTposterior}
  p_T(\pvec|\data) \propto p(\pvec|H)L(\pvec)^{\frac{1}{T}} \;.
\end{equation}

Increasing temperatures above $T=1$ reduces the contrast of the likelihood
surface, broadening peaks, with the posterior approaching the prior in the high
temperature limit.
Parallel tempering exploits this ``flattening'' of the posterior with increasing
temperature by constructing an ensemble of tempered chains with temperatures
spanning $T=1$ to some
finite maximum temperature $T_\text{max}$.
Chains at higher temperatures sample a distribution closer to the prior, and are
more likely to explore parameter space and move between isolated modes.
Regions of high posterior support found by the high-temperature chains are then
passed down through the temperature ensemble by periodically proposing swaps in
the locations of
adjacent chains.  Such swaps are accepted at a rate $r_s = \text{min}(1,
\omega_{ij})$, where
\begin{equation}
  \label{eqn:PTacceptance}
  \omega_{ij} =
\left(\frac{L(\pvec_j)}{L(\pvec_i)}\right)^{\frac{1}{T_i}-\frac{1}{T_j}}\;,
\end{equation}
with $T_i < T_j$. 

For non-trivial posteriors this technique greatly increases the sampling
efficiency of the $T=1$ chain, but does so at a cost. 
In our implementation, samples with $T>1$ are not used in construction of the
final posterior distribution, but they are kept for calculation of evidence
integrals via thermodynamic integration in post-processing~\ref{sec:MCMCevidence}.

All samples from chains
with $T>1$ are ultimately discarded, as they are not drawn from the target
posterior. From a computational perspective however, each chain can run in
parallel and not affect the total run time of the analysis.
The MCMC implementation of \li, {\tt LALInferenceMCMC},  uses the Message Passing
Interface (MPI)~\cite{Forum:1994:MMI:898758} to achieve this parallelization.
In our calculations, the temperatures~$T_i$ are distributed logarithmically. 
Chains are not forced to be in sync, and each chain proposes a swap in location
with the chain above it (if one exists) every 100 samples.

\subsection{Nested Sampling} 

\begin{figure}
\centering
\includegraphics[width=\columnwidth]{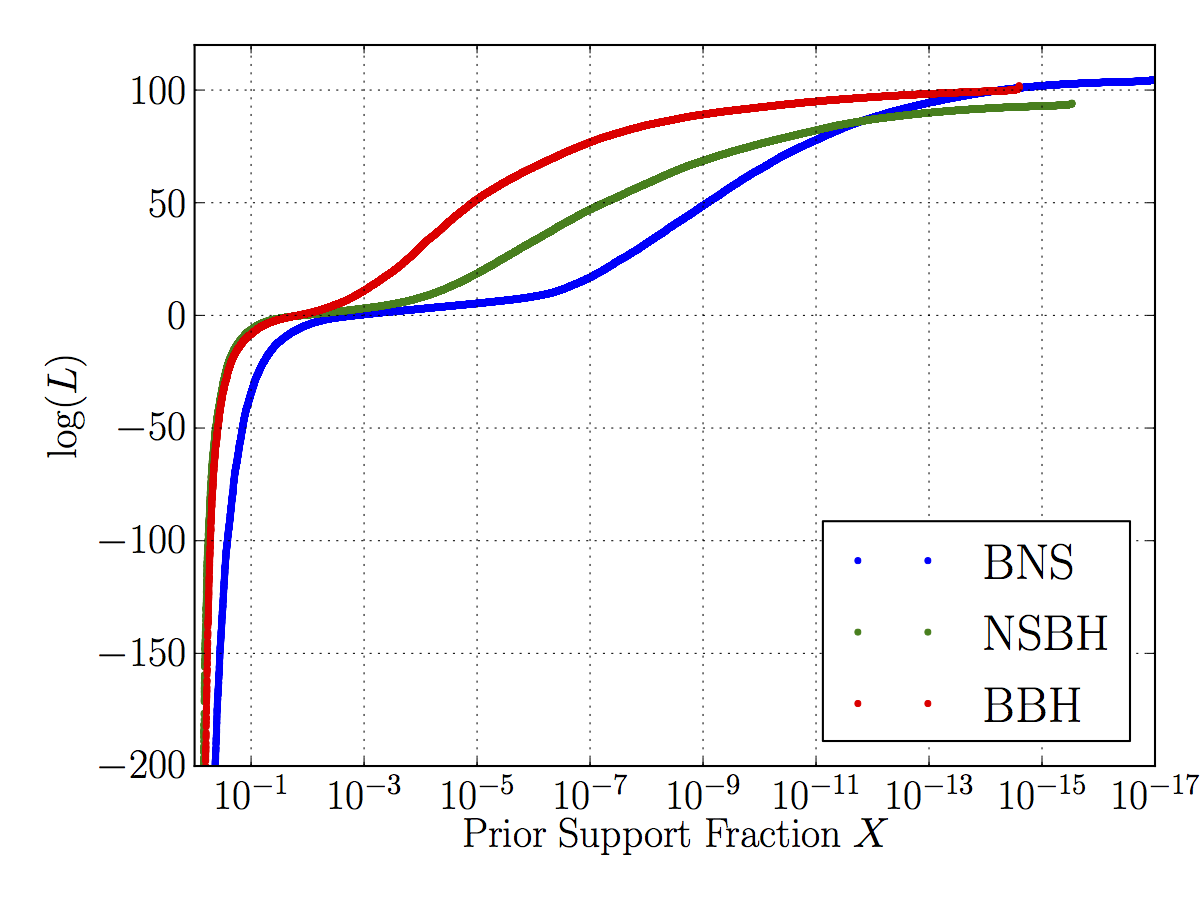}
\caption{The profile of the likelihood function for each of the injections in
Table \ref{tab:injections}, mapped onto the fractional prior support parameter
$X$ (see Eq.~\eqref{eq:X}). The algorithm proceeds from left (sampling entire
prior) to right (sampling a tiny restricted part of the prior). The
values of $\log(L)$ are normalised to the likelihood of the noise model.
}\label{fig:logLlogX}
\end{figure}

Nested sampling is a Monte Carlo technique introduced by
Skilling~\cite{Skilling:2006} for the computation of the Bayesian evidence that
will also provide samples from the posterior distribution. This is done by
transforming the multi-dimensional integral of Equation~\eqref{eq:evidence} into
a one-dimensional integral over the prior volume. The prior volume is defined as
$X$ such that $\D X = \D\pvec p(\pvec|H)$. Therefore,
\begin{equation}
\label{eq:nestedpriorvol}
X(\lambda) = \int_{p(d|\pvec,H)>\lambda} \D\pvec p(\pvec|H).
\end{equation}
This integral computes the total probability volume contained within a
likelihood contour defined by $p(d|\pvec,H)=\lambda$. With this in hand,
Equation~\eqref{eq:evidence} can now be written as
\begin{equation}
\label{eq:nestedevidence}
Z = \int_0^1 L(X) \D X,
\end{equation}
where $L(X)$ is the inverse of Equation~\eqref{eq:nestedpriorvol} and is a
monotonically decreasing function of $X$ (larger prior volume enclosed implies
lower likelihood value). By evaluating the likelihoods $L_i = L(X_i)$ associated
with a monotonically decreasing sequence of prior volumes $X_i$,
\begin{equation}
0 < X_M < \ldots < X_2 < X_1 < X_0 = 1,\label{eq:X}
\end{equation}
the evidence can be easily approximated with the trapezium rule,
\begin{equation}
Z = \sum_{i=1}^M \frac{1}{2}(X_{i-1} - X_{i+1}) L_i.
\end{equation}
Examples of the function $L(X)$ for CBC sources are shown in figure
\ref{fig:logLlogX}.

Applying this technique follows a fundamental set of steps. First, a set of
initial `live' points are sampled from the entire prior distribution. The point
with the lowest likelihood value is then removed and replaced by a new sample
with higher likelihood. This removal and replacement is repeated until a
stopping condition has been reached. By default, the loop continues while
$L_{max}X_i / Z_i > e^{0.1}$, where $L_{max}$ is the maximum likelihood so far
discovered by the sampler, $Z_i$ is the current estimate of the total evidence,
and $X_i$ is the fraction of the prior volume inside the current contour line.
In short, this is checking whether the evidence estimate would change by more
than a factor of $\sim0.1$ if all the remaining prior support were at the
maximum likelihood.
 Posterior samples can then be produced by re-sampling the chain of removed
points and current live points according to their posterior probabilities:
\begin{equation}
\label{eq:nsposteriorsamples}
p(\pvec|d,H) = \frac{\tfrac{1}{2}(X_{i-1} - X_{i+1}) L_i}{Z}.
\end{equation}
The estimation of the prior volume and method for efficiently generating new
samples varies between implementations. In \li~we have included two such
implementations, one based on an MCMC sampling of the constrained prior
distribution, and the other on the {\sc MultiNest} method, with extensions.
These are described in the following two sections \ref{sec:LINestdescription}
and \ref{sec:BAMBIdescription}.

\subsubsection{LALInferenceNest}\label{sec:LINestdescription}

The primary challenge in implementing the nested sampling algorithm is finding
an efficient means of drawing samples from the limited prior distribution
\begin{align}\label{eq:limitedprior}
p'(\pvec|H_S)\propto \begin{cases}p(\pvec|H_S) & \quad
L(\data|\pvec)>L_{\rm min} \\ 0 & \quad \text{otherwise} \end{cases} .
\end{align}
In \li~we build on the previous \texttt{inspnest} implementation described in
\cite{Veitch:2010}, with several enhancements described here.
This uses a short MCMC chain (see section \ref{sec:MCMC}) to generate each new
live point, which is started from a randomly-selected existing live point.

\begin{figure}
\centering
\includegraphics[width=\columnwidth]{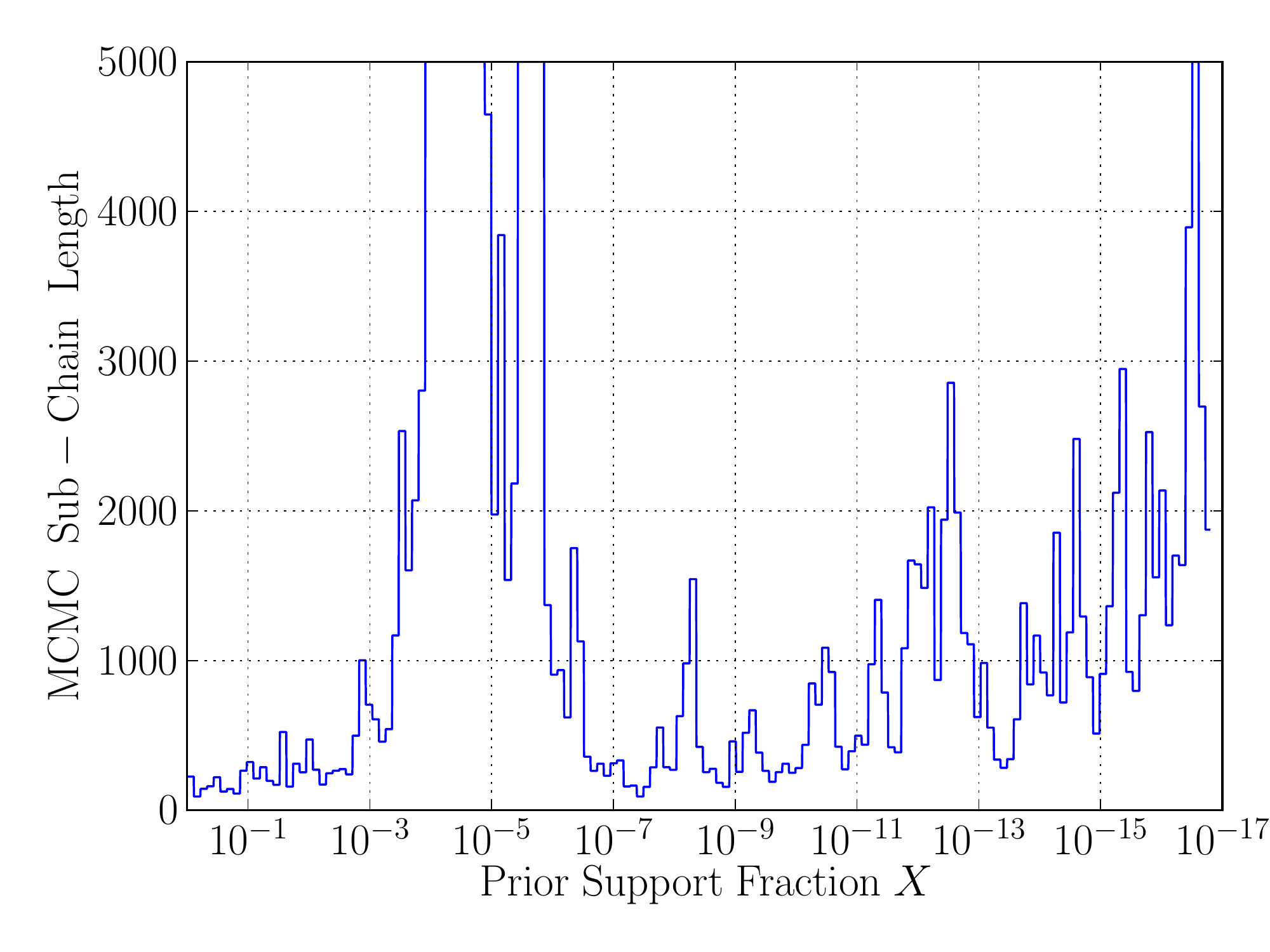}
\caption{\label{fig:autoNMCMC}Length of MCMC sub-chain for nested sampling analysis of the BNS system
(as in Table \ref{tab:injections}) as a function of prior scale. As the run
progresses, the length of the MCMC sub-chain used to generate the next live
point automatically adapts to the current conditions, allowing it to use fewer
iterations where possible. The chain is limited to a maximum of 5000
iterations.}
\end{figure}

 We use proposals of the same form as described in \ref{sec:proposals} with slight
differences: the differential evolution proposal is able to use the current set
of live points as a basis for drawing a random difference vector, and for
empirically estimating the correlation matrix used in the eigenvector proposal.
This ensures that the scale of these jumps adapts automatically to the current
concentration of the remaining live points.
In contrast to Eq.~\eqref{eqn:acceptance}, the target distribution that we are sampling
is the limited prior distribution $p'$ of Eq.~\eqref{eq:limitedprior}, so the acceptance ratio
is
\begin{align}
  \label{eqn:NSacceptance}
  \alpha &= \frac{Q(\pvec|\pvec')p'(\pvec'|H)}{Q(\pvec'|\pvec)p'(\pvec|H)}.
\end{align}
Furthermore, we have introduced additional features which help to reduce the
amount of manual tuning required to produce a reliable result.
\paragraph{Autocorrelation adaptation}
In \cite{Veitch:2010} it was shown that the numerical error on the evidence
integral was dependent not only on the number of live points $N_{\rm live}$ and
the information content of the data (as suggested by Skilling), but also on the
length of the MCMC sub-chains $N_{\rm MCMC}$ used to produce new samples (this
is not included in the idealised description of nested sampling, since other
methods of drawing independent new samples are also possible, see section
\ref{sec:BAMBIdescription}).
In \texttt{inspnest}, the user would specify this number at the start of the
run, depending on their desire for speed or accuracy. The value then remained
constant throughout the run. This is inefficient, as the difficulty of
generating a new sample varies with the structure of the $p'(\pvec|H_S)$
distribution at different values of $L_{\rm min}$. For example, there may be
many secondary peaks which are present up to a certain value of $L_{\rm min}$,
but disappear above that, making the distribution easier to sample.
To avoid this inefficiency (and to reduce the number of tuning parameters of the
code), we now internally estimate the required length of the sub-chains as the
run progresses.
To achieve this, we use the estimate of the autocorrelation timescale $\tau_i$
(defined as in Eq.~\ref{eq:ACL}) for parameter $i$ of a sub-chain generated from a randomly selected live point.
The sum is computed up to the lag $M_i$ which is the first time the correlation
drops below $0.01$, i.e $\hat{c}_i(M_i)\le0.01$.
The timescale is computed for each parameter being varied in the model, and the
longest autocorrelation time is used as the number of MCMC iterations
($M=\mathrm{max}(M_1,\ldots,M_i)$ for subsequent sub-chains until it is further
updated after $N_{\rm live}/4$ iterations of the nested sampler.
As the chain needed to compute the autocorrelation timescale is longer than the
timescale itself, the independent samples produced are cached for later use. We
note that as the nested sampling algorithm uses many live points, the
correlation between subsequent points used for evaluating the evidence integral
will be further diluted, so this procedure is a conservative estimate of the
necessary chain thinning.
The adaptation of the sub-chain length is shown in figure \ref{fig:autoNMCMC},
where the algorithm adapts to use $<1000$ MCMC steps during the majority of the
analysis, but can adjust its chain length to a limit of 5000 samples for the
most difficult parts of the problem.

\paragraph{Sloppy sampling}
For the analysis of CBC data, the computational cost of a likelihood evaluation
completely dominates that of a prior calculation,
since it requires the generation of a trial waveform and the calculation of an
inner product (with possible FFT into the frequency domain).
The task of sampling the likelihood-limited prior $p'(\pvec|H)$ is performed by
sampling from the prior distribution, rejecting any points that fall beneath the
minimum threshold $L_{\rm min}$. During the early stages of the run, the $L_{\rm
min}$ likelihood bound encloses a large volume of the parameter space, which may
take many iterations of the sub-chain to cross, and a proposed step originating
inside the bound is unlikely to be rejected by this cut. We are free to make a
shortcut by not checking the likelihood bound at each step of the sub-chain,
allowing it to continue for $ME$ iterations, where $E$ is the fraction of
iterations where the likelihood check is skipped. Since the calculation of the
prior is essentially free compared to that of the likelihood, the computational
efficiency is improved by a factor of $~(1-E)^{-1}$. The likelihood bound is
always checked before the sample is finally accepted as a new live point.

Since the optimal value of $E$ is unknown, and will vary throughout the run as
the $L_{\rm min}$ contour shrinks the support for the $p'(\pvec|H)$
distribution, we adaptively adjust it based on a target for the acceptance of
proposals at the likelihood-cut stage.
Setting a target acceptance rate of 0.3 at the likelihood cut stage, and having
measured acceptance rate $\alpha$, we adjust $E$ in increments of $5\%$ upward
when $\alpha>0.3$ or downward when $\alpha<0.3$, with a maximum of $1$. This
procedure allows the code to calculate fewer likelihoods when the proposal
distribution predominantly falls inside the bounds, which dramatically improves
the efficiency at the start of the run.

\begin{figure}
\centering
\includegraphics[width=\columnwidth]{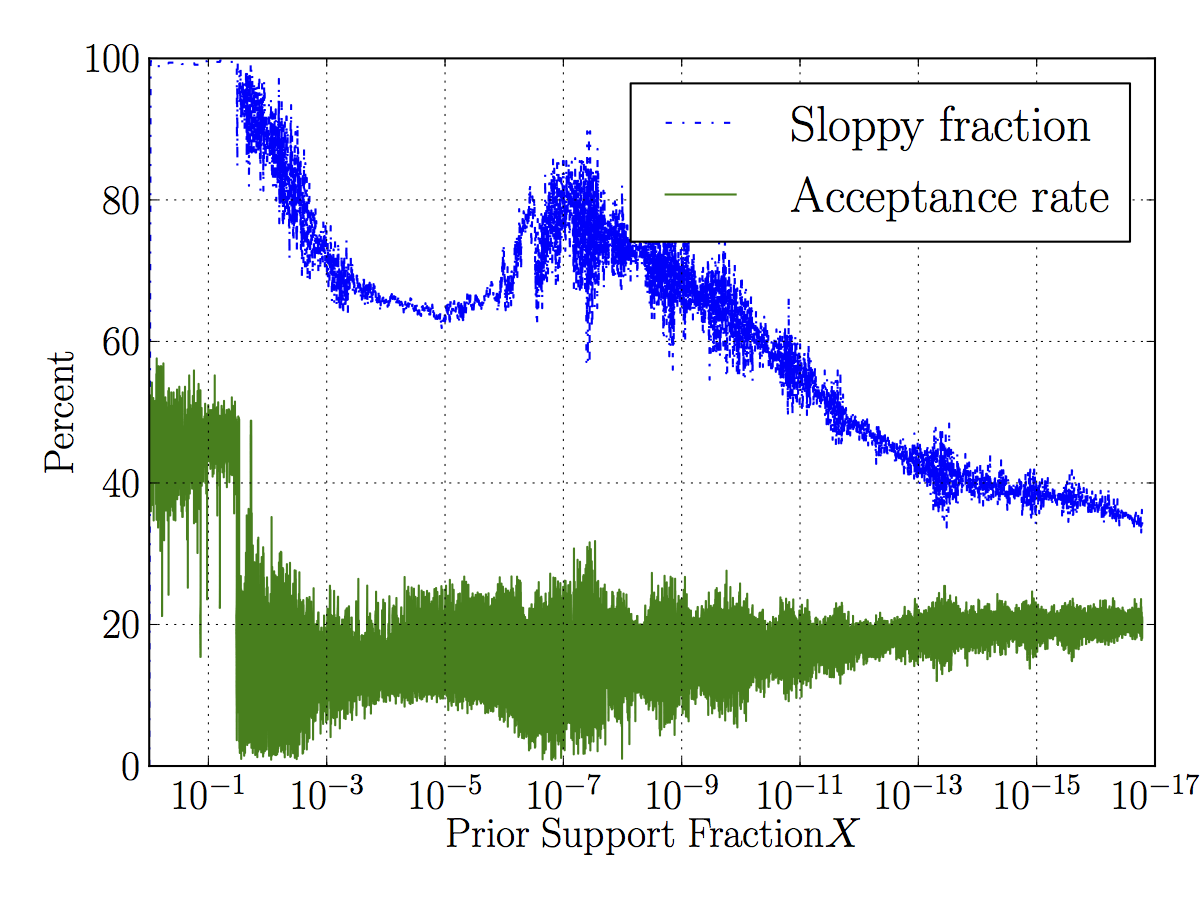}
\caption{Acceptance ratio and fraction of sloppy jumps for nested sampling
analysis of a BNS system. The dashed blue line shows the automatically
determined fraction of proposals for which the likelihood calculation is
skipped. The solid green line shows the overall acceptance rate for new live
points, which thanks to the adaptive jumps remains at a healthy level despite
the volume of the sampled distribution changing by 17 orders of magnitude
throughout the run.}
\end{figure}

\paragraph{Parallelisation}\label{sec:LINestparallel}
Although the nested sampling algorithm itself is a sequential method, we are
able to exploit a crude parallelisation method to increase the number of
posterior samples produced. This involves performing separate independent runs
of the algorithm on different CPU cores, and then combining the results weighted
by their respective evidence.
Consider a set of nested sampling runs indexed by $i$, with each iteration
indexed by $j=1 \ldots \xi_i$, where $\xi_i$ is the number of iterations in run
$i$ before it terminates, and $Z_i$ denotes the evidence estimate from that run.
Our implementation also outputs the $N_\mathrm{live}$ live points at the time of
algorithm termination, which are indexed
$\xi_{i+1}\ldots\xi_{i+N_\mathrm{live}}$. These last samples are treated
separately since they are all drawn from the same prior volume. The runs must
all be performed with identical data and models, but with different random seeds
for the sampler.

For each sample $\pvec_{ij}$ we calculate the posterior weight $w_{ij}=L_{ij}
V_{ij} / Z_i$, where $\log V_{ij} = -j/N_\mathrm{live}$ for the points up to
$j\le\xi_i$ and $V_{ij}=-\xi_i/N_\mathrm{live}$ for the final points $j>\xi_i$.
By resampling any individual chain according to the weights $w_{ij}$ we can
produce a set of samples from the posterior.
The resulting sets of posteriors for each $i$ are then resampled according to
the evidence $Z_i$ calculated for each chain. This ensures that chains which
fail to converge on the global maximum will contribute proportionally fewer
samples to the final posterior than those which do converge and produce a higher
$Z_i$ estimate.
The resampling processes can be performed either with or without replacement,
where the latter is useful in ensuring that no samples are repeated. In this
paper independent samples are used throughout, as repeated samples will distort
the tests of convergence by artificially lowering the KS test statistic.

In practice, this procedure reduces the wall time necessary to produce a given
number of posterior samples, as the chains can be spread over many CPU cores.

\subsubsection{{\sc MultiNest} \& BAMBI}\label{sec:BAMBIdescription} 
{\sc MultiNest}~\cite{Feroz:2008,Feroz:2009,Feroz:2013} is a generic algorithm
that implements the nested sampling technique.
It uses a model-based approach to generate samples within the volume $X$ enclosed
by the likelihood contour $L(X)>L_\mathrm{min}$.
The set of live points is
enclosed within a set of (possibly overlapping) ellipsoids and a new
point is then drawn uniformly from the region enclosed by these ellipsoids. The
volume of ellipsoids is used in choosing which to sample from and points are
tested to ensure that if they lie in multiple ($N$) ellipsoids they are accepted
as a sample only the corresponding fraction of the time ($1/N$). The ellipsoidal
decomposition of the live point set is chosen to minimize the sum of volumes of
the ellipsoids. This method is well suited to dealing with posteriors that have
curving degeneracies, and allows mode identification in multi-modal posteriors.
If there are various subsets of the ellipsoid set that do not overlap in 
parameter space, these are identified as distinct modes and subsequently evolved
independently.

{\sc MultiNest} is able to take advantage of parallel computing architectures by
allowing each CPU to compute a new proposal point. As the run progresses, the
actual sampling efficiency (fraction of accepted samples from total samples
proposed) will drop as the ellipsoidal approximation is less exact and the
likelihood constraint on the prior is harder to meet. By computing $N$ samples
concurrently, we can obtain speed increases of up to a factor of $N$ with the
largest increase coming when the efficiency drops below $1/N$.

The user only needs to tune a few parameters for any specific implementation in
addition to providing the log-likelihood and prior functions. These are the
number of live points, the target efficiency, and the tolerance. The number of
live points needs to be enough that all posterior modes are sampled (ideally
with at least one live point in the initial set) and we use from $1000$ to
$5000$ for our analyses. The target efficiency affects how conservatively the
ellipsoidal decomposition is made and a value of $0.1$ ($10\%$) was found to be
sufficient; smaller values will produce more precise posteriors but require more
samples. Lastly, a tolerance of $0.5$ in the evidence calculation is
sufficiently small for the run to converge to the correct result.

{\sc MultiNest} is implemented for \li~within the Blind Accelerated
Multimodal Bayesian Inference (BAMBI) algorithm~\cite{Graff:2012}. BAMBI
incorporates the nested sampling performed by {\sc MultiNest} along with the
machine learning of {\sc SkyNet}~\cite{Graff:2014} to learn the likelihood
function on-the-fly. Use of the machine learning capability requires further
customisation of input settings and so is not used for the purposes of this
study.

\subsection{Jump Proposals}\label{sec:proposals}
For both the MCMC sampler and the MCMC-subchains of the Nested Sampler,
efficiently exploring the parameter space is essential to optimising performance
of the algorithms.
Gaussian jump proposals are typically sufficient for unimodal posteriors and
spaces without strong correlations between parameters, but
there are many situations where strong parameter correlations exist and/or
multiple isolated modes appear spread across the multi-dimensional parameter
space. 
When parameters are strongly correlated, the ideal jumps would be along these
correlations, which makes 1D jumps in the model parameters very inefficient.
Furthermore to sample between isolated modes, a chain must make a large number
of improbable jumps through regions of low probability. 
To solve this problem we have used a range of jump proposals, some of which are
specific to the CBC parameter estimation problem and some of which are more
generally applicable to multimodal or correlated problems.

To ensure that an MCMC equilibrates to the target distribution, the
jump proposal densities in Eq.~\eqref{eqn:acceptance} must be computed
correctly.  Our codes achieve this using a ``proposal cycle.''  At the
beginning of a sampling run, the proposals below are placed into an
array (each proposal may be put multiple times in the array, according
to a pre-specified weight factor).  The order of the array is then permuted
randomly before sampling begins.  Throughout the run, we cycle through
the array of proposals (maintaining the order), computing and applying the jump proposal
density for the chosen proposal at each step as in
Eq.~\eqref{eqn:acceptance}.  This procedure ensures that there is only a
single proposal ``operating'' for each MCMC step, simplifying the
computation of the jump proposal density, which otherwise would have
to take into account the forward and reverse jump probabilities for
all the proposals simultaneously.

\subsubsection*{Differential Evolution}

Differential evolution is a generic technique that attempts to solve the
multimodal sampling problem by leveraging information gained previously in the
run~\citep{differentialEvo,TerBraak:2008}. It does so by drawing two previous samples
$\pvec_1$ and $\pvec_2$ from the chain (for MCMC) or from the current set of
live points (nested sampling), and proposing a new sample $\pvec'$ according to:
\begin{equation}
  \label{eqn:DE}
  \pvec' = \pvec + \gamma(\pvec_2-\pvec_1) \;,
\end{equation}
where $\gamma$ is a free coefficient. 
50\% of the time we use this as a mode-hopping proposal, with $\gamma=1$. 
In the case where $\pvec_1$ and $\pvec$ are in the same mode, this proposes a
sample from the mode containing $\pvec_2$. The other
50\% of the time we choose $\gamma$ according to 
\begin{equation}
  \label{eqn:DE-gamma-scale}
  \gamma \sim N\left(0, 2.38/\sqrt{2 N_\mathrm{dim}}\right),
\end{equation}
where $N_\mathrm{dim}$ is the number of parameter space dimensions.
The scaling of the distribution for $\gamma$ is suggested in
\citet{TerBraak:2008} following \citet{Roberts:2001} for a good
acceptance rate with general distributions.  The differential
evolution proposal in this latter mode proves useful when linear
correlations are encountered in the distribution, since the jump
directions tend to lie along the principal axes of the posterior
distribution.  However, this proposal can perform poorly when the
posterior is more complicated.

Drawing from the past history of the chain for the MCMC differential
evolution proposal makes the chain evolution formally non-Markovian.
However, as more and more points are accumulated in the past history,
each additional point accumulated makes a smaller change to the
overall proposal distribution.  This property is sufficient to make
the MCMC chain asymptotically Markovian, so the distribution of
samples converges to the target distribution; in the language of
\citet{Roberts:2007}, Theorem 1, $D_n \to 0$ in probability as $n \to
\infty$ for this adaptive proposal, and therefore the posterior is the
equilibrium distribution of this sampling.

\subsubsection*{Eigenvector jump}
The variance-covariance matrix of a collection of representative points drawn
from the target distribution (the
current set of nested sampling live points)
can be used as an automatically self-adjusting proposal distribution. In our
implementation, we calculate the eigenvalues and eigenvectors of the estimated
covariance matrix,
and use these to set a scale and direction for a jump proposal. This type of
jump results in a very good acceptance rate when the underlying distribution is
approximately Gaussian, or is very diffuse (as in the early stages of the nested sampling run).
In the nested sampling algorithm, the covariance matrix is updated every $N_{live}/4$
iterations to ensure the jump scales track the shrinking scale of the target
distribution. Within each sub-chain the matrix is held constant to ensure
detailed balance.

\subsubsection*{Adaptive Gaussian}
We also use a 1 dimensional Gaussian jump proposal, where the jump for
a single parameter $\theta_k$ is $\theta_k' = \theta_k +
N(0,\sigma_k)$.  The width of the proposal is scaled to achieve a
target acceptance rate of $\xi \simeq 0.234$ by adjusting 
\begin{align}
  \sigma_k \leftarrow \sigma_k + s_\gamma \frac{1-\xi}{100} \Delta
\end{align}
when a step is accepted, where $s_\gamma$ is a scaling factor and
$\Delta$ is the prior width in the $k$th parameter, and adjusting
\begin{align}
  \sigma_k \leftarrow \sigma_k - s_\gamma \frac{\xi}{100} \Delta
\end{align}
when a step is rejected.  For the MCMC, the adaptation phase lasts for
100,000 samples, and $s_\gamma = 10\left(t-t_0\right)^{-1/5} - 1$
during this phase; otherwise $s_\gamma = 0$.  The nested sampling
algorithm has $s_\gamma = 1$.

\subsubsection*{Gravitational-wave specific proposals}

We also use a set of jump proposals specific to the CBC parameter estimation
problem addressed in this work.
These proposals are designed to further improve the sampling efficiency by
exploring known structures in the CBC posterior distribution,
primarily in the sky location and extrinsic parameter sub-spaces.
\paragraph*{Sky location}
Determining the sky position of the CBC source is an important issue for
followup observations of any detected sources. The position, parameterised by
$(\alpha,\delta,\dL)$,
is determined along with the other parameters by the \li~code, but it
can present difficulties due to the highly structured nature of the posterior
distribution.
Although the non-uniform amplitude response of a single detector allows some
constraint of the sky position of a source, the use of a network of detectors
gives far better resolution
of the posterior distribution. This improvement is heuristically due to the
ability to resolve the difference in time of arrival of the signal at each
detector, which
allows triangulation of the source direction. The measured amplitude of the
signal and the non-uniform prior distribution further constrain the posterior,
but the major structure in the likelihood can be derived by considering
the times of arrival in multiple detectors. This leads us to include two
specific jump proposals similar to those outlined in \cite{Veitch:2010}, which
preserve the times of arrival in two and three detector networks respectively.
\begin{description}
\item[Sky Reflection]
In the case of a three-detector network, the degeneracy of the ring based on
timing is broken by the presence of a third detector. In this case, there
are two solutions to the triangulation problem which correspond to the true
source location, and its reflection in the plane containing the three detector
sites.
If the normal vector to this plane is $\hat{n}$, the transition (in Cartesian
coordinates with origin at the geocentre) between the true point $\hat{x}$ and
its reflection $\hat{x}'$ is written
\begin{align}
\hat{x}'=\hat{x}-2\hat{n}|\hat{n}.(\hat{x}-\hat{x}_i)|
\end{align}
where $\hat{x}_i$ is the unit vector pointing in the direction of one of the
detector sites. The resulting point is then projected back onto the unit sphere
parameterised by $\alpha,\delta$.
To ensure detailed balance, the resulting point is perturbed by a small random
vector drawn from a 3D Gaussian in $(t,\alpha,\delta)$
The time parameter is updated in the same way as for the sky rotation proposal
above. As in the two-detector case, the degeneracy between these points can be
broken by consideration of the signal amplitudes observed in the detector,
however this is not always the case as the secondary mode can have a similar
likelihood.

\end{description}

\paragraph*{Extrinsic parameter proposals}\label{sec:proposals:extrinsic}
\begin{description}
\item[Extrinsic Parameter proposal] There exist a correlation between the
inclination, distance, polarization and the sky location dues to the sensitivity
of the antenna beam patterns of the detectors. This correlation makes the two
solutions for the sky location from the thee-detector network (described above)
correspond to different values of inclination, distance and polarization. We
solve analytically the values of those parameters when trying to jump between
the two sky reflections. The equations are detailed in \cite{Raymond:2014uha}.
\item[Polarization and Phase correlation] There exists a degeneracy between the
$\phi$ and $\psi$ parameters when the orbital plane is oriented perpendicular to
the line of signal, i.e. $\iota=\{0,\pi\}$. In general these parameters tend to
be correlated along the axes $\alpha=\psi+\phi$ and $\beta=\psi-\phi$. We
propose jumps which choose a random value of either the $\alpha$ or $\beta$
parameter (keeping the other constant) to improve the sampling of this
correlation.
\end{description}

\paragraph*{Miscellaneous proposals}
\begin{description}
\item[Draw from Prior] A proposal that generates samples from the prior
distribution (see section \ref{sec:prior}) by rejection sampling. This is mostly useful
for improving the mixing of high-temperature MCMC chains, as it does not depend on the previous iteration.
\item[Phase reversal] Proposes a change in the orbital phase parameter
$\phi_{j+1} = (\phi_j + \pi) \pmod{2\pi}$, which will keep the even harmonics
of the signal unchanged, but will flip the sign of the odd harmonics.
Since the even harmonic $l=m=2$ dominates the signal, this is useful for
proposing jumps between multiple modes which differ only by the relatively
small differences in the waveform generated by the odd harmonics.

\item[Phase and polarization reversal] Proposes a simultaneous change of the
orbital phase and polarisation parameters $\phi_{j+1} = (\phi_j +
\pi)\,\pmod{2\pi}$ and $ \psi_{j+1} = (\psi_j + \pi/2)\,\pmod{\pi}$.

\item[Gibbs Sampling of Distance] 
  The conditional likelihood of the distance parameter $d_L$ follows a known form,
  which allows us to generate proposals
from this distribution independently of the previous iteration, reducing the
correlation in the chains.
  As the signal amplitude scales proportionally to ${d_L}^{-1}=u$, the logarithm
of the likelihood function (Equation \eqref{eq:L}), constrained to only distance
variations, is quadratic in $u$,
\begin{align}
\log L(u)&=A+Bu+Cu^2,
\end{align}
which in our case yields a Gaussian distribution with mean 
$\mu=-B/2C$ and variance $\sigma^2=1/2C$.
By calculating the value of $\log L$ at three different distances, the quadratic
coefficients are found and a new proposed distance can be generated from the
resulting Gaussian distribution.
\end{description}

\section{Post-processing} 
\label{sec:post_processing}

The main data products of all the above algorithms are a set of `samples'
assumed to be drawn independently from the posterior probability distribution
$p(\pvec|d,I)$ (as defined in Equation~\eqref{eq:Bayes}) and, for the nested
sampling algorithms, an approximation to the evidence $Z=P(d|I)$ (for MCMC, evidence
computation is performed in post-processing, see section \ref{sec:MCMCevidence}).  Each
algorithm initially produces outputs which are different in both their form and
relation to these quantities. A suite of Python scripts has been specifically
developed for the purpose of converting these outputs to a common results format
in order to facilitate comparisons between the algorithms and promote
consistency in the interpretation of results. At the time of writing these
scripts (and associated libraries) can be found in the open-source LALsuite
package \cite{LAL}. The end result of this process is a set of
web-ready HTML pages containing the key meta-data and statistics from the
analyses and from which it should be 
possible 
to reproduce any results produced by the codes. In this section we outline in
more detail the steps needed to convert or \textit{post-process} the output of
the different algorithms to this common results format and important issues
related to interpreting these results and drawing scientific conclusions.

\subsection{MCMC}
The MCMC algorithm in \li~produces a sequence of $O(10^6)$ - $O(10^8)$
samples, depending on the number of source parameters in the model, the number of
interferometers used, and the bandwidth of the signal. Each sample consists of a
set of source parameters $\{\pvec\}$ and associated values of the likelihood
function $L(d|\pvec)$ and prior $p(\pvec)$. 
We cannot immediately take this output sequence to be our 
posterior samples as we cannot assume that all the samples were drawn
independently from the actual posterior distribution. 

In order to generate a set of independent posterior samples the post-processing
for the MCMC algorithm first removes a number of samples at the beginning of the
chain -- the so-called `burn-in' -- where the MCMC will not yet be sampling from
the posterior probability density function.  For a $d$-dimensional parameter
space, the distribution of the log-likelihood is expected to be close to
$\mathcal{L}_\text{max} - X$, where $\mathcal{L}_\text{max}$ is the maximum
achievable log-likelihood, and $X$ is a random variable following a Gamma($d/2$,
1) distribution \cite{Raftery07estimatingthe}.  Thus, we consider the burn-in to
end when a chain samples log-likelihood values that are within $d/2$ of the
highest log-likelihood value found by the chain.  Once we have discarded these
samples, the set of remaining samples is then `down-sampled'; the chain is
re-sampled randomly at intervals inversely proportional to the the autocorrelation
length to produce a sub-set of samples which are assumed to be drawn
independently 
from the posterior distribution. See section \ref{sec:MCMC} above for more
details.

\subsection{Nested sampling}
The output of both of the nested sampling algorithms in \li~are a list
(or lists in the case of parallel runs) of the live points
sampled from the prior distribution for a particular model and data set and
consisting of a set of parameters and their associated $\log(L_{ij})$ and $Z_{ij}$.
These live points approximately lie on the contours enclosing the nested prior
volumes and each has associated with it some fraction of the evidence assumed to
be enclosed within said contour.  The post-processing step takes this
information and uses it to
generate posterior samples from the
list of retained live points using Eq.~\ref{eq:nsposteriorsamples} for
single runs, along with the procedure described in section
\ref{sec:LINestparallel} for parallel runs.

\subsection{Evidence calculation using MCMC outputs}\label{sec:MCMCevidence}
Whilst the nested sampling algorithms in \li~directly produce
an approximation to the value of the evidence $Z$ (and produce
posterior samples as a by-product), we can also use the output from
the MCMC algorithms to calculate independent
estimates of $Z$ in post-processing.  We have tested several methods of computing the
evidence from posterior samples, including the harmonic mean
\citep{NewtonRaftery:1994,Chib:1995,vanHaasteren:2009}, direct
integration of an estimate of the posterior density
\citep{2009arXiv0911.1777W}, and thermodynamic integration (see
e.g.\ \cite{Neal:1993,2009PhRvD..80f3007L}).  We have found that only
thermodynamic integration permits reliable estimation of the evidence
for the typical number and distribution of posterior samples we obtain
in our analyses.

\begin{figure}
  \includegraphics{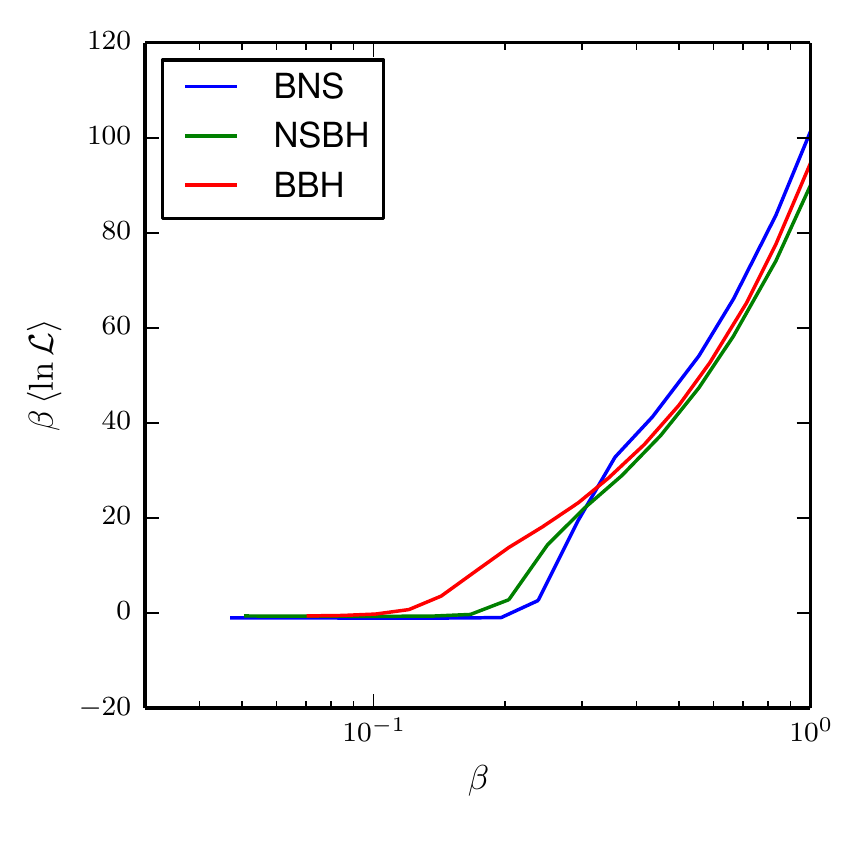}
  \caption{\label{fig:mcmc-evidence} The integrand of the evidence
    integral (Eq.~\eqref{eq:evidence-beta-integral}) versus $\beta$
    for the analyses of synthetic \ac{GW} signals in
    \S~\ref{sec:injections}.  The evidence is given by the area under
    each curve.  Table \ref{tab:AnalyticEvidence} gives the results of
    the integration together with the estimated error in the
    quadrature, following the procedure described in
    \S~\ref{sec:MCMCevidence}.  The jaggedness of the curves
    illustrates that the temperature spacing required for convergent
    MCMC simulations is larger than that required for accurate
    quadrature to compute the evidence; the flatness at small $\beta$
    illustrates that, for these simulations, the high-temperature
    limit is sufficient for convergence of the evidence integral. }
\end{figure}

Thermodynamic integration considers the evidence as a function of the
temperature, $Z(\beta|H)$, defined as
\begin{align}
Z(\beta|H) &\equiv \int \D\pvec p(d|H,\pvec,\beta)p(\pvec|H) \nonumber \\
&= \int \D\pvec p(d|H,\pvec)^{\beta}p(\pvec|H) 
\end{align}
where $\beta=1/T$ is the inverse temperature of the chain.
Differentiating with respect to $\beta$, we find 
\begin{align}\label{eq:partition}
\frac{d}{\D\beta}\ln Z(\beta|H) = \langle \ln p(d|H,\pvec)\rangle _{\beta}
\end{align}
where $\langle \ln p(d|H,\pvec)\rangle _{\beta}$ is the expectation value of the
log likelihood for the chain with temperature $1/\beta$.  We can now integrate
\eqref{eq:partition} to find the logarithm of the evidence
\begin{align}
  \label{eq:evidence-beta-integral}
\ln Z = \int_0^1\D\beta\ \langle \ln p(d|H,\pvec)\rangle _{\beta}.
\end{align}
It is straightforward to compute $\langle \ln p(d|H,\pvec)\rangle
_{\beta}$ for each chain in a parallel-tempered analysis; the integral
in Eq.\ \eqref{eq:evidence-beta-integral} can then be estimated using
a quadrature rule.  Because our typical temperature spacings are
coarse, the uncertainty in this estimate of the evidence is typically
dominated by discretisation error in the quadrature.  We estimate that
error by performing the quadrature twice, once using all the
temperatures in the chain and once using half the temperatures.  To
achieve very accurate estimates of the evidence, sometimes $\sim20$ to
$\sim 30$ temperatures are needed, out to a maximum of $\beta^{-1}\sim
10^5$, which adds a significant cost over the computations necessary
for parameter estimation; however, reasonably accurate estimates of
the evidence can nearly always be obtained from a standard run setup
with $\sim 10$ chains.  Figure \ref{fig:mcmc-evidence} plots the
integrand of Eq.~\eqref{eq:evidence-beta-integral} for the synthetic
GW signals analysed in \S~\ref{sec:injections}, illustrating both the
coarse temperature spacing of the runs and the convergence of the
evidence integral at high temperature.

\subsection{Generation of statistics and marginal posterior distributions}\label{sec:credibleIntervals}
Whilst the list of posterior samples contains all the information about the
distribution of the source parameters obtained from the analysis, we need to
make this more intelligible by summarising it in an approximate way.
We have developed a number of different summary statistics which provide
digested information about the posterior
distributions, which are applied in post-processing to the output samples.

The simplest of these are simply the mean and standard deviation of the
one-dimensional marginal distributions
for each of the parameters. These are estimated as the sample mean, standard
deviation, etc., over the samples, which converge on their continuous distribution
equivalents~\eqref{eq:expectations} in the limit of large numbers of samples. These
are particularly useful for giving simple measures of the compatibility of
the results with the true values, if analysing a known injection.

However, estimators are not always representative of the much larger amount of
information contained in the marginal posterior distributions on each of the
parameters (or combinations of them).
For summarising one- or two-dimensional results we
create plots of the marginal posterior probability density function by
binning the samples in the space of the parameters and normalising the resulting
histogram by the number of samples.

We are also interested in obtaining estimates of the precision of the resulting
inferences, especially when comparing results from a large number of simulations
to obtain an expectation of parameter estimation performance under various
circumstances.
We quantify the precision in terms of `credible intervals', defined for a
desired level of credibility (e.g. $P_\mathrm{cred}=95\%$ probability that the parameter lies
within the interval), with the relation
\begin{equation}
  \mathrm{credible \phantom{a}level} = \int_{\mathrm{credible\phantom{a}
interval}} \hspace{-50pt}
 p(\vec{\theta}|d) d \vec{\theta}.
\end{equation}

The support of the integral above is the credible interval, however this is not
defined uniquely by this expression.
In one dimension, we can easily find a region enclosing a fraction $x$ of the
probability by sorting the samples by their parameter values and choosing
the range from $[N(1-x)/2,N(1+x)/2]$ where $N$ is the number of independent
samples in the posterior distribution. The statistical error on the fraction $x$
of the true distribution enclosed, caused by the approximation with discrete samples
is $\approx\sqrt{x(1-x)/N}$. To achieve a $1\%$ error in the $90\%$ region we therefore
require 900 independent samples. Typically we collect a few thousand samples,
giving an error $<1\%$ on the credible interval.

We are also interested in the \emph{minimum} credible interval, which is
the smallest such region that encloses the desired fraction of the posterior.
In the case of a unimodal one-dimensional posterior this leads to the highest
posterior density interval.

To find estimates of the minimum credible intervals
we use a number of techniques that have
different regimes of usefulness, depending primarily on the number of
samples output from the code and the number of parameters we are
interested in analysing conjointly.

When we are considering the one-dimensional marginal posterior
distributions, we simply compute a histogram for the parameter
of interest using equally-sized bins.
This directly tells us the probability associated with that
region of the parameter space: the probability density is approximately equal to the
fraction of samples in the bin divided by the bin width.
This simple histogram method involves an appropriate choice of
the bin size.
We must be careful to choose a bin size small enough that we have good
resolution and can approximate the density as piecewise constant within
each bin, but large enough so that the sampling error within each bin
does not overwhelm the actual variations in probability between bins.

To recover the minimum credible interval we apply a greedy algorithm to the
histogram bins. This orders the bins by probability, and starting from the highest
probability bin, works its way down the list of bins until the required total
probability has been reached.
Although this procedure generally yields satisfactory results, it is subject to
bias due to the discrete number of samples per bin. To see this, consider a uniform
probability distribution that has been discretely sampled. The statistical variation
of the number of samples within bins will cause those where the number fluctuates upward
to be chosen before those where it fluctuates downward. The credible interval estimated
by this method will therefore be smaller than the true interval containing the desired
proportion of the probability.
In \cite{Sidery:2014} we investigate several methods of overcoming this problem.

\section{Validation of results}
\label{sec:validation}
To confirm the correctness of the sampling algorithms, we performed cross-comparisons
of recovered posterior distributions for a variety of known distributions and 
example signals.
The simplest check we performed was recovery of the prior distribution, described in
section \ref{sec:prior}. The one-dimensional distributions output by the codes
were compared using a Kolmogorov-Smirnov test, where the comparisons between the three codes
on the 15 marginal distributions were all in agreement with p-values above $0.02$.
We next analysed several known likelihood functions, where we could perform cross-checks between
the samplers.
These were a unimodal $15$-dimensional correlated Gaussian, a bimodal correlated
Gaussian distribution, and the Rosenbrock banana function.
For the unimodal and bimodal distributions we can also compare the results of the samplers
to the analytical marginal distributions to confirm they are being sampled correctly.

\subsection{Analytic likelihoods} 
\label{sec:analytic-likelihood}

\begin{table}
\begin{tabular}{c|cccc}
 ~ & ~ & Nested & ~ & MCMC  \\
 Distribution & Analytic &  Sampling & BAMBI & thermo. \\
\hline
Unimodal & -21.9 & $-21.8\pm0.1$ & $-21.8\pm0.12$ & $-20.3 \pm 1.9$ \\
Bimodal & -30.0 & $-30.0\pm0.1$ & $-29.9\pm0.14$ & $-26.7 \pm 3.0$ \\
Rosenbrock & -- & $-70.9\pm0.2$ & $-69.1\pm0.2$ & $-63.0 \pm 7.6$ \\
BNS & -- & $68.7\pm0.34$ & $69.98\pm0.17$ & $68.2 \pm 1.1$ \\
NSBH & -- & $62.2\pm0.27$ & $63.67\pm0.16$ & $63.40 \pm 0.72$ \\
BBH & -- & $71.4\pm0.18$ & $72.87\pm0.15$ & $72.44 \pm 0.11$
\end{tabular}
\caption{\label{tab:AnalyticEvidence}The log evidence estimates for
  the analytic likelihood distributions (\S
  \ref{sec:analytic-likelihood}) and the simulated signals (\S
  \ref{sec:injections}) calculated with the three methods, with
  estimated uncertainty. For the thermodynamic integration
  method we used 16 steps on the temperature ladder, except for the Rosenbrock
  likelihood which required 64. For distributions that permit an analytic
  computation of evidence, the samplers produce evidence estimates
  consistent with the true value. For the others, the estimates
  produced by the samplers are not consistent, indicating that there remains
  some systematic error in the evidence calculation methods for the more difficult
  problems.
  }
\end{table}

The multivariate Gaussian distribution was specified by the function
\begin{align}
\log L_\mathrm{MV} &= - \frac{1}{2}(\hat{\theta}_i-\theta_i) C^{-1}_{ij}
(\hat\theta_j-\theta_j).
\end{align}
where $C_{ij}$ is a covariance matrix of dimension 15, and the mean values
$\hat\theta_i$ are
chosen to lie within the usual ranges, and have the usual scales, as in the \ac{GW} 
case. $C_{ij}$ was chosen so that its eigenvectors do not lie parallel to the
axes defined by the parameters $\theta_i$,
and the ratio of the longest to shortest axis was $\sim 200$. The evidence
integral of this distribution can be
computed to good approximation over a prior domain bounded at $5\sigma$ using
the determinant of the covariance matrix and the prior volume $V$,
$Z_{MV}=V^{-1}(2/\pi)^{15/2}\det{C_{ij}}^{-1/2}\approx e^{-21.90}$.

\begin{figure*}
\centering
\includegraphics[width=0.3\textwidth]{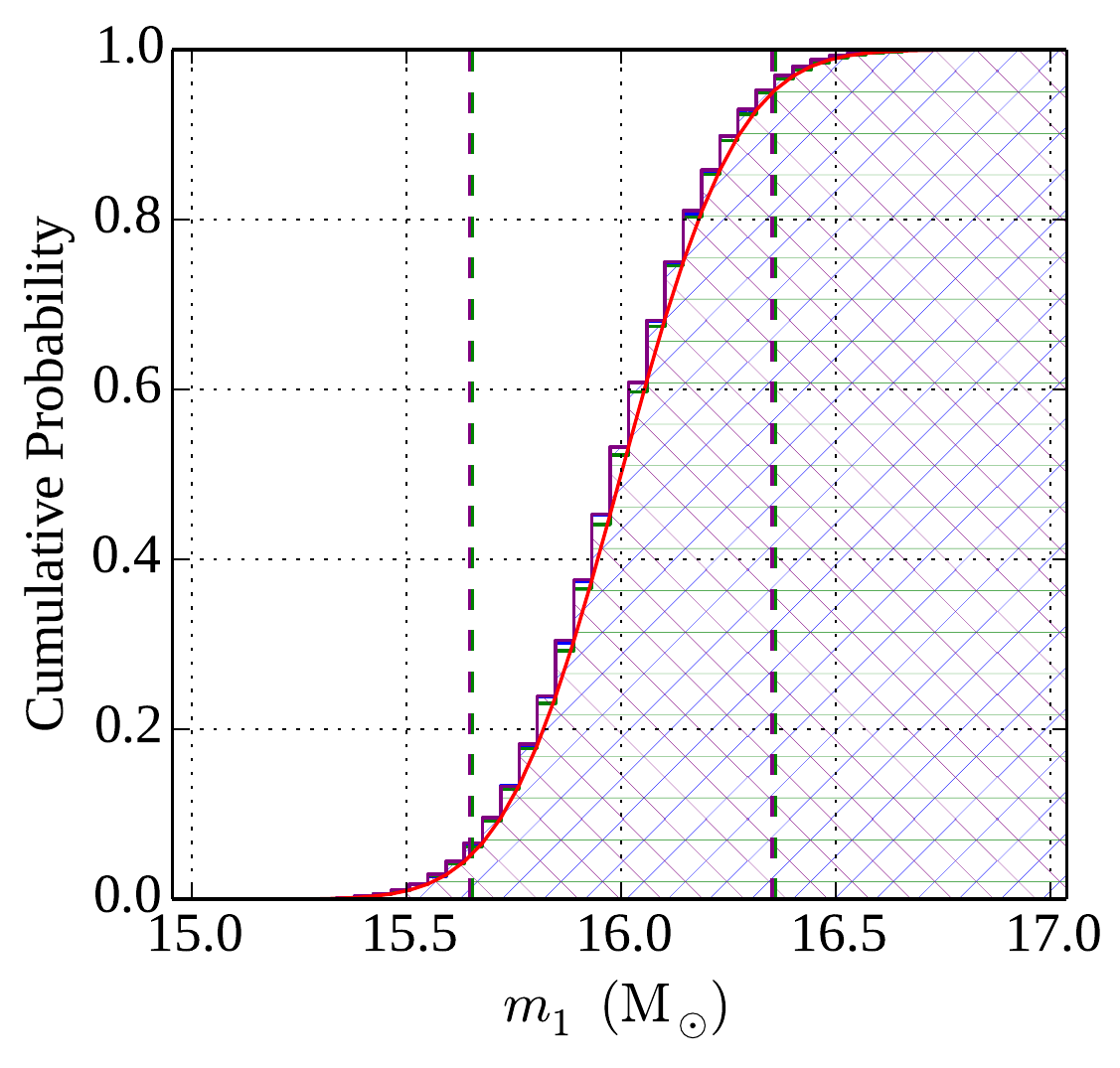}
\includegraphics[width=0.3\textwidth]{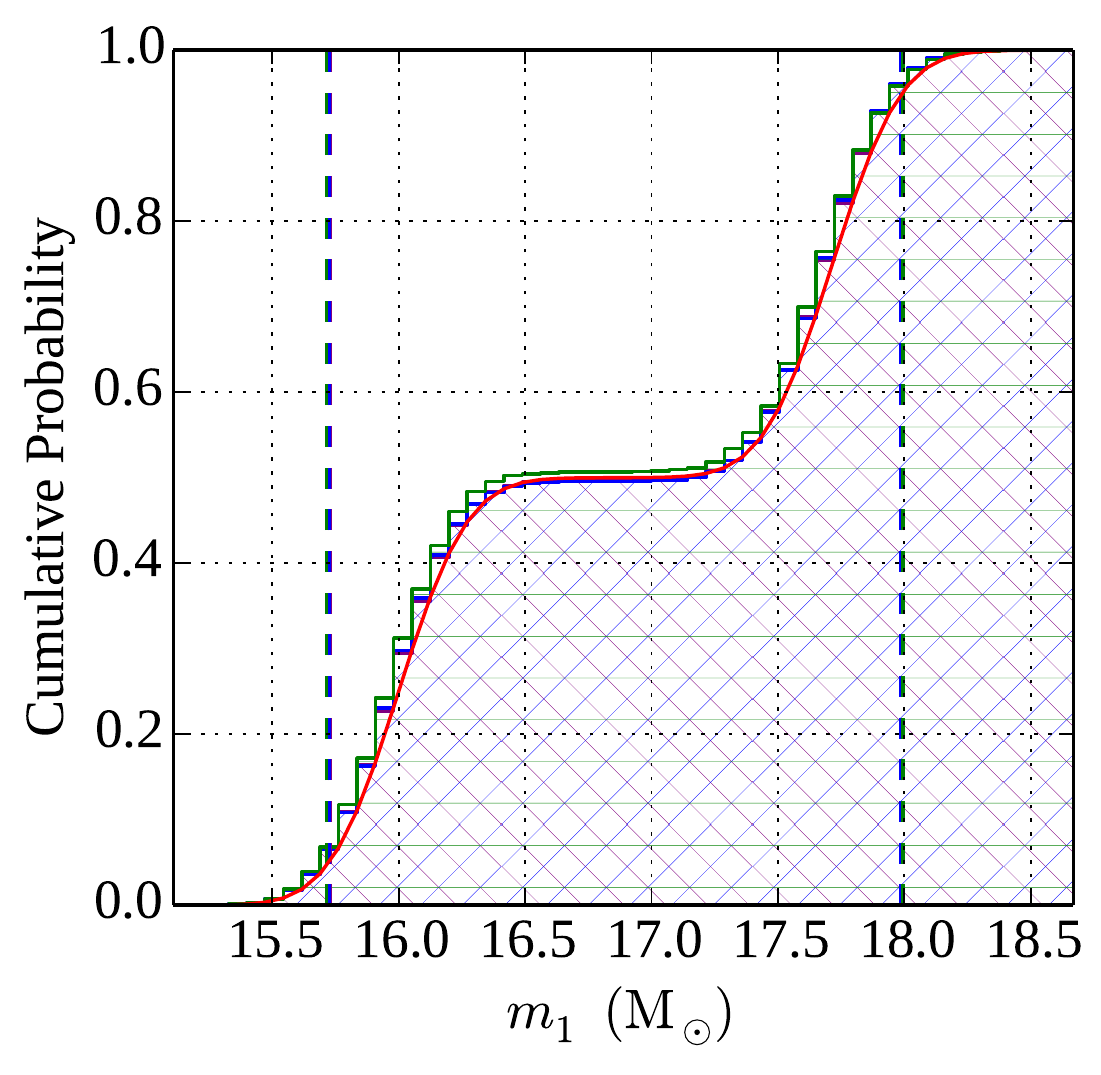}
\includegraphics[width=0.3\textwidth]{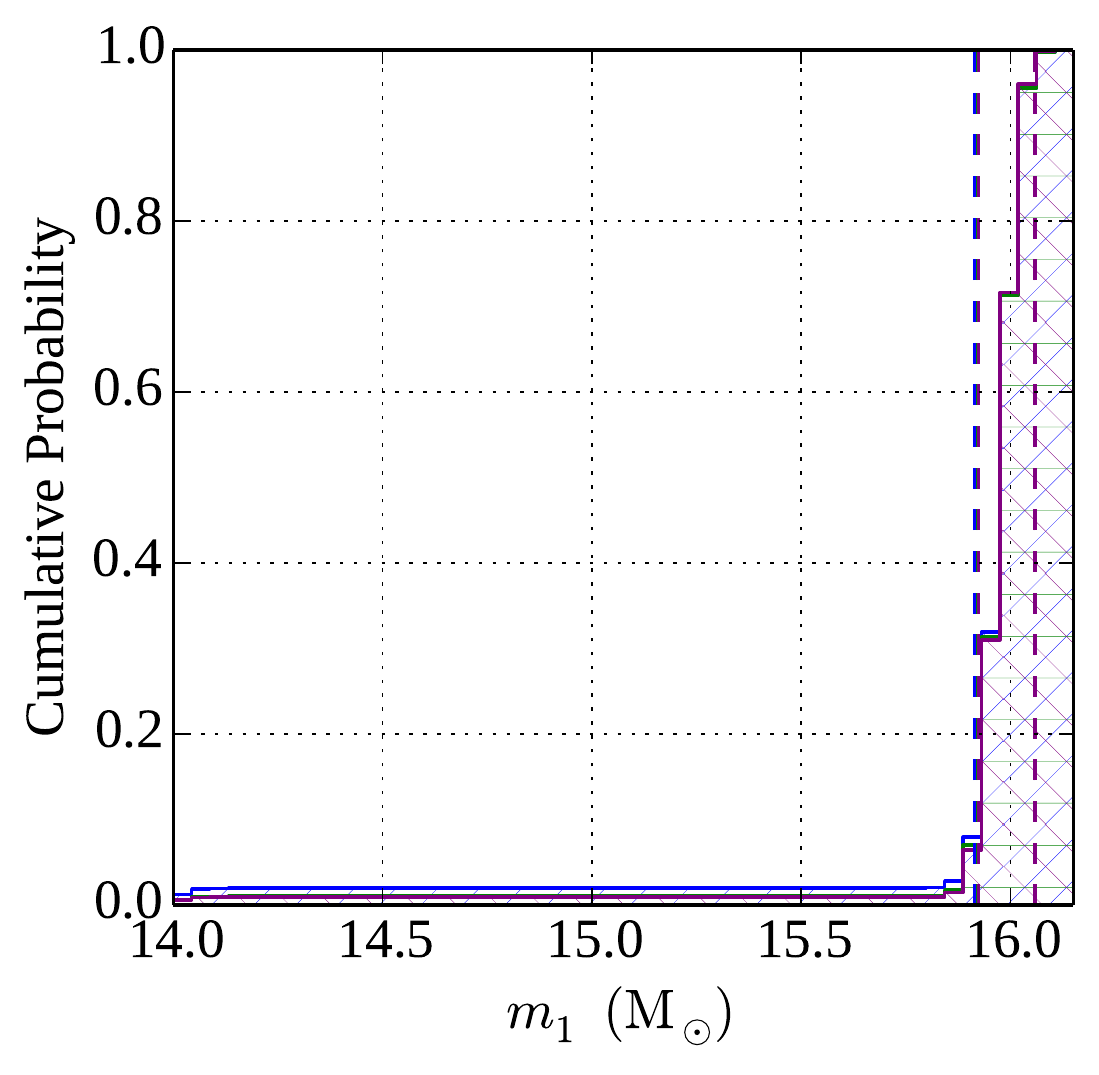}
\caption{\label{fig:analyticL} Example comparing cumulative
distributions for the analytic likelihood functions for each sampler for the
(arbitrary) $m_1$ parameter for the three test likelihood functions.
The samplers are shown as  Nest:purple left hatches, MCMC: green horizontal hatches BAMBI: blue right hatches,
with the true cumulative distributions shown in red where available. (left) uni-modal multivariate
Gaussian distribution (middle) bimodal distribution (right) Rosenbrock
distribution. The different methods show good agreement with each other, and
with the known analytic distributions. Vertical dashed lines indicate the 5\%--95\%
credibility interval for each method.}
\end{figure*}

The bimodal distribution was composed of two copies of the unimodal multivariate
Gaussian
used above, with two mean vectors $\hat\theta_i$ and $\hat\lambda_i$ separated
by $8\sigma$,
as defined by $C_{ij}$.
Using a bounding box at $\pm9\sigma$ about the mid-point of the two modes, the
evidence is calculated as $Z_{BM}'\approx e^{-30.02}$.

The Rosenbrock ``banana'' function is a commonly used test function for optimisation
algorithms~\cite{RosenbrockBanana}.
For this distribution, we do not have analytic one-dimensional marginal
distributions to compare to, or known evidence values, so we were only able to do cross-comparisons
between the samplers.

Each sampler was run targeting these known distributions, and the recovered posterior distributions
and evidences were compared. The posterior distributions agreed for all parameter as expected,
and an example of one parameter is shown in figure \ref{fig:analyticL}.

The recovered evidence values are shown in table \ref{tab:AnalyticEvidence}. For the MCMC sampler the quoted errors come from the thermodynamic integration quadrature error estimates described in \S\,\ref{sec:MCMCevidence}; for the nested samplers the quoted errors are estimated by running the algorithm multiple times and computing
the standard deviation of the results.
For the simplest unimodal and bimodal distributions we see excellent agreement between
the sampling methods, which agree within the $1\sigma$ statistical error estimates.
The more difficult Rosenbrock likelihood results in a statistically significant disagreement between
the nested sampling and BAMBI algorithms, with BAMBI returning the higher evidence estimate. 
To highlight the difficulty, for this problem
the thermodynamic integration methods used with MCMC required 64 temperature ladder steps to reach
convergence to $\beta\langle \log L\rangle=0$ at high temperatures, as opposed to the 16 used in the other problems.
This pattern is repeated in the evidence for the signals, where there is a difference of several standard
deviations between the methods.

\subsection{Simulated GW signals}\label{sec:injections} 

As an end-to-end test, we ran all three sampling flavours of \li~({\it MCMC},
section~\ref{sec:MCMC}; {\it Nest}, section~\ref{sec:LINestdescription} and {\it BAMBI},
section~\ref{sec:BAMBIdescription}) on three test signals, described in
table~\ref{tab:injections}. These signals were injected into coloured Gaussian noise
of known power spectrum and recovered with the same approximant used in generating
the injection, listed in table \ref{tab:injections}.
Since we used inspiral-only waveforms models for both injection and recovery, there
is a sharp cutoff in the signal above the waveform's termination frequency. It has
been shown that in some circumstances the presence of this cutoff provides an artificially sharp feature
which can improve parameter estimation beyond that of a realistic signal~\cite{Mandel:2014tca}.
Nonetheless, since the focus of this study is the consistency of the algorithms, we can proceed
to use the sharply terminating waveforms for comparison purposes.

\begin{table*}
\begin{tabular}{|c|c|c|c|c|c|c|c|c|c|c|c|}
\hline
Fig. & Name & Approximant & $m_1$ & $m_2$ & $a_1$ & $a_2$ & $t_1$ & $t_2$ & $\iota $ & distance &
Network SNR\\
& & & $(\Msun)$ & $(\Msun)$ & & & (\mbox{Rad}) & (\mbox{Rad})  & $(\mbox{Rad})$ & (Mpc) & \\
\hline
\ref{fig:BNS} & BNS & TaylorF2 3.5PN & 1.3382 & 1.249 & 0 & 0 & - & - & 2.03 & 135 & 13 \\
\ref{fig:NSBH} & NSBH & SpinTaylorT4 3.5PN & 15 & 1.35 & 0.63 & 0 & 0 & - & 1.02 & 397 & 14 \\
\ref{fig:BBH} & BBH & SpinTaylorT4 3.5PN & 15 & 8 & 0.79 & 0.8 & 3.1 & 1.44 & 2.307 & 500 & 15 \\
\hline
\end{tabular}
\caption{\label{tab:injections}Details of the injected signals used in section \ref{sec:injections}, showing
the waveform approximant used with the masses ($m_{\{1,2\}}$), spin magnitudes and tilt angles ($a_{\{1,2\}},t_{\{1,2\}}$),  and the distance and inclination ($\iota$).}
\end{table*}

\begin{figure*}
\centering
\includegraphics[width=0.32\textwidth]{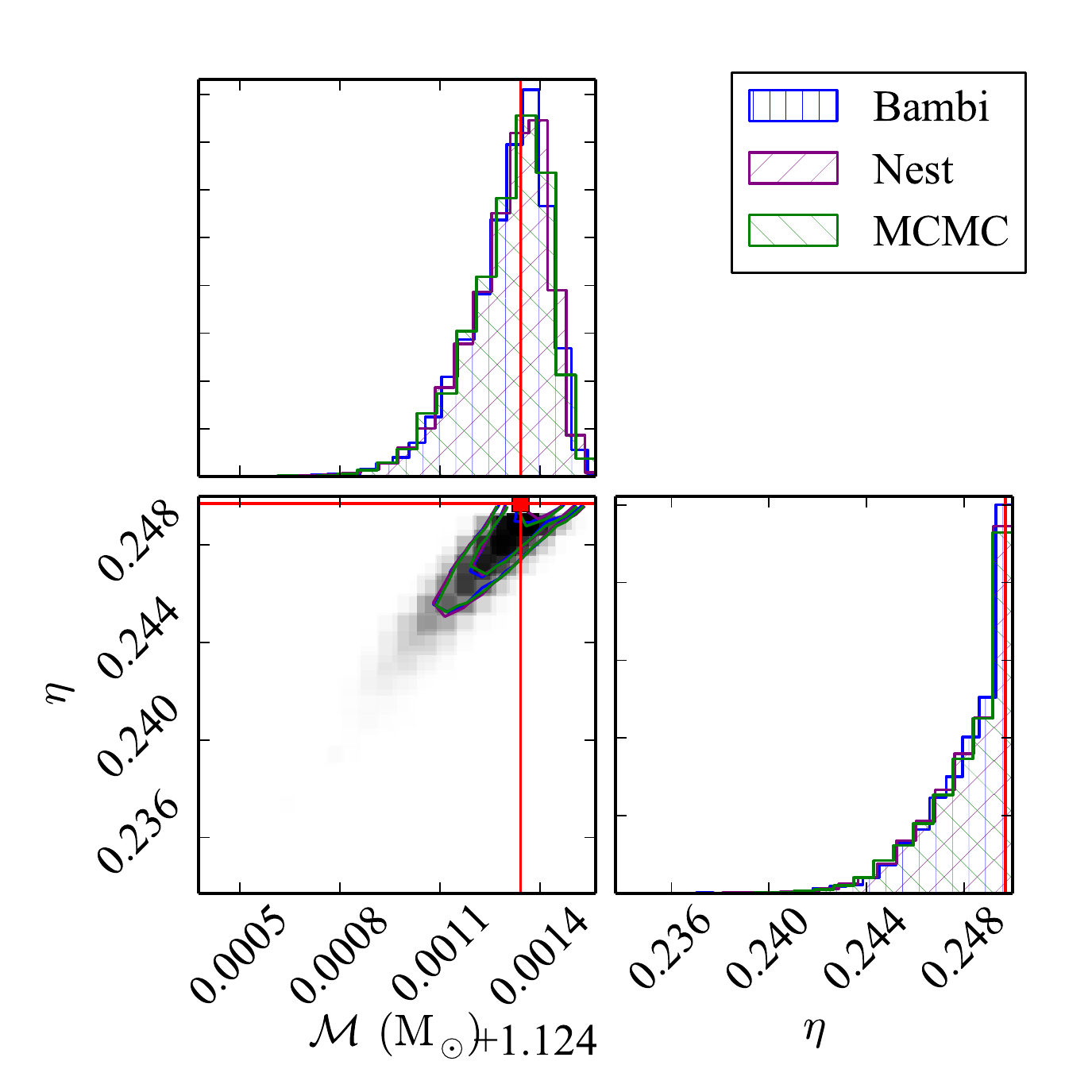}
\includegraphics[width=0.32\textwidth]{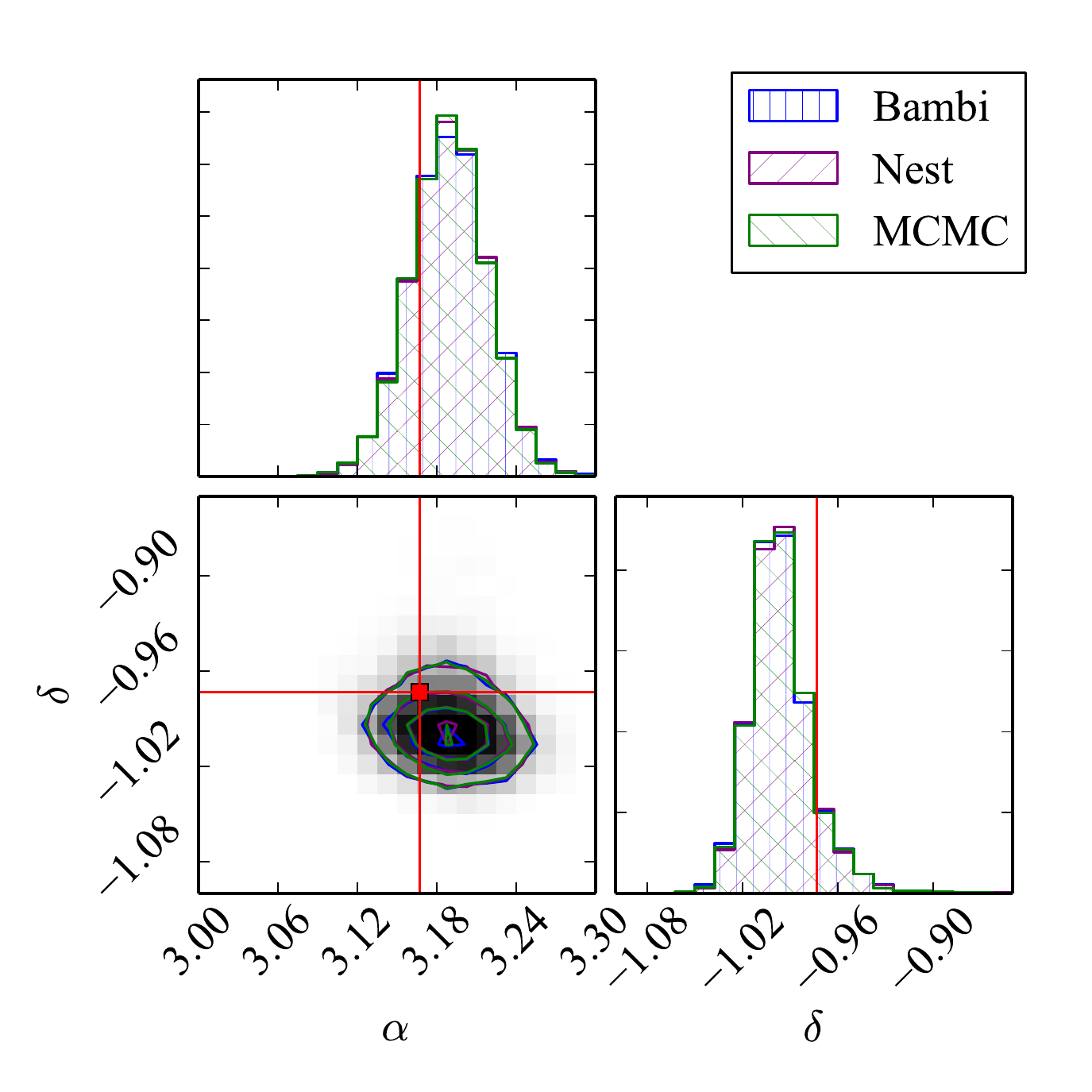}
\includegraphics[width=0.32\textwidth]{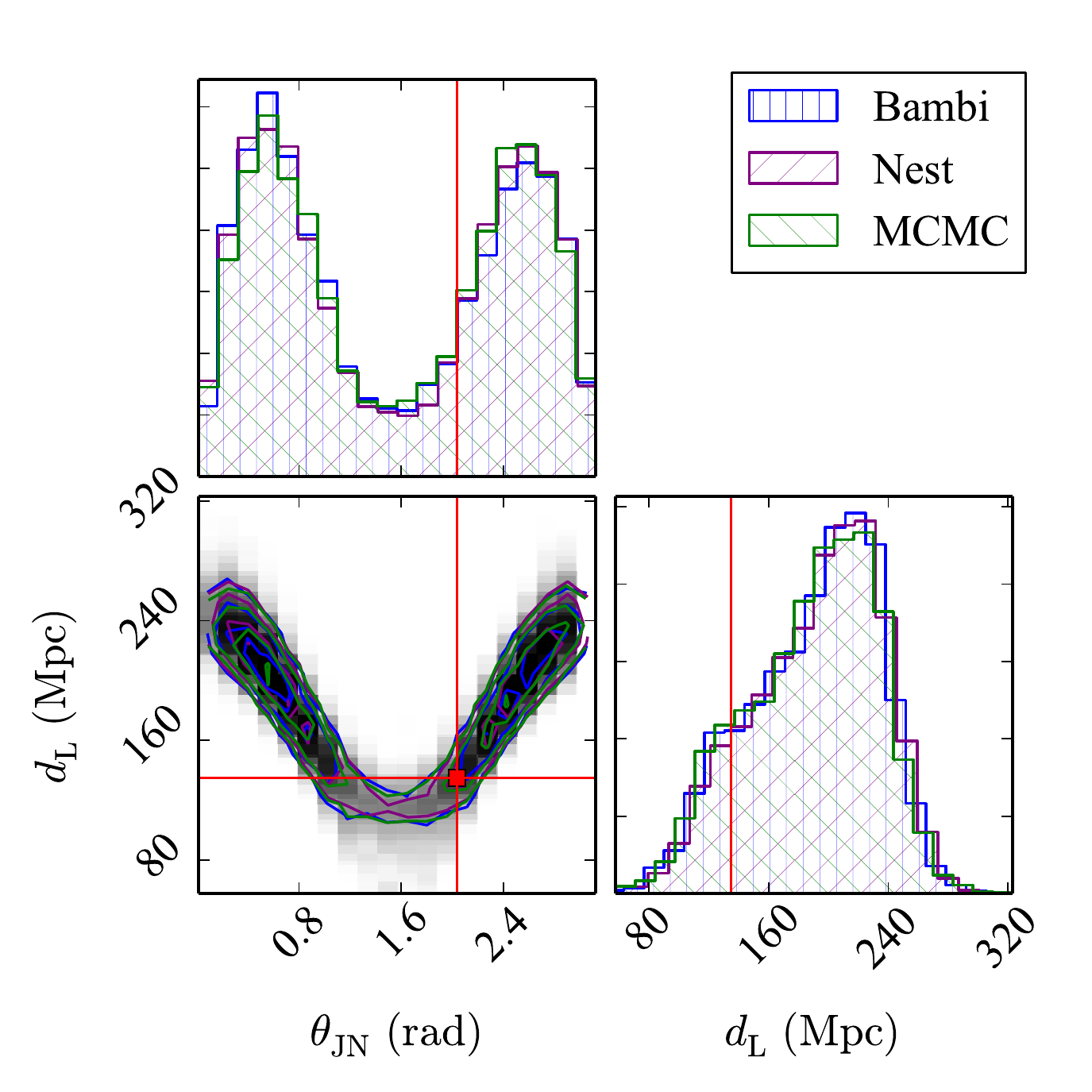}
\caption{\label{fig:BNS} Comparison of probability density 
functions for the BNS signal (table \ref{tab:injections}) as determined by each
sampler. Shown are selected 2D posterior density functions in greyscale, with
red cross-hairs indicating the true parameter values, and contours indicating
the 90\% credible region as estimated by each sampler. On the axes are superimposed the one-dimensional
marginal distributions for each parameter, as estimated by each sampler, and the true
value indicated by a vertical red line.
The colours correspond to blue:
Bambi, magenta: Nest, green: MCMC. (left) The mass posterior distribution 
parametrized by chirp mass and symmetric mass ratio. (centre) The location of the source on the sky. (right)
The distance $d_L$ and inclination $\theta_{JN}$ of the source showing the
characteristic V-shaped degeneracy.
}
\end{figure*}

\begin{figure*}
\centering
\includegraphics[width=0.32\textwidth]{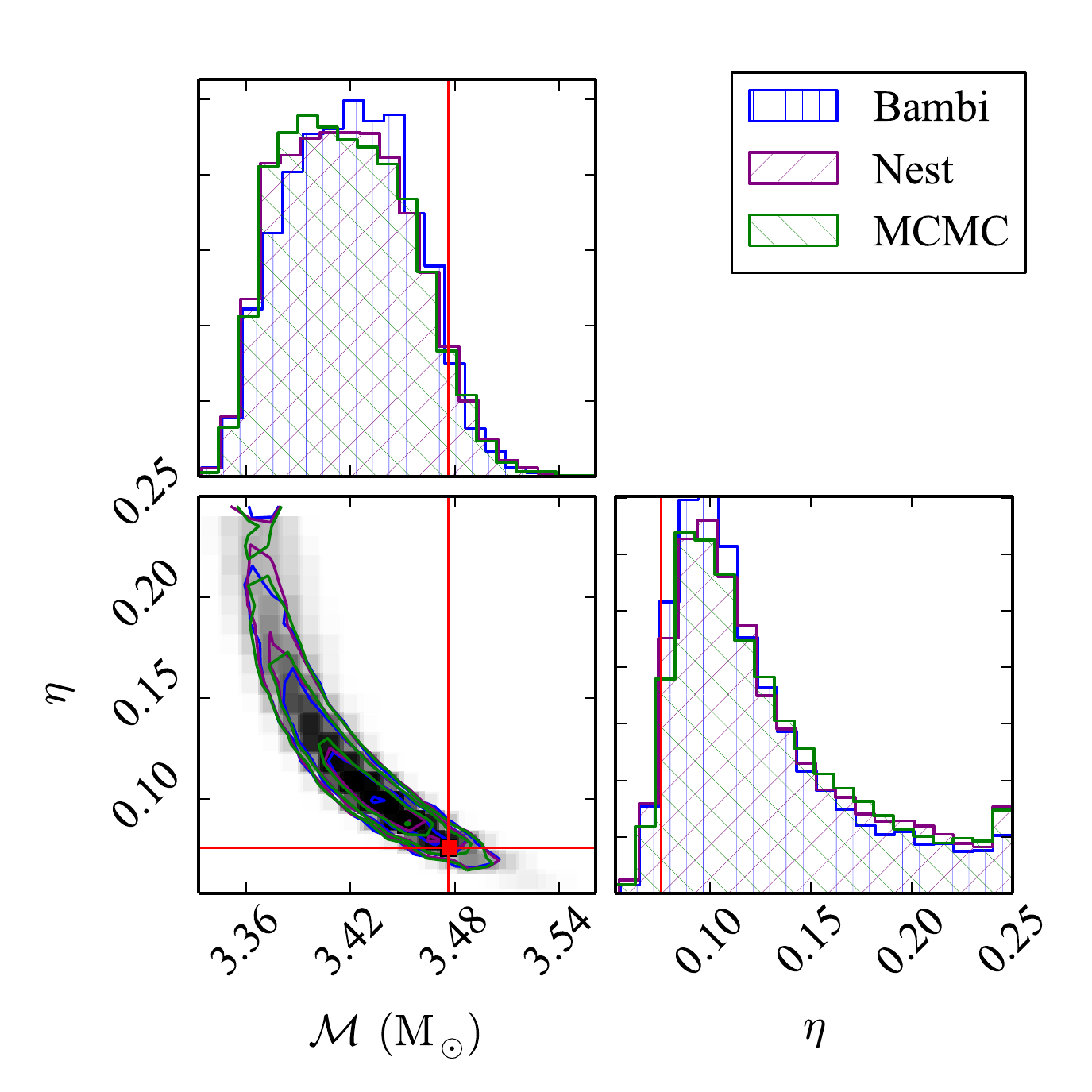}
\includegraphics[width=0.32\textwidth]{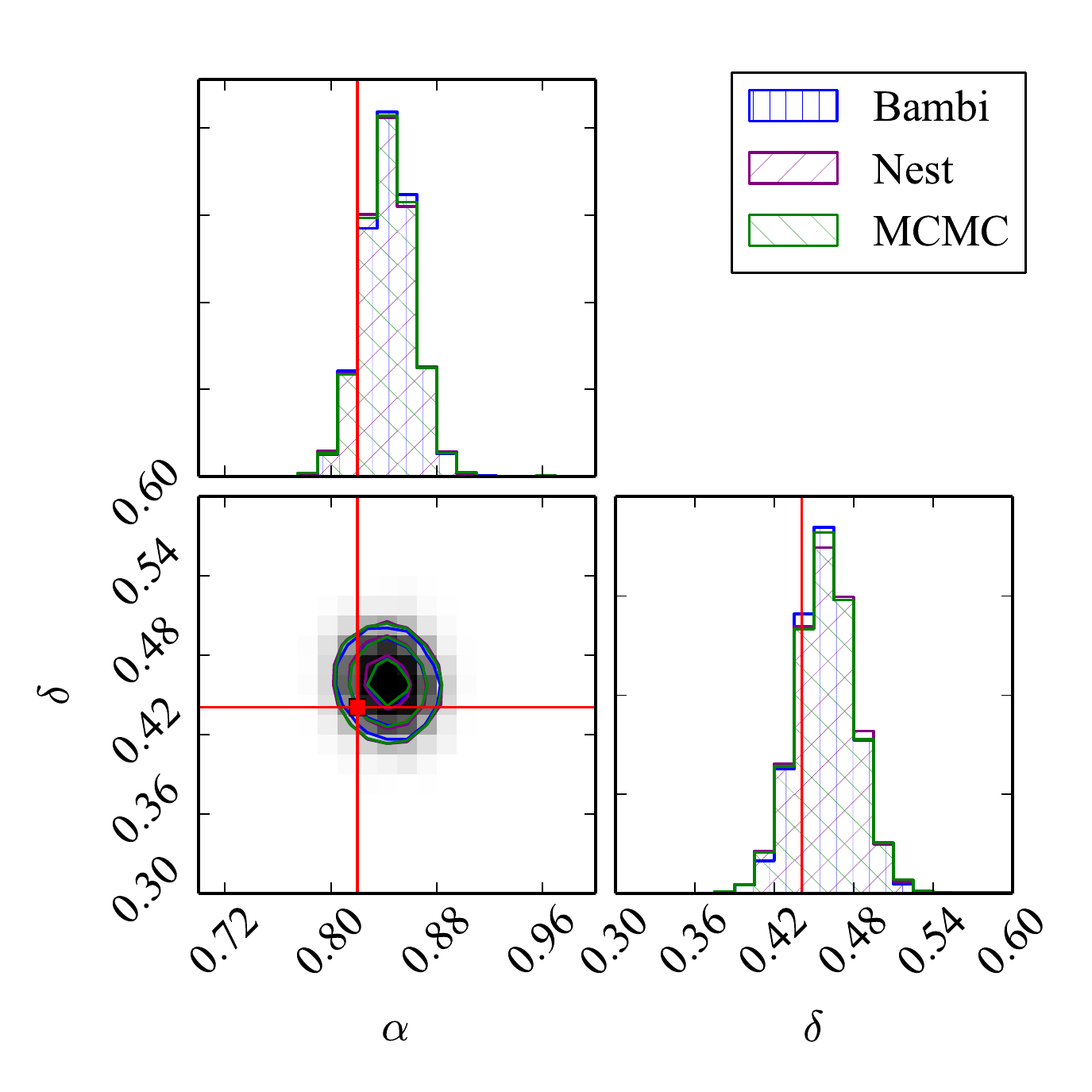}
\includegraphics[width=0.32\textwidth]{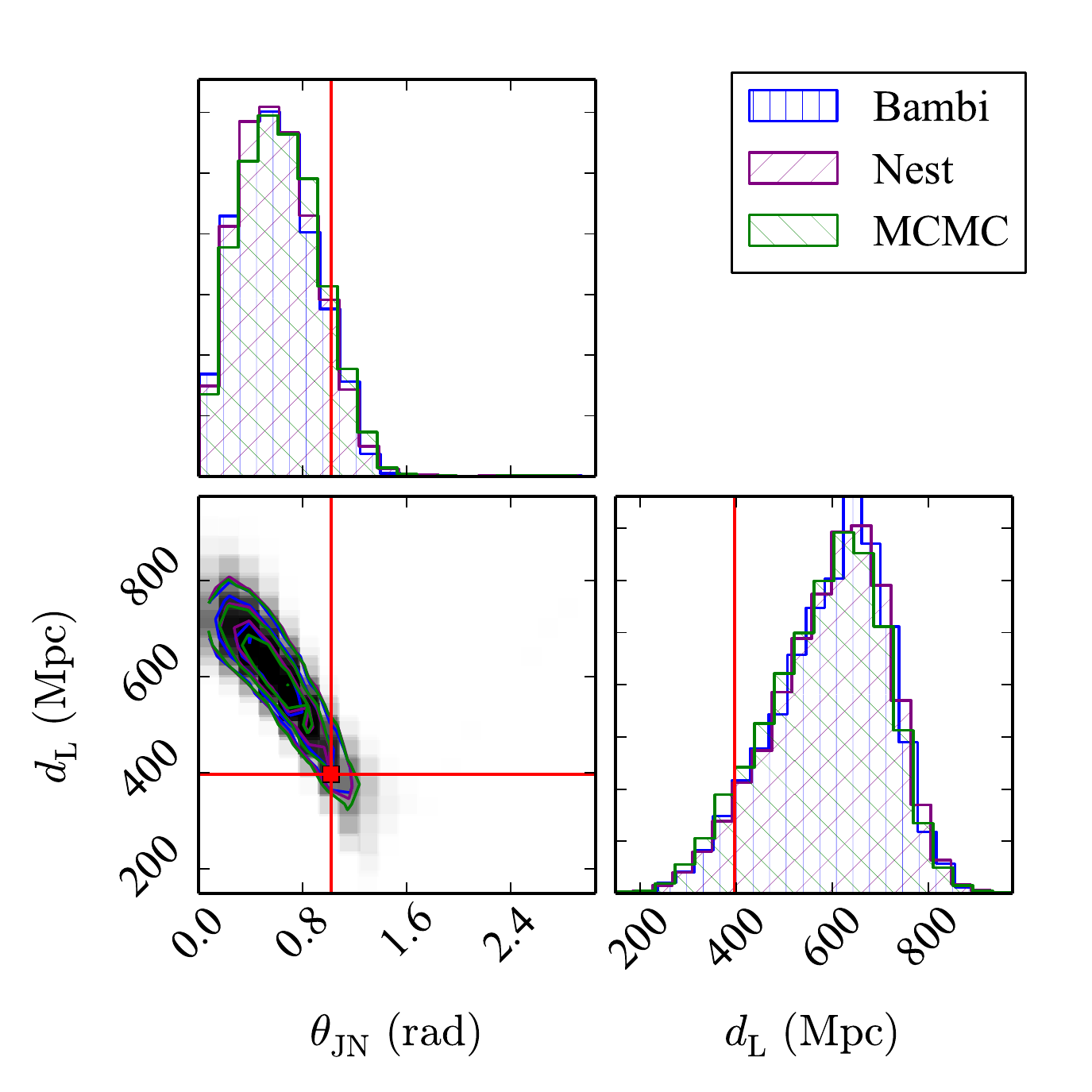}
\includegraphics[width=0.32\textwidth]{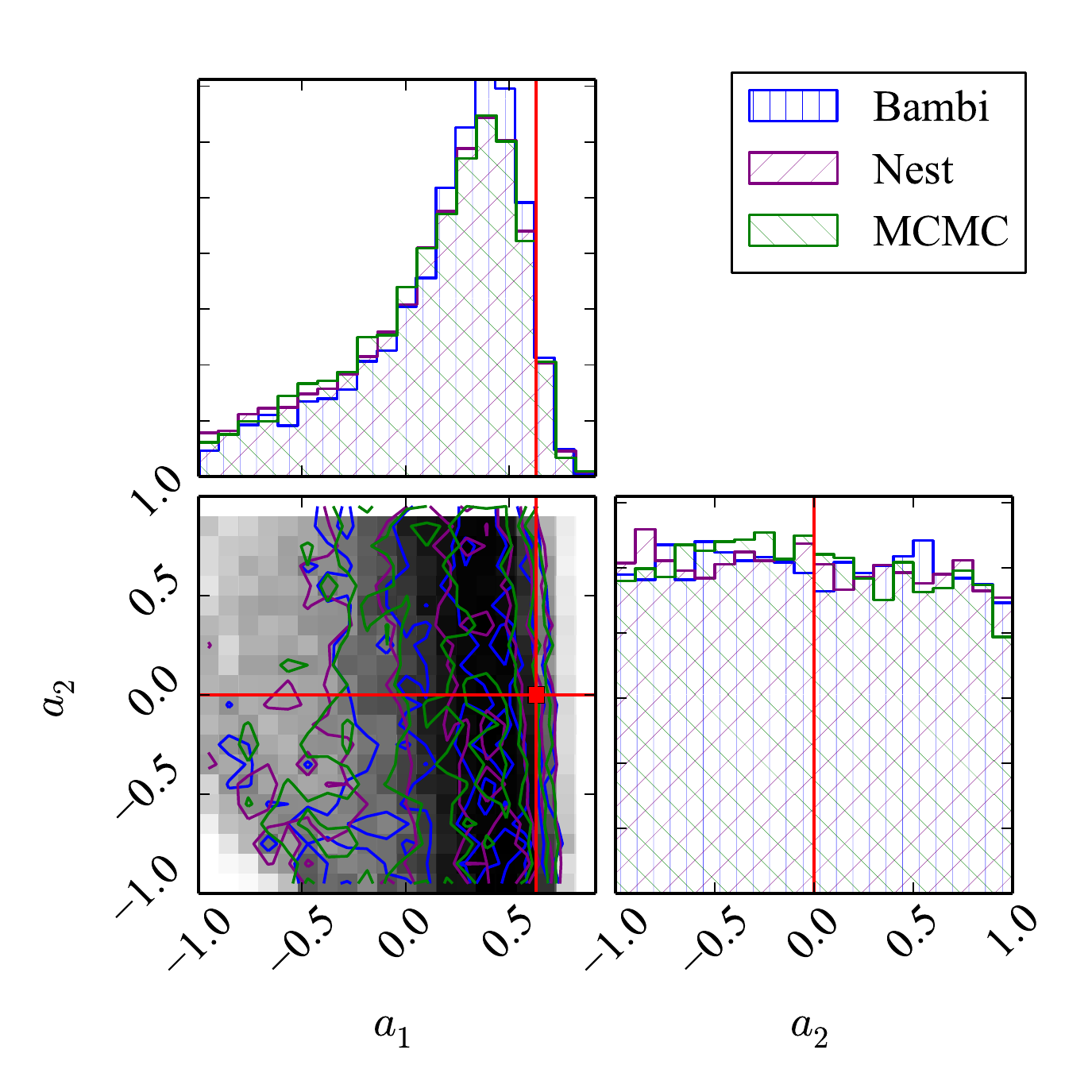}
\includegraphics[width=0.32\textwidth]{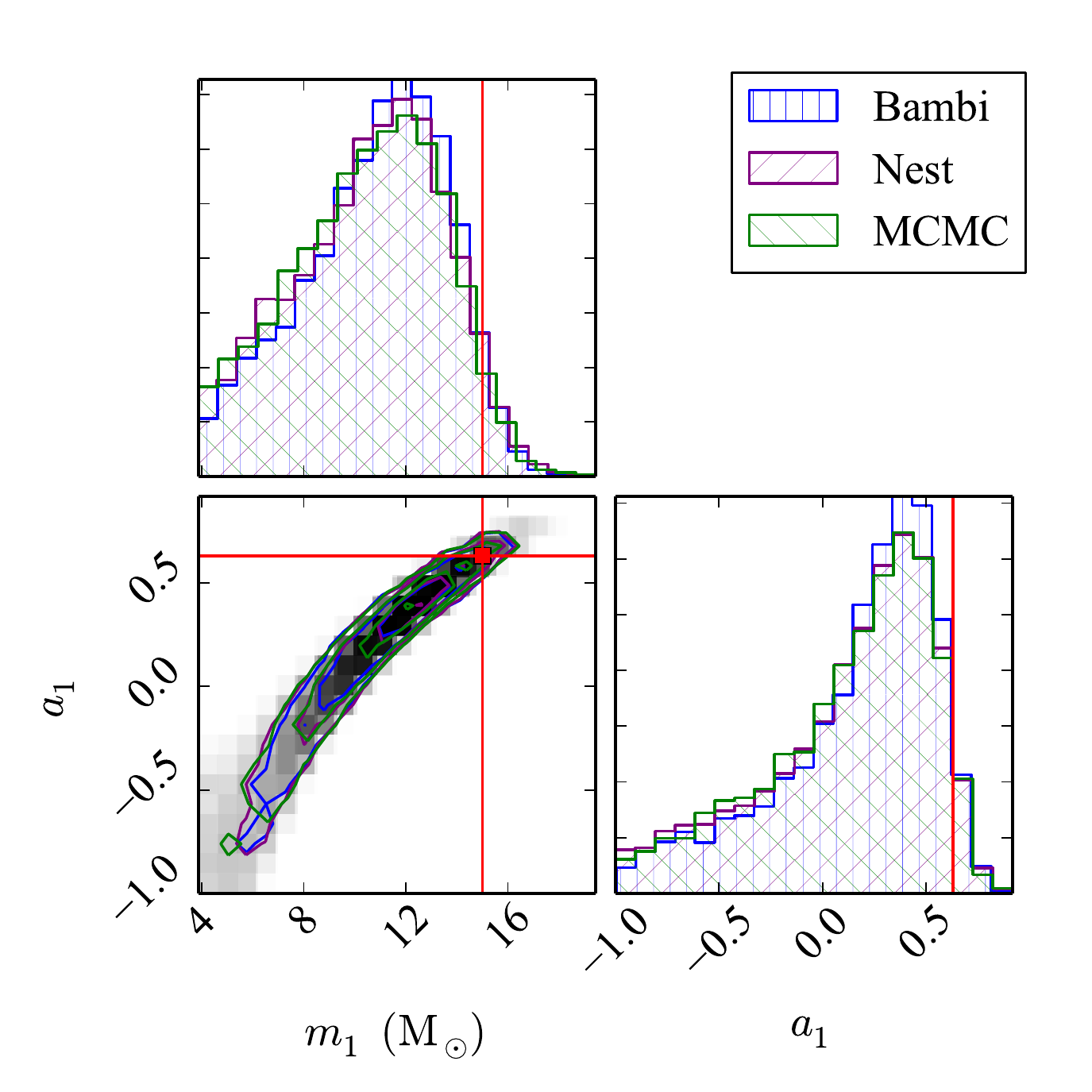}
\includegraphics[width=0.32\textwidth]{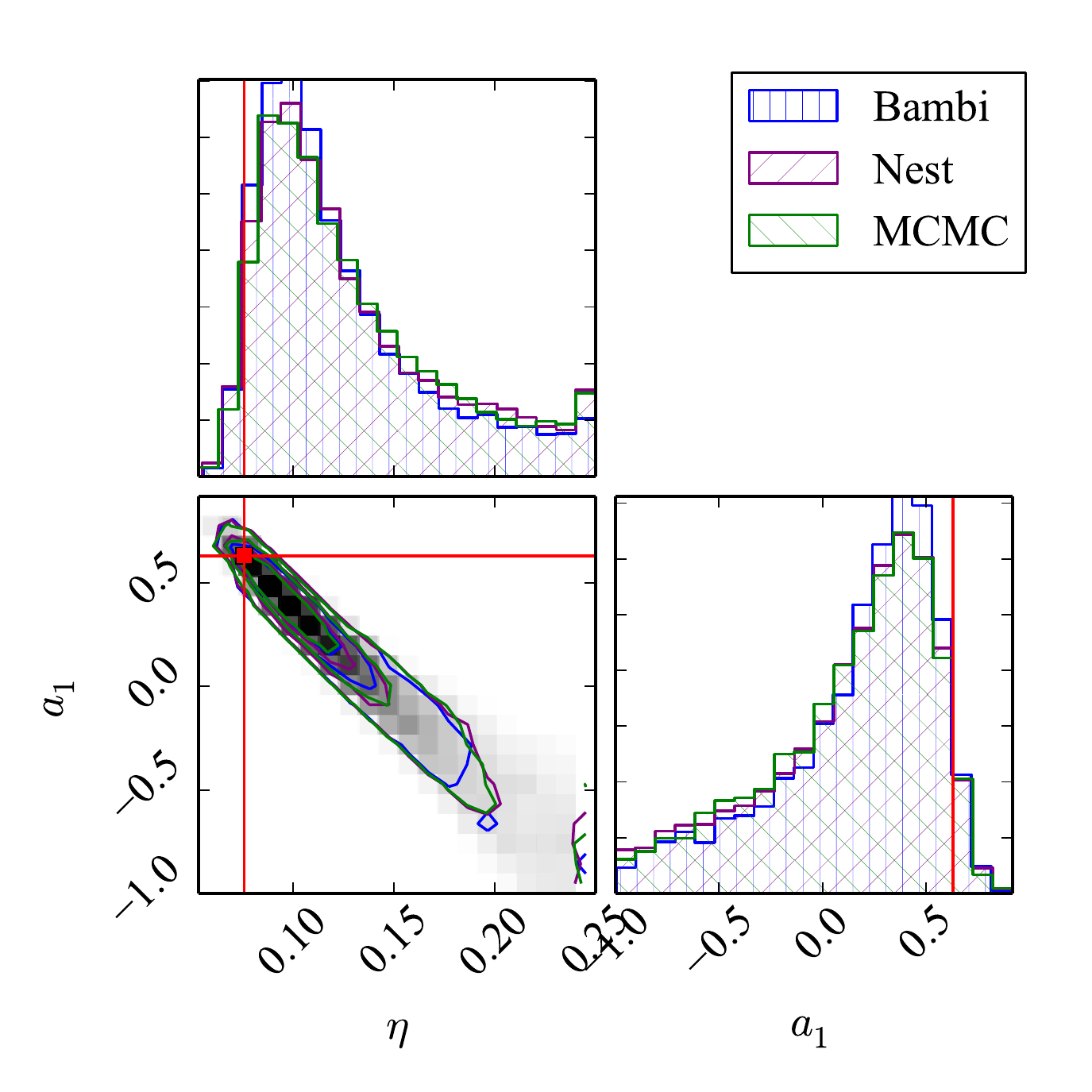}
\caption{\label{fig:NSBH} Comparison of probability density 
        functions for the NSBH signal (table \ref{tab:injections}), with same color scheme as fig~\ref{fig:BNS}.
        (first row left) The mass posterior distribution 
        parametrized by chirp mass and symmetric mass ratio. (first row centre) The location of the source on the sky. (first row right)
        The distance $d_L$ and inclination $\theta_{JN}$ of the source. In this case the V-shaped degeneracy is broken,
        but the large correlation between $d_L$ and $\theta_{JN}$ remains. (second row left) The spin magnitudes posterior distribution. 
        (second row centre) The spin and mass of the most massive member of the binary 
        illustrating the degeneracy between mass and spin. (second row right) The spin and symmetric mass ratio.}
\end{figure*}

\begin{figure*}
\centering
\includegraphics[width=0.32\textwidth]{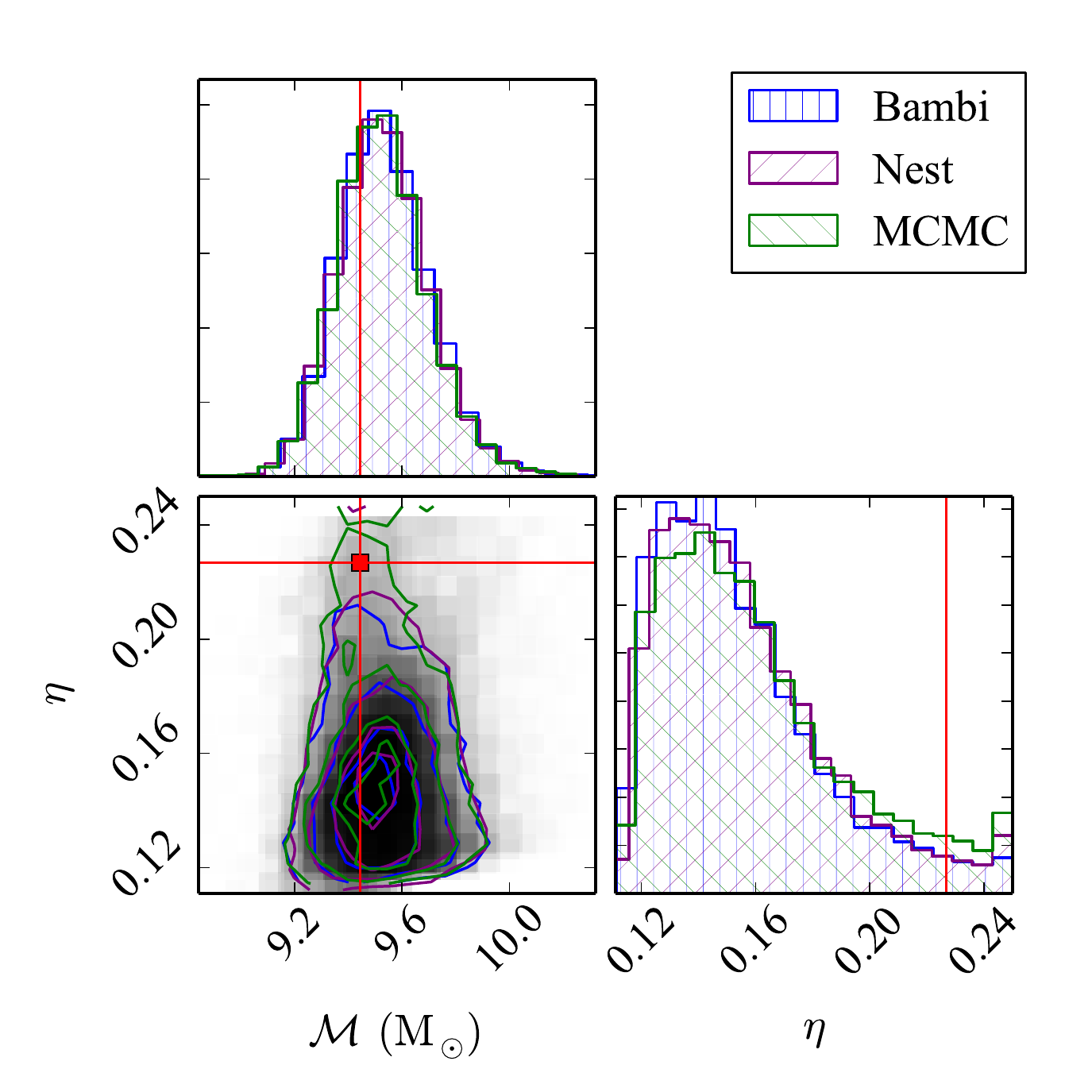}
\includegraphics[width=0.32\textwidth]{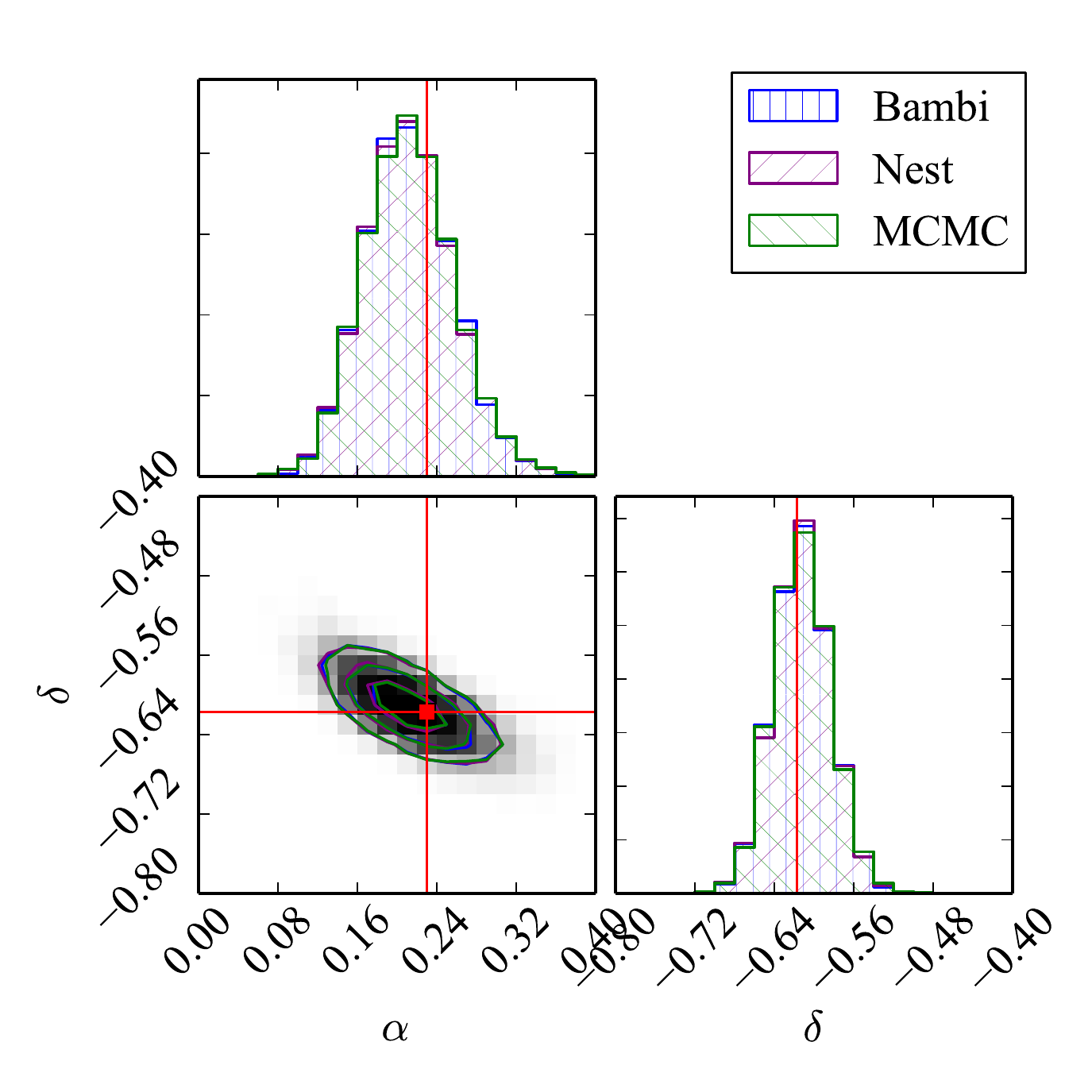}
\includegraphics[width=0.32\textwidth]{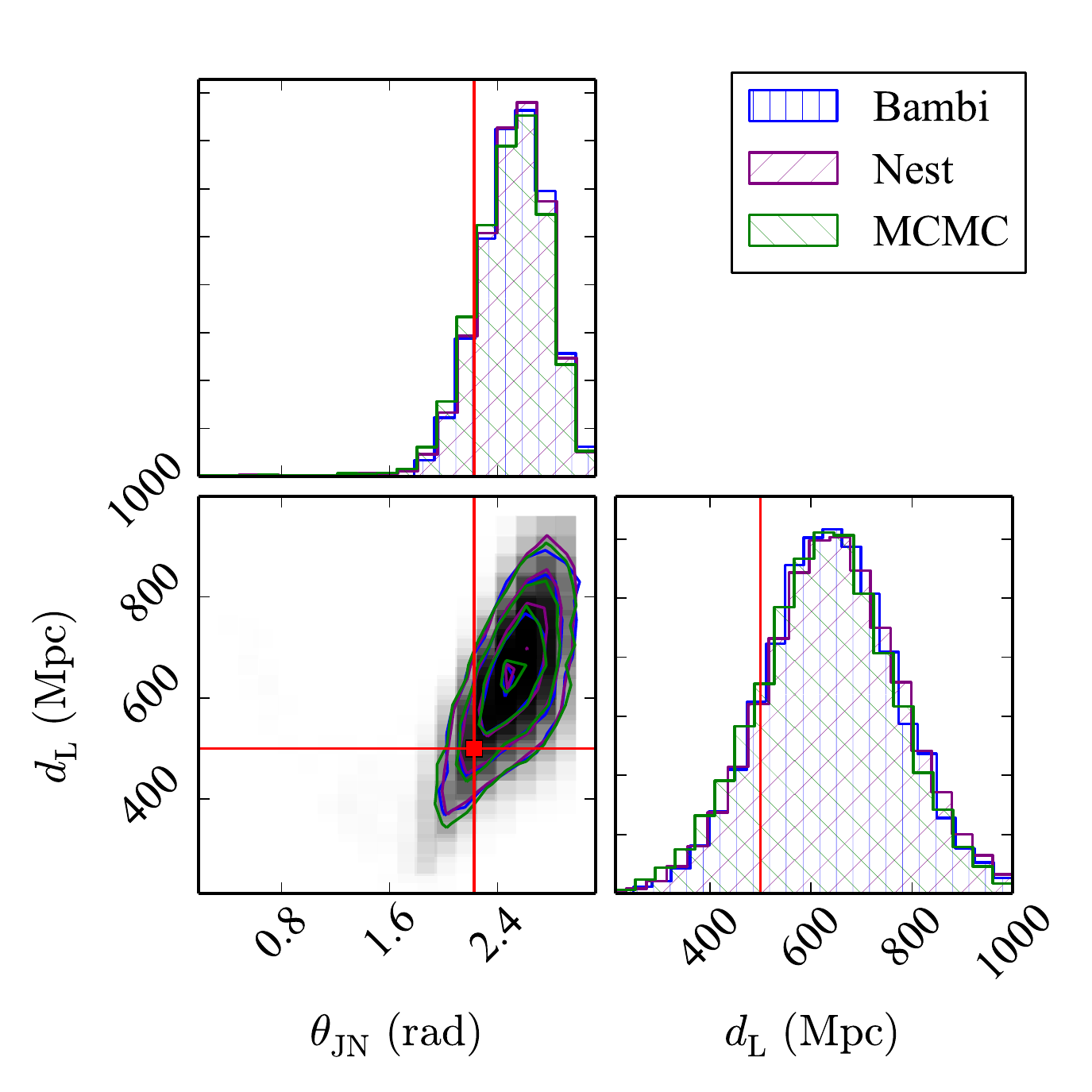}
\includegraphics[width=0.32\textwidth]{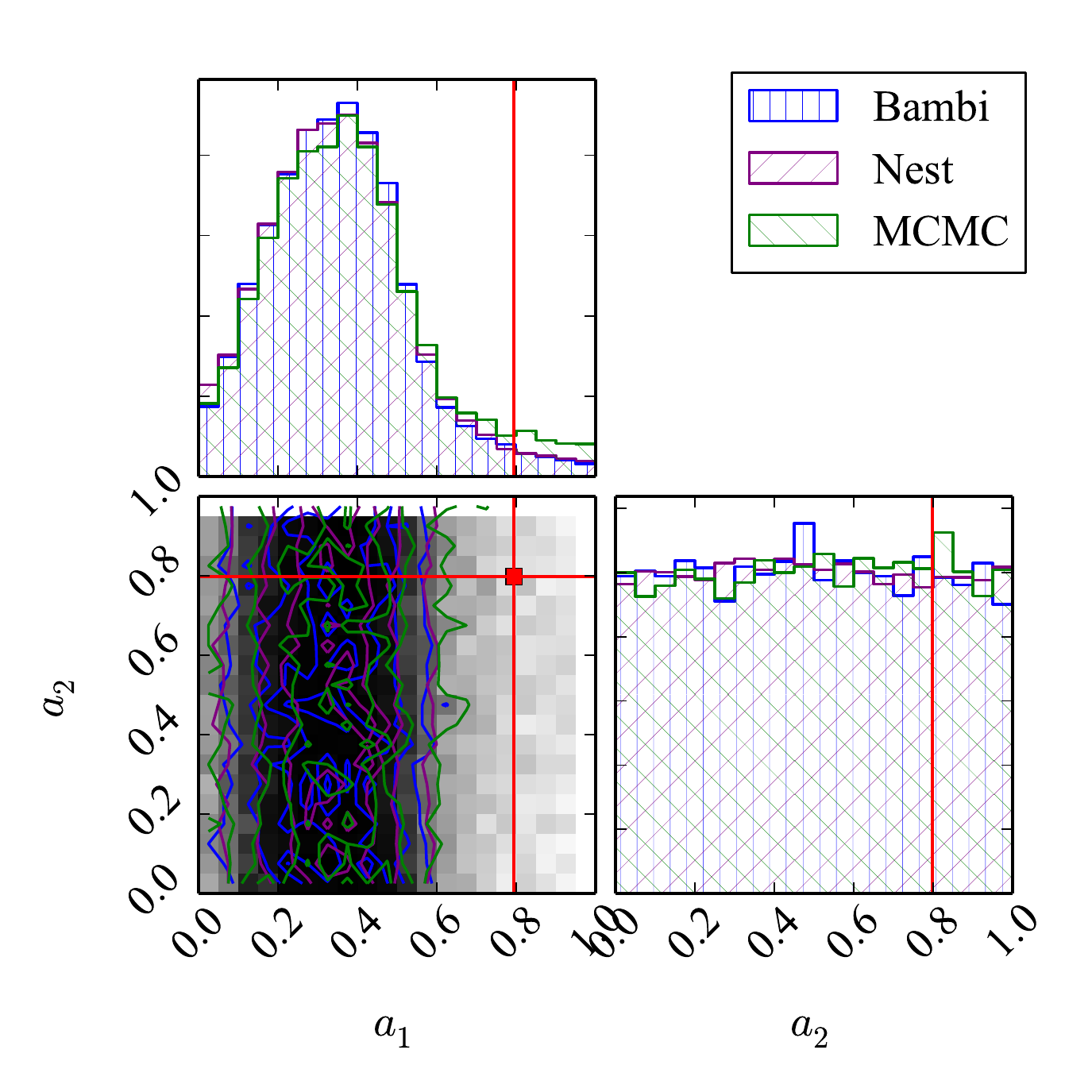}
\includegraphics[width=0.32\textwidth]{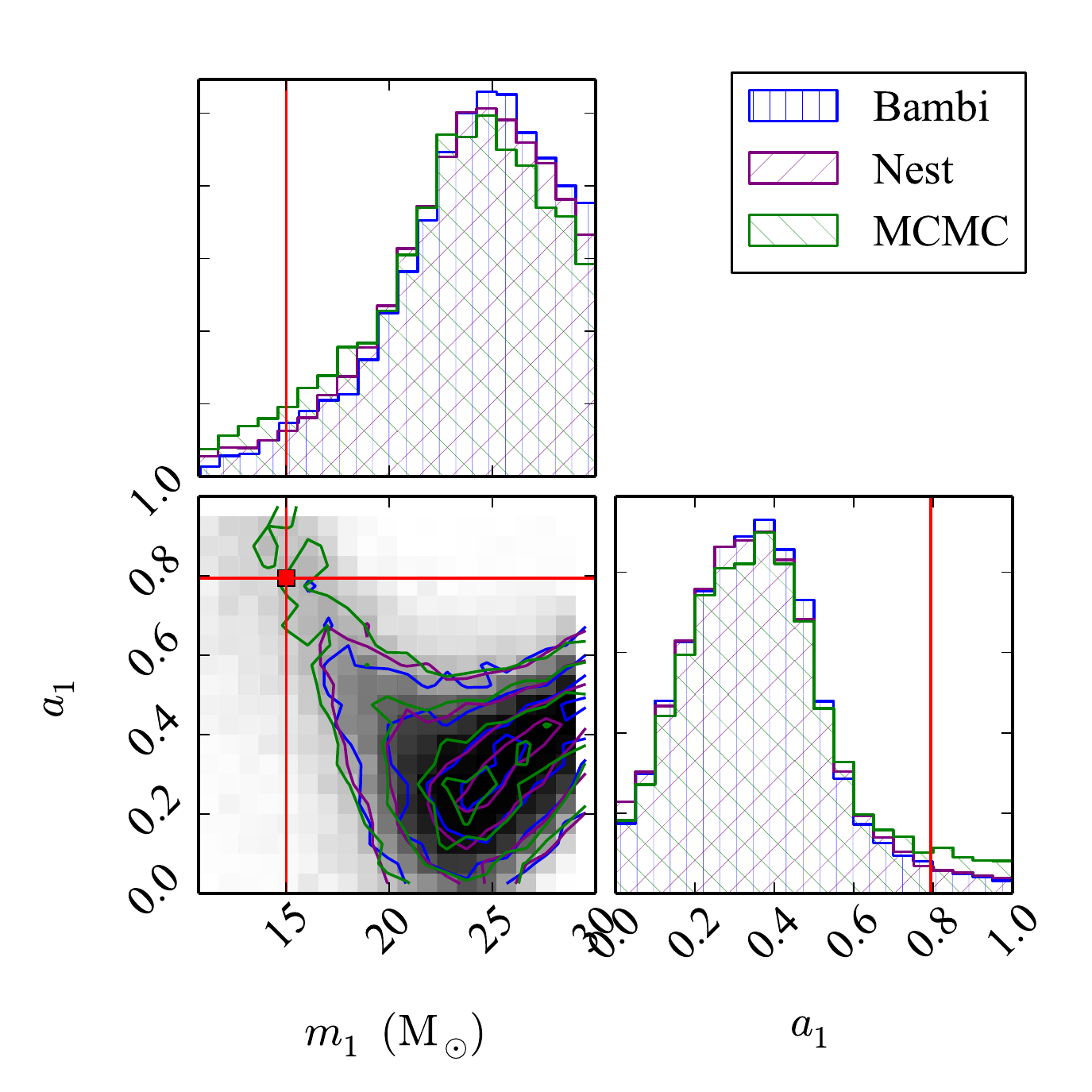}
\includegraphics[width=0.32\textwidth]{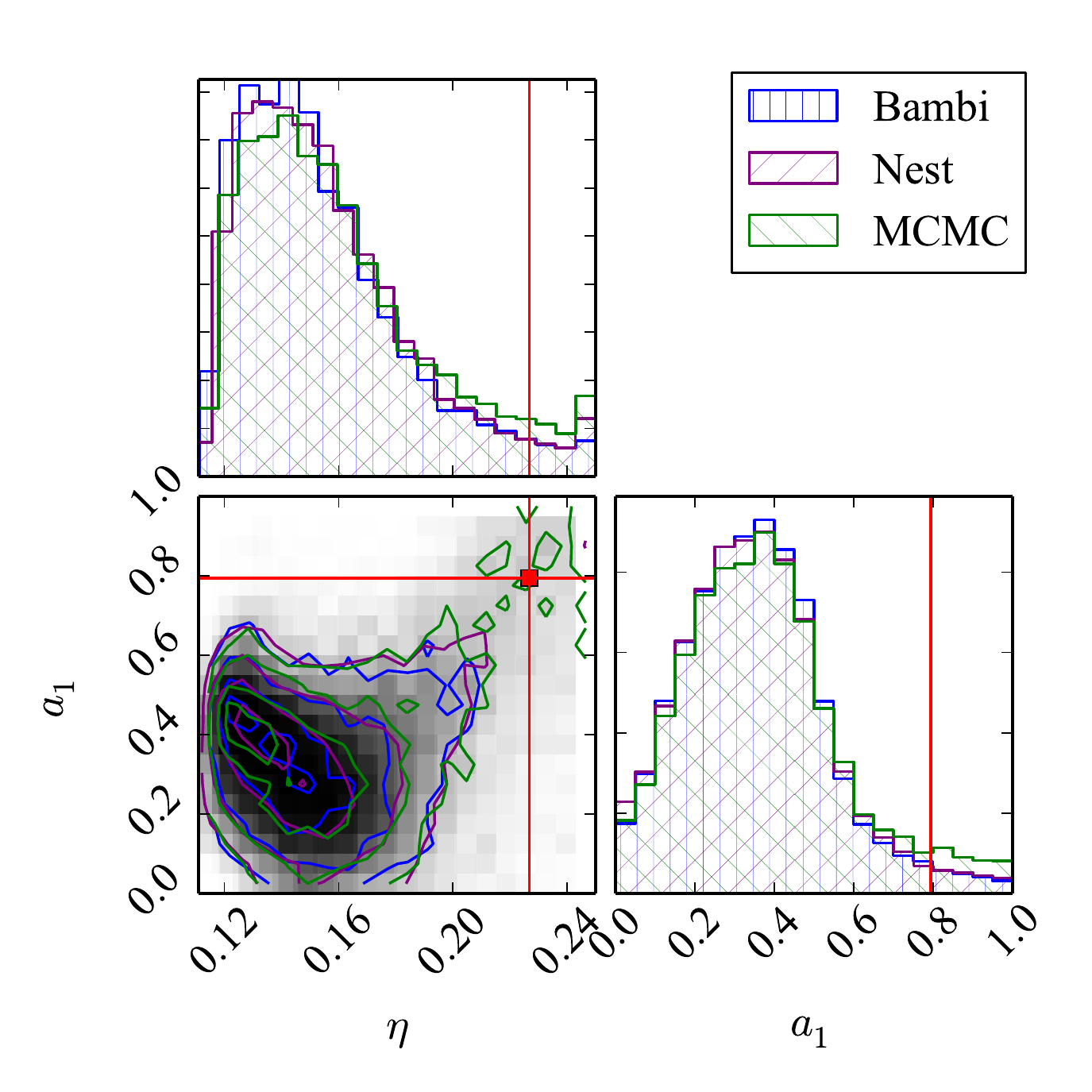}
\includegraphics[width=0.32\textwidth]{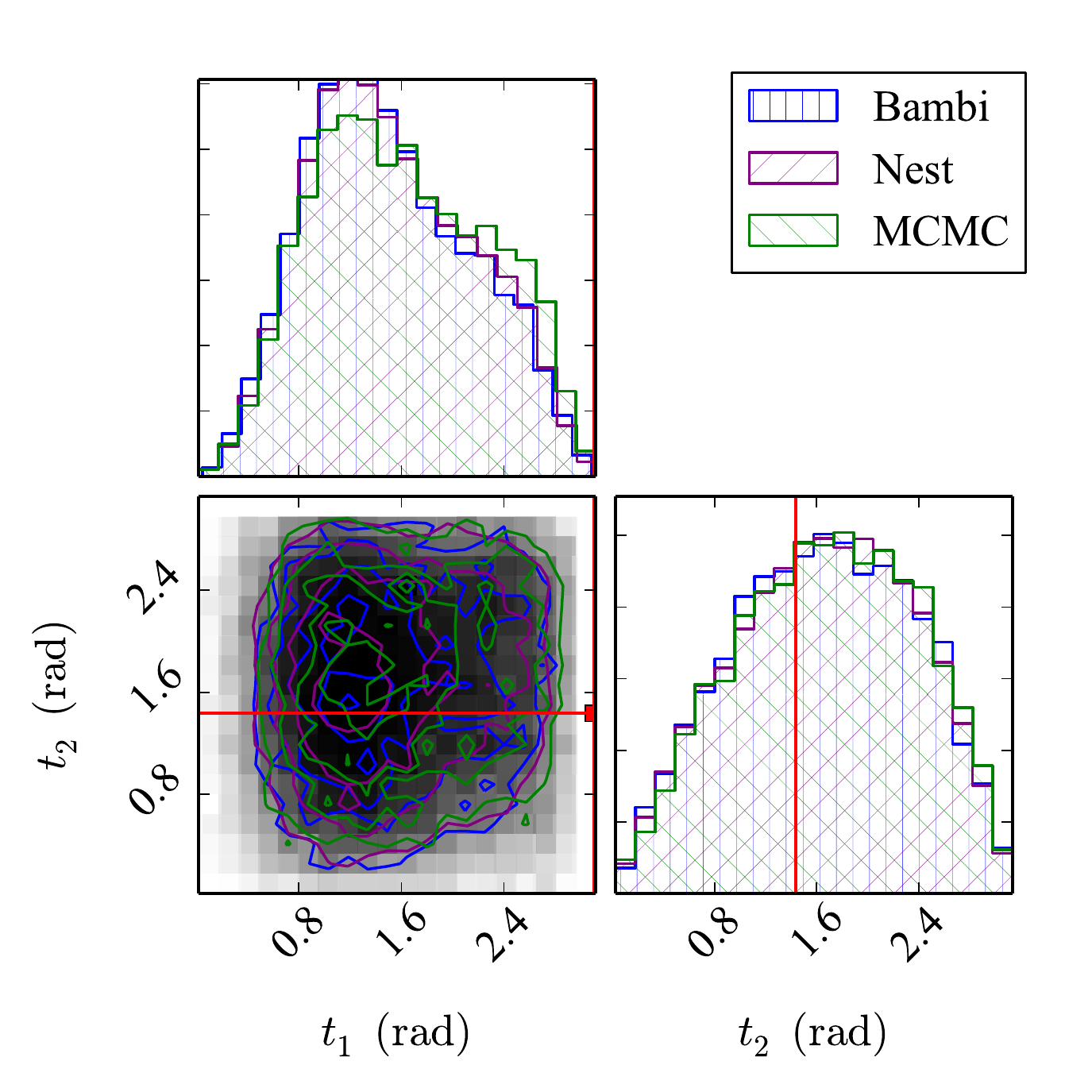}
\includegraphics[width=0.32\textwidth]{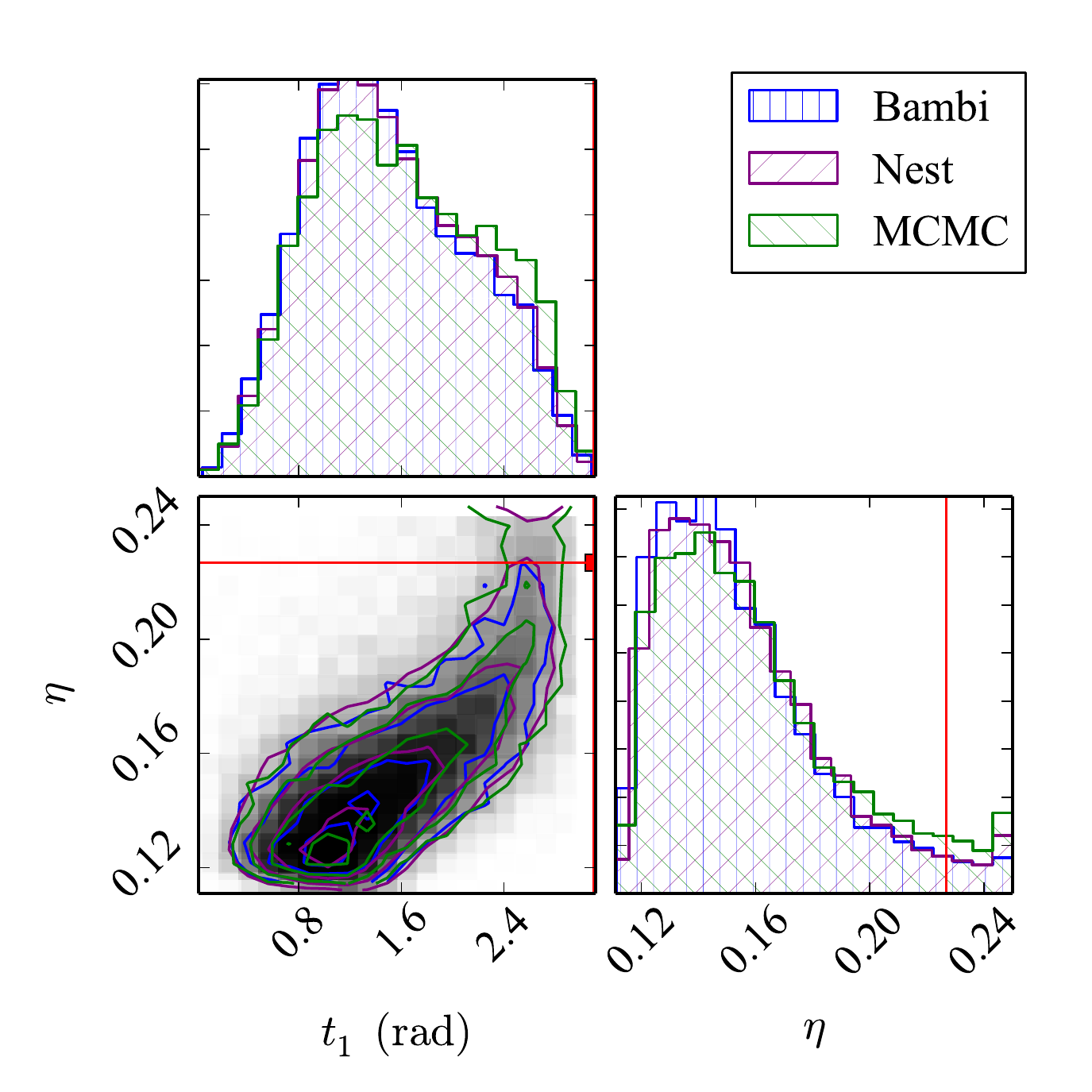}
\includegraphics[width=0.32\textwidth]{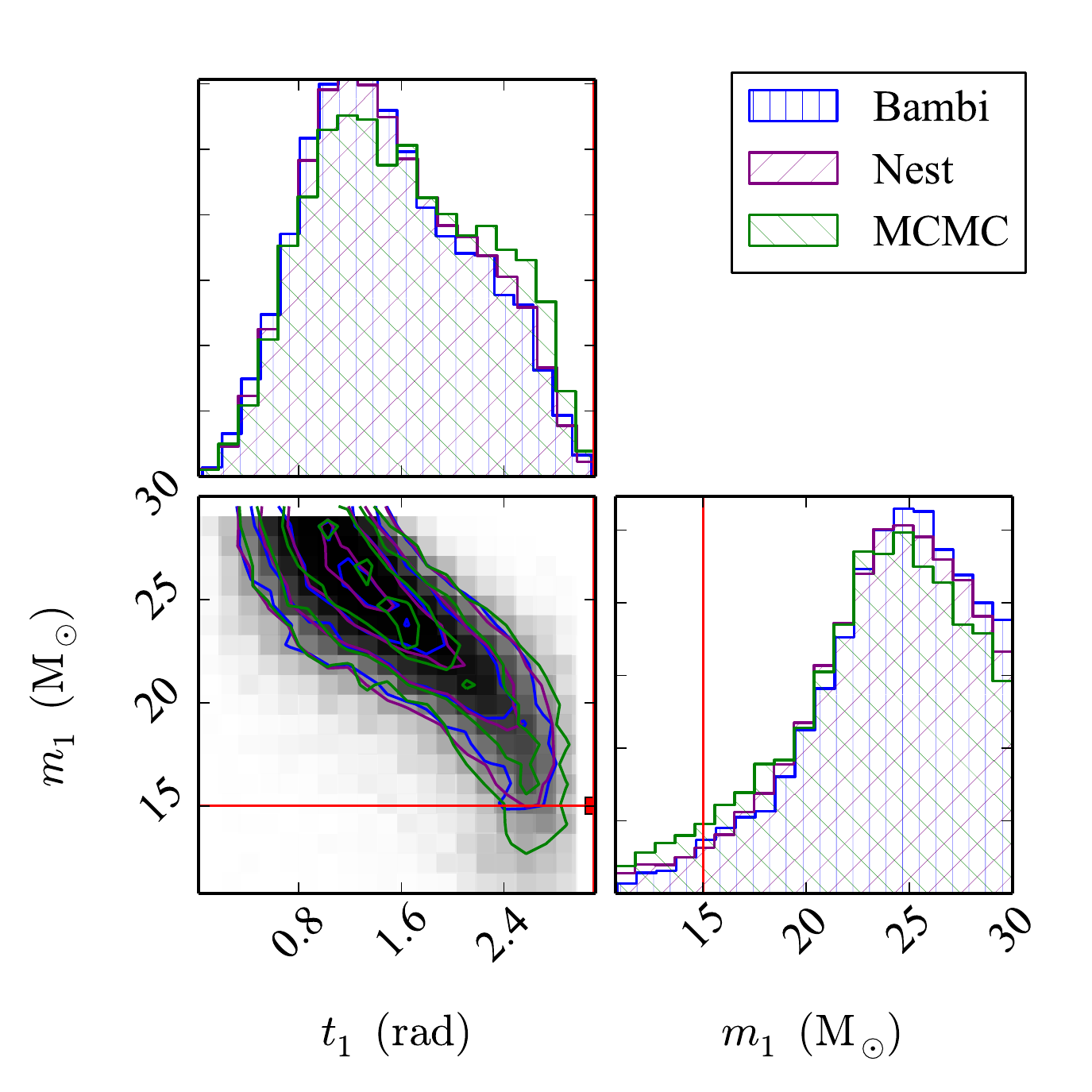}
\caption{\label{fig:BBH} Comparison of probability density 
        functions for the BBH signal (table \ref{tab:injections}), with same color scheme as fig~\ref{fig:BNS}.
        (first row left) The mass posterior distribution 
        parametrized by chirp mass and symmetric mass ratio. (first row centre) The location of the source on the sky. (first row right)
        The distance $d_L$ and inclination $\theta_{JN}$ of the source showing the degeneracy is broken, as in the NSBH case.
        (second row left) The spins magnitude posterior distribution. 
        (second row centre) The spin and mass of the most massive member of the binary 
        illustrating the degeneracy between mass and spin. (second row right) The spin and symmetric mass ratio.
        (third row left) The spins tilt posterior distribution. 
        (third row centre) The spin tilt of the more massive member of the binary and the symmetric mass ratio. 
        (third row right) The spin tilt and mass of the most massive member of the binary.}
\end{figure*}

Figures \ref{fig:BNS}, \ref{fig:NSBH} and \ref{fig:BBH} show two-dimensional
90\% credible intervals obtained by all three samplers on various combinations of
parameters.
Figure \ref{fig:BNS} (see table \ref{tab:injections}) shows the typical posterior structure for a \ac{BNS}
system. We show only three two-dimensional slices through the nine-dimensional (non-spinning)
parameter space, highlighting the most relevant parameters for an astrophysical analysis.
Selected one-dimensional 90\% credible intervals are shown in table \ref{tab:BNS}.
This is the least challenging of the three example signals, since we restrict the model to
non-spinning signals only. The posterior PDFs show excellent agreement between the sampling methods.
In the leftmost panel we show the recovered distribution of the masses, parametrised
by the chirp mass and symmetric mass ratio. This shows the high accuracy to which the chirp
mass can be recovered compared to the mass ratio, which leads to a high degree of correlation
between the estimated component masses. The domain of the prior ends at a maximum of $\eta=0.25$,
which corresponds to the equal mass configuration.
In the central panel we show the estimated sky location, which is well determined here thanks
to the use of a three-detector network.
In the rightmost panel, the correlation between the distance and inclination angle is visible,
as both of these parameter scale the effective amplitude of the waveform.
The reflection about the $\theta_{JN}=\pi/2$ line shows the degeneracy which is sampled
efficiently using the extrinsic parameter jump proposals~\ref{sec:proposals:extrinsic}.

Similarly to Figure \ref{fig:BNS}, Figure \ref{fig:NSBH} (see table \ref{tab:injections}) shows the posterior for a \ac{NSBH} 
system. This signal was recovered using a spin-aligned waveform model, and we show six two-dimensional slices of this 
eleven-dimensional parameter space. Selected one-dimensional 90\% credible intervals are shown in table \ref{tab:NSBH}.
The top-left panel shows the $\Mc-\eta$ distribution; in comparison to Figure \ref{fig:BNS} the mass ratio
is poorly determined. This is caused by the correlation between the $\eta$ parameter and the aligned spin magnitudes,
which gives the model greater freedom in fitting $\eta$, varying $a_1$ and $a_2$ to compensate. This correlation
is visible in the bottom-right panel. The other panels on the bottom row illustrate other correlations between the intrinsic
parameters.
The top-right panel shows the correlation between distance and inclination, where in this case the spins help break the degeneracy about the $\theta_{JN}=\pi/2$ line.

Lastly, figure \ref{fig:BBH} (see table \ref{tab:injections}) shows the posterior for a \ac{BBH} system, recovered taking into account precession effect from two independent spins. We show nine two-dimensional slices of this fifteen-dimensional parameter space. One-dimensional 90\% credible intervals are shown in table \ref{tab:BBH}.
In addition to the features similar to figure \ref{fig:BNS} in the top row, correlations with spin magnitudes (middle row) and tilt angles (bottom row) are shown. Note that the injected spin on the first component is almost anti-aligned with the orbital angular momentum, such that the tilt angle $t_1=3.1$, an unlikely random choice. This angle has a low prior probability, and as a result the injected value lies in the tails of the posterior distribution. This has repercussions in the recovered distributions for the spin magnitude and mass ratio, since they are partially degenerate in their effect on the phase evolution of the waveform, which results in the true value also being located in the tails of these distributions.

In all three cases, the three independent sampling algorithms converge on the same posterior distributions, indicating that the algorithms can reliably determine the source parameters,
even for the full 15-dimensional spinning case.

\begin{table*}
\begin{tabular}{|l||c|c|c|c|c|c|c|}
        \hline & $\mathcal{M} $ (\Msun)  & $\eta$ & $m_1$ (\Msun) & $m_2$ (\Msun) & $d$ (Mpc) & $\alpha$ (rad) & $\delta$ (rad) \\ 
 \hline \hline Nest & $1.1253^{1.1255}_{1.1251}$  & $0.2487^{0.25}_{0.2447}$  & $1.4^{1.5}_{1.3}$  & $1.2^{1.3}_{1.1}$  & $197^{251}_{115}$  & $3.19^{3.24}_{3.14}$  & $-0.997^{-0.956}_{-1.02}$ \\ 
MCMC & $1.1253^{1.1255}_{1.1251}$  & $0.2487^{0.25}_{0.2447}$  & $1.4^{1.5}_{1.3}$  & $1.2^{1.3}_{1.1}$  & $195^{250}_{113}$  & $3.19^{3.24}_{3.14}$  & $-0.998^{-0.958}_{-1.02}$\\ 
BAMBI & $1.1253^{1.1255}_{1.1251}$  & $0.2487^{0.25}_{0.2449}$  & $1.4^{1.5}_{1.3}$  & $1.2^{1.3}_{1.1}$  & $196^{251}_{114}$  & $3.19^{3.24}_{3.14}$  & $-0.998^{-0.958}_{-1.02}$ \\ \hline
Injected & 1.1253 & 0.2497 & 1/3382 & 1.249 & 134.8 & 3.17 & -0.97 \\
\hline 
 \end{tabular}
 \caption{\label{tab:BNS}
     \ac{BNS} recovered parameters. Median values and $5\% - 95\%$ credible interval for a selection of parameters for each of the sampling algorithms.}

\begin{tabular}{|l||c|c|c|c|c|c|c|c|c|}
        \hline & $\mathcal{M} (\Msun) $ & $\eta$ & $m_1$ (\Msun)  & $m_2$ (\Msun)  & $d$ (Mpc) & $a_1$ & $a_2$ & $\alpha$ (rad) & $\delta$ (rad) \\ 
 \hline \hline Nest & $3.42^{3.48}_{3.36}$  & $0.11^{0.23}_{0.076}$  & $11^{15}_{5.3}$  & $1.7^{2.9}_{1.4}$  & $612^{767}_{383}$  & $0.36^{0.75}_{0.041}$  & $0.49^{0.95}_{0.046}$  & $0.843^{0.874}_{0.811}$  & $0.459^{0.495}_{0.422}$ \\ 
MCMC & $3.42^{3.48}_{3.36}$  & $0.12^{0.23}_{0.077}$  & $11^{15}_{5.3}$  & $1.7^{2.9}_{1.4}$  & $601^{763}_{369}$  & $0.35^{0.73}_{0.038}$  & $0.48^{0.94}_{0.045}$  & $0.843^{0.874}_{0.812}$  & $0.459^{0.496}_{0.422}$ \\ 
BAMBI & $3.42^{3.48}_{3.37}$  & $0.11^{0.22}_{0.075}$  & $11^{15}_{5.8}$  & $1.6^{2.7}_{1.3}$  & $609^{767}_{378}$  & $0.36^{0.72}_{0.042}$  & $0.49^{0.95}_{0.044}$  & $0.843^{0.874}_{0.811}$  & $0.459^{0.495}_{0.422}$ \\ \hline
        Injected & 3.477 & 0.076 & 15 & 1.35 & 397 & 0.63 & 0.0  & 0.82 & 0.44 \\
\hline 
 \end{tabular} \caption{\label{tab:NSBH}\ac{NSBH} recovered parameters, defined as above.}

\begin{tabular}{|l||c|c|c|c|c|c|c|c|c|}
        \hline & $\mathcal{M}$ (\Msun)  & $\eta$ & $m_1$ (\Msun)  & $m_2$ (\Msun)  & $d$ (Mpc) & $a_1$ & $a_2$ & $\alpha$ (rad) & $\delta$ (rad) \\ 
 \hline \hline Nest & $9.5^{9.8}_{9.3}$  & $0.15^{0.217}_{0.12}$  & $24.3^{29.3}_{16.3}$  & $5.5^{7.7}_{4.7}$  & $647^{866}_{424}$  & $0.34^{0.66}_{0.082}$  & $0.48^{0.95}_{0.049}$  & $0.21^{0.29}_{0.14}$  & $-0.612^{-0.564}_{-0.659}$  \\ 
MCMC & $9.5^{9.8}_{9.3}$  & $0.15^{0.23}_{0.12}$  & $23.8^{29.1}_{14.8}$  & $5.5^{8.2}_{4.7}$  & $630^{847}_{404}$  & $0.36^{0.78}_{0.092}$  & $0.51^{0.95}_{0.05}$  & $0.21^{0.3}_{0.14}$  & $-0.612^{-0.563}_{-0.658}$  \\ 
BAMBI & $9.5^{9.8}_{9.3}$  & $0.149^{0.216}_{0.12}$  & $24.5^{29.2}_{16.3}$  & $5.4^{7.5}_{4.7}$  & $638^{859}_{428}$  & $0.35^{0.69}_{0.087}$  & $0.49^{0.94}_{0.049}$  & $0.21^{0.29}_{0.14}$  & $-0.612^{-0.565}_{-0.659}$ \\ \hline
        Injected & 9.44 & 0.227 & 15 & 8 & 500 & 0.79 & 0.77 & 0.230 & -0.617 \\
\hline 
 \end{tabular}
 \caption{\label{tab:BBH}\ac{BBH} recovered parameters, defined as above.}

\end{table*}

We also computed the evidence for each signal, relative to the Gaussian
noise hypothesis, using each sampler, with errors computed as in \S\,\ref{sec:analytic-likelihood}.
The results in table \ref{tab:AnalyticEvidence}
show that the two flavours of nested sampling produce more precise estimates,
		according to their own statistical error estimates, but they disagree in the
		mean value. The thermodynamic integration method used with the MCMC algorithm
		(with 16 steps on the temperature ladder), produces a larger statistical error
estimate, which generally encloses both the nested sampling and BAMBI estimates.
These results indicate that there remains some systematic disagreement between the
different methods of estimating evidence values, despite the good agreement between
the posteriors. The BAMBI method generally produces a higher evidence estimate compared
to the nested sampling approach, by around a factor of $e$. This indicates that further
improvement is necessary before we can rely on these methods to distinguish models which
are separated by evidence values lower than this factor.

\subsection{Confidence intervals} 

Having checked the agreement of the posterior distributions on three
selected injections, we performed a further check to ensure that the
probability distributions we recover are truly representative of the
confidence we should hold in the parameters of the signal.
In the ideal case that our noise and waveform
model matches the signal
and noise in the data, and our prior distribution matches the set of
signals in the simulations, then the recovered credible regions should
match the probability of finding the true signal parameters within that
region.
By setting up a large set of test signals in simulated noise
we can see if this is statistically true by determining the frequency
with which the true parameters lie within a certain confidence level.
This allows us to check that our credible intervals are well calibrated,
in the sense of~\cite{Dawid:1982}.

For each run we calculate credible intervals from the
posterior samples, for each parameter. We can then examine the number
of times the injected value falls within a given credible interval.
If the posterior samples are an unbiased estimate of the true probability,
then $10\%$ of the runs should find the injected values within a $10\%$
credible interval, $50\%$ of runs within the $50\%$ interval, and so on.

We perform a KS-test on whether the results match the expected 1 to 1
relation between the fraction of signals in each credible region, and
the level associated with that region.

For 1 dimensional tests our credible regions are defined as the connected region
from the lowest parameter value to the value where the integrated probability
reaches the required value. In practice we order the samples by parameter value
and query what fraction of this list we count before passing the signal value.

To perform this test, we drew 100 samples from the prior distribution of section \ref{sec:prior},
providing a set of injections to use for the test. This was performed using
the TaylorF2 waveform approximant for both injection and recovery, with simulated
Gaussian data using the initial LIGO and Virgo noise curves and 3 detector sites.

\begin{figure}
\centering
\includegraphics[width=\columnwidth]{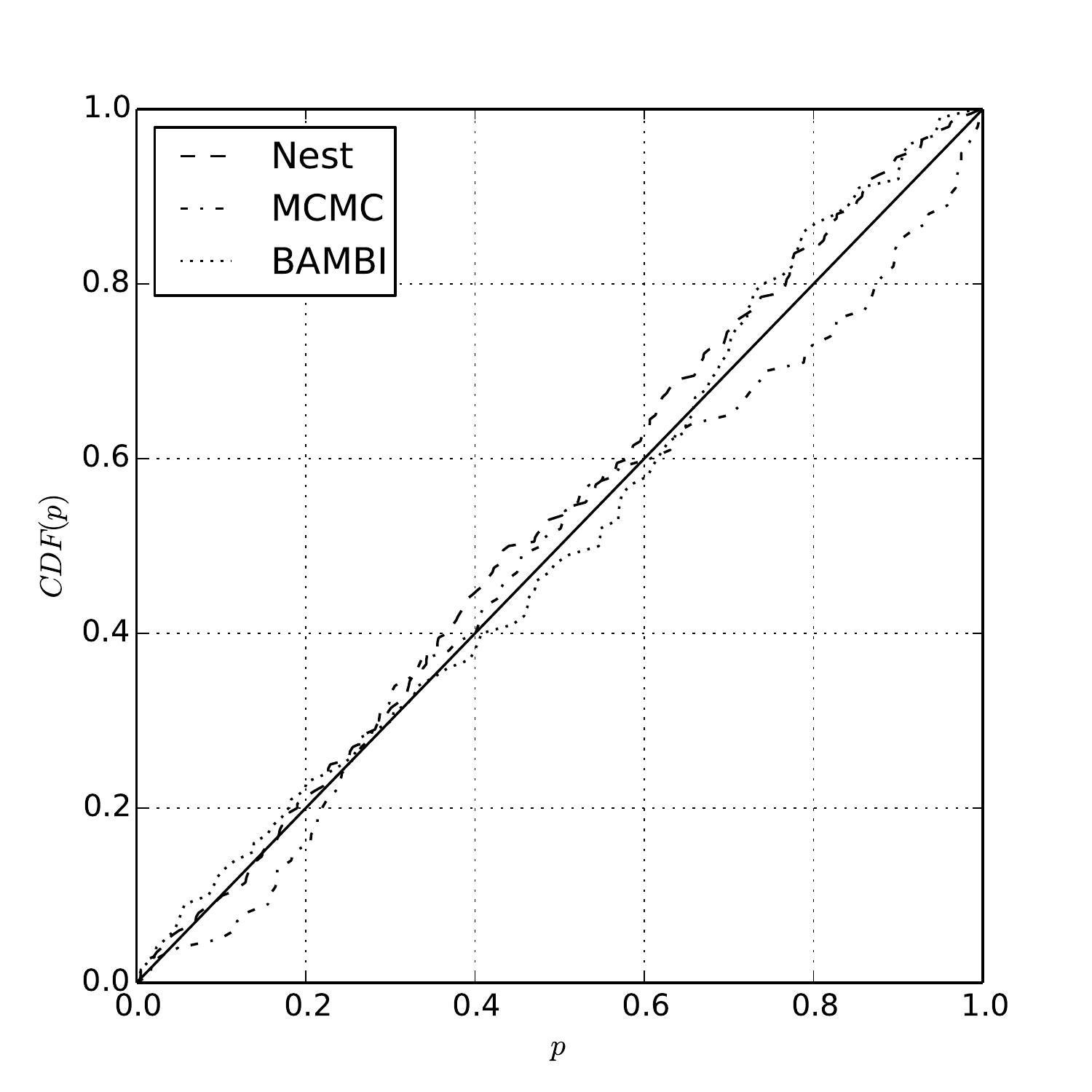}
\caption{\label{fig:PP} P vs P plot for the distance parameter. On the $x$ axis is the probability $p$ contained in a credible interval, and on the $y$ axis the fraction of true values which lay inside that interval. The diagonal line indicates the ideal distribution where credible intervals perfectly reflect the frequency of recovered injections. For all three sampling algorithms the results are statistically consistent with the diagonal line, with the lowest KS statistic being $0.25$.}
\end{figure}

We calculated the cumulative distribution of the number of times the true value
for each parameter was found within a given credible interval $p$, as a function
of $p$, and compared the result to a perfect $1-1$ distribution using a KS test.
All three codes passed this test for all parameters, indicating that our sampling
and post-processing does indeed produce well-calibrated credible intervals.
Figure \ref{fig:PP} shows an example of the cumulative distribution of p-values
produced by this test for the distance parameter. Similar plots were obtained
for the other parameters.

\section{Computational Performance}\label{sec:performance} 

We have benchmarked the three samplers using the three \ac{GW} 
events described in section \ref{sec:injections}. Although the specific performances
listed are representative only of these signals, they do provide a rough idea of
the relative computational performance of the sampling methods and the relative difficulty in
the BNS, NSBH and BBH analyses, when running in a typical configuration.
The computational cost of a parameter estimation run is strongly dependent on two main 
factors: the waveform family used (see sec.~\ref{sec:waveforms}) and the 
structure of the parameter space.
Profiling of the codes show that computation of waveforms is the dominating factor,
as the calculation of the phase evolution at each frequency bin is relatively expensive compared
to the computation of the likelihood once the template is known.

The computationally easiest waveform to generate is TaylorF2, where an analytic
expression for the waveform in the frequency domain is available.
For the BNS signal simulated here, around 50 waveforms can be 
generated per second at our chosen configuration ($32\,$s of data sampled at $4096\,$Hz).
On the other hand, more sophisticated waveforms, like SpinTaylorT4 with 
precessing spins, require solving differential equations in the time 
domain, and a subsequent FFT (the likelihood is always calculated in the 
frequency domain), which raises the CPU time required to generate a 
single waveform by an order of magnitude.

The structure of the parameter space affects the length of a run in 
several ways.
The first, and most obvious, is through the number of dimensions: when 
waveforms with precessing spins are considered a 15-dimension parameter 
space must be explored, while in the simpler case of non-spinning 
signals the number of dimensions is 9.
The duration of a run will also depend on the correlations present in 
the parameter space, e.g. between the distance and inclination 
parameters~\cite{2013PhRvD..88f2001A}. Generally speaking runs where 
correlations are stronger will take longer to complete as the codes will 
need more template calculations to effectively sample the parameter 
space and find the region of maximum likelihood.

Table \ref{tab:RunTimes} shows a comparison of the efficiency of each
code running on each of the simulated signals in terms of the cost in
CPU time, wall time, and the CPU/wall time taken to generate each sample which ended
up in the posterior distribution. These numbers were computed using
the same hardware, Intel Xeon E5-2670 2.6\,GHz processors.

We note that at the time of writing the three samplers have different level of parallelization, 
which explains the differences between codes of the ratio CPU time to wall time.

\begin{table}
\begin{tabular}{c|cccc}
BNS & Bambi & Nest & MCMC \\
\hline
posterior samples & 6890 & 19879 & 8363 \\
CPU time (s.) & 3317486 & 1532692 & 725367 \\
wall time (s.) & 219549 & 338175 & 23927 \\
CPU seconds/sample & 481.5 & 77.1 & 86.7 \\
wall seconds/sample & 31.9 & 17.0 & 2.9 \\
\hline
NSBH & Bambi & Nest & MCMC \\
\hline
posterior samples & 7847 & 20344 & 10049 \\
CPU time (s.) & 2823097 & 9463805 & 4854653 \\
wall time (s.) & 178432 & 2018936 & 171992 \\
CPU seconds/sample & 359.8 & 465.2 & 483.1 \\
wall seconds/sample & 22.7 & 99.2 & 17.1 \\
\hline
BBH & Bambi & Nest & MCMC \\
\hline
posterior samples & 10920 & 34397 & 10115 \\
CPU time (s.) & 2518763 & 7216335 & 5436715 \\
wall time (s.) & 158681 & 1740435 & 200452 \\
CPU seconds/sample & 230.7 & 209.8 & 537.5 \\
wall seconds/sample & 14.5 & 50.6 & 19.8 \\
\end{tabular}
\caption{\label{tab:RunTimes} Preformance of all three sampling methods 
on the three signals from table \ref{tab:injections}. The time quoted in the ``CPU time'' line is the
cumulative CPU-time across multiple cores, while the time quoted in the ``wall time'' line is the actual 
time taken to complete the sampling. The difference is an indication of the varying degrees of parallelism
in the methods.}
\end{table}

\section{Conclusions and future goals}
\label{sec:conclusions}

In this paper we have described the application of three stochastic sampling
algorithms to the problem of compact binary parameter estimation and model
selection. Their implementation in the \li\ package provides a flexible
and open-source toolkit which builds upon much previous work to give reliable results~ \cite{RodriguezEtAl:2013,2008ApJ...688L..61V,PhysRevLett.112.251101,Christensen:2001cr,MCMC:2004,2006CQGra..23.4895R,
2007PhRvD..75f2004R, 2008ApJ...688L..61V, 2008CQGra..25r4011V,
2009CQGra..26k4007R, 
2009CQGra..26t4010V, 2010CQGra..27k4009R,Veitch:2010,Feroz:2009,Feroz:2013, Graff:2012}.
The independent sampling methods have allowed us to perform
detailed cross-validation of the results of inference on a range
of \ac{GW} signals from compact binary coalescences, such as will be observed by future gravitational-wave detectors.
We have also performed internal consistency checks of the recovered posterior
distributions to ensure that the quoted credible intervals truly represent
unbiased estimates of the parameters under valid prior assumptions.

The release of the \li\ toolkit as part of the open-source \LAL\
package, available from \cite{LAL}, has already provided a base for developing methods for testing
general relativity~\cite{2011PhRvD..83h2002D, 2012PhRvD..85h2003L, Agathos:2013upa} and performing parameter estimation on a variety of other GW
sources\cite{Pitkin:2012yg,Logue:2012zw}. In the future we intend to further develop the implementation
to accommodate more sophisticated noise models for data analysis in the advanced detector
era. This will enable us to provide parameter estimation results which are robust
against the presence of glitches in the data, against time-dependent fluctuations
in the noise spectrum\cite{RoeverMeyerChristensen2011,Littenberg:2013,Littenberg:2014oda}, and will allow us to incorporate uncertainty in the calibration of the instruments.

Work is also ongoing in improving inference to incorporate systematic
uncertainties in the waveform models which affect estimates of intrinsic
parameters~\cite{S6pepaper}.  

Meanwhile, recent advances in reduced order modelling of the
waveforms and developments of surrogate models for the most expensive waveforms
should result in a dramatic improvement in the speed of parameter estimation~\cite{Field:2013cfa,Canizares:2013ywa,Purrer:2014fza,Canizares:2014fya}.  More intelligent proposal distributions also have the potential to reduce the
autocorrelation timescales in the MCMC and Nested Sampling algorithms, further
improving the efficiency of these methods.

The work described here should serve as a foundation
for these further developments, which will be necessary to fully exploit the science capabilities
of the advanced generation of gravitational-wave detectors, and produce parameter
estimates in a timely manner.

\section*{Acknowledgements}
The authors gratefully acknowledge the support of the LIGO-Virgo Collaboration
in the development of the \li\ toolkit, including internal
review of the codes and results. We thank Neil Cornish and Thomas Dent for
useful feedback on the manuscript.
The results presented here were produced using the computing
facilities of the LIGO DataGrid and XSEDE, including:
the NEMO computing cluster at the Center for Gravitation and
Cosmology at UWM under NSF Grants PHY-0923409 and
PHY-0600953; the Atlas computing cluster at the Albert Einstein Institute,
Hannover; the LIGO computing clusters at Caltech, Livingston and Hanford;
and the ARCCA cluster at Cardiff University.
\Cref{fig:BNS,fig:NSBH,fig:BBH} were produced with the help of triangle.py~\cite{triangle}.

JV was supported by
the research programme of the Foundation for Fundamental
Research on Matter (FOM), which is partially supported by
the Netherlands Organisation for Scientific Research (NWO),
and by the UK Science and Technology Facilities Council (STFC)
grant ST/K005014/1.
VR was supported by a Richard Chase Tolman fellowship at the California Institute of Technology (Caltech)
PG was supported by an appointment to the NASA Postdoctoral Program at the Goddard Space Flight Center,
administered by Oak Ridge Associated Universities through a contract with NASA.
MC was supported by the National Science Foundation Graduate Research Fellowship
Program, under NSF grant number DGE 1144152.
JG's work was supported by the Royal Society.
SV acknowledges the support of the National Science Foundation and the LIGO Laboratory. LIGO was constructed by the California Institute of Technology and Massachusetts Institute of Technology with funding from the National Science Foundation and operates under cooperative agreement PHY-0757058.
NC's work was supported by NSF grant PHY-1204371.
FF is supported by a Research Fellowship from Leverhulme and Newton Trusts.
TL, VK and CR acknowledge the support of the NSF LIGO grant, award PHY-1307020.
RO'S acknowledges the support of NSF grants PHY-0970074 and PHY-1307429, and the UWM Research Growth Initiative.
MP is funded by STFC under grant ST/L000946/1.

This is LIGO document number P1400152.

\bibliography{cbc-group,lalinf_ref}

\end{document}